\documentclass[%
twocolumn,
 superscriptaddress,
 amsmath,amssymb,
prb,
]{revtex4-1}

\usepackage{graphicx}% Include figure files
\usepackage{dcolumn}% Align table columns on decimal point
\usepackage{bm}% bold math
\usepackage{hyperref}% add hypertext capabilities
\usepackage{color}
\usepackage{braket}
\usepackage{subcaption}
\usepackage{mathtools}
\usepackage{bbold}

\begin{document}

\preprint{}

\title{Ab-initio Van der Waals electrodynamics: polaritons and electron scattering from plasmons and phonons in BN-capped graphene}

\author{Francesco Macheda}
\affiliation{Dipartimento di Fisica, Università di Roma La Sapienza, Piazzale Aldo Moro 5, I-00185 Roma, Italy}
%Dipartimento di Fisica, Università di Roma La Sapienza, Roma, Italy }%
\author{Francesco Mauri}
\affiliation{Dipartimento di Fisica, Università di Roma La Sapienza, Piazzale Aldo Moro 5, I-00185 Roma, Italy}
\affiliation{Istituto Italiano di Tecnologia, Graphene Labs, Via Morego 30, I-16163 Genova, Italy}%
\author{Thibault Sohier}
\affiliation{Laboratoire Charles Coulomb (L2C), Université de Montpellier, CNRS, Montpellier, France}


\begin{abstract}
% general
Plasmons and polar phonons are elementary electrodynamic excitations of matter. In 2d and at long wavelengths, they couple to light and act as the system polaritons.  They also dictate the scattering of charged carriers. Van der Waals heterostructures offer the opportunity to couple excitations from different layers via long-range Coulomb interactions, modifying both their dispersion and their scattering of electrons. Even when the excitations do not couple, they are still influenced by the screening from all layers, leading to complex dynamical interactions between electrons, plasmons and polar phonons.
% Model
We develop an efficient ab initio model to solve the dynamical electric response of Van der Waals heterostructures, accompanied by a formalism to extract relevant spectroscopic and transport quantities. Notably, 
we obtain scattering rates for electrons of the heterostructure coupling remotely with electrodynamic excitations.
We apply those developments to BN-capped graphene, in which polar phonons from BN couple to plasmons in graphene. We study the nature of the coupled excitations, their dispersion and their coupling to graphene's electrons.
% results
Regimes driven by either phonons or plasmons are identified, as well as a truly hybrid regime corresponding to the plasmon-phonon-polariton at long wavelengths. Those are studied as a function of the graphene's Fermi level and the number of BN layers.  In contrast with standard descriptions in terms of surface-optical phonons, we find that the electron-phonon interaction stems from several different modes. 
Moreover, the dynamical screening of the coupling between BN's LO phonons and graphene's electrons crosses over from inefficient to metal-like depending on the relative value of the phonons' frequency and the energetic onset of interband transitions.
While the coupling is significant in general, the associated scattering of graphene's carriers is found to be negligible in the context of electronic transport.
\end{abstract}

\maketitle

\section{Introduction}

Van der Waals heterostructures \cite{Castellanos-Gomez2022,Novoselov2016} (vdWH) have emerged as some of the most promising  ways to explore and exploit the properties of materials at the nanoscale \cite{Liu2016}. Stacking different layers of 2d materials offers the opportunity to tailor novel properties. Unsurprisingly, this comes with some challenges on the modelling side.
The electrodynamic response of a material, i.e. the response to a momentum and frequency-dependent electric potential $V(\mathbf{q}, \omega)$, dictates a wide range of physical properties. This is evident from the fundamental importance of the inverse screening function\cite{Keldysh2014} in describing optical and electronic properties.
The electrodynamic elementary excitations of the system are intrinsic, self-sustained collective modes driven by long-range electromagnetic interactions \cite{Nozieres1999}. 
%They obviously couple to external electric perturbations.
This work will focus on plasmons and polar phonons, that is collective oscillations of electronic or atomic plasma.
%This work will focus on plasmons, that is self-sustained plasma oscillations of electrons, and polar phonons, that is collective harmonic oscillations of atoms generating polarization. 
Other excitations such as excitons or magnons are out of scope. 

In 2d and at long wavelengths, longitudinal plasmons and phonons couple to light\cite{Basov2016,Low2017,Rivera2019}. The resulting light-matter quasiparticles,  polaritons, are usually discussed at momenta close to the light cone ($q_0 \gtrsim \omega /c $, where $c$ is the speed of light). However, highly-confined plasmon- and phonon-polaritons\cite{Chen2012,Fei2012,Yan2013,Brar2014,Woessner2015,Yoxall2015,Li2021b,Chen2023} in 2d materials have been observed at momenta $10$ to $100$ times larger. Those can be studied without accounting for relativistic retardation effects, as will be done here.
Electrodynamic collective modes in general are also a major source of scattering, via remote interactions between layers\cite{Tielrooij2018,Yang2018}.
A good understanding and predictive models for the emerging interlayer couplings between collective modes thus lead to the exciting prospect of engineering light-matter interactions\cite{Basov2016,Low2017} and electron scattering\cite{Sohier2021a} in future photonic and electronic devices based on vdWH.

To obtain polariton dispersions, one can resort to empirical and analytical approaches such as continuum electrodynamics \cite{Dai2014, PhysRev.140.A2076,PhysRev.182.539}, ultimately relying on parametrized models for the dielectric functions.
As a consequence, electron scattering involving coupled plasmons and phonons systems \cite{Kim1978,Sanborn1995,Fratini2008,Ong2012,Hauber2017a} is not treated from a fully microscopic point of view. 

In order to be predictive in complex systems, a microscopic ab-initio approach that directly simulates the electrodynamic response of vdWH is highly desirable. 
Unfortunately, direct ab initio calculations of vdWH including plasmonic and phononic excitations are costly. Indeed, lattice mismatches imply large supercells which in addition to the number of layers increases considerably the number of atoms in the simulation cell. 
Another strategy is to coarse-grain the layer properties and focus on interlayer interactions. The single layer quantities are easily obtained from first principles, and used in a model to build the properties of the multilayer. In the current framework, this amounts to extracting single layer dynamical responses from ab-initio, and model the multilayer response.

We build on a previous work based on this strategy in the static limit \cite{Sohier2021a} and add dynamical electronic and atomic contributions. With respect to similar efforts in the literature \cite{Andersen2015,Gjerding2020,Sponza2020,intVeld_2023}, the Van der Waals electrodynamics (VED) model presented here offers both technical and conceptual improvements.
%, additional physical insights and access to a wider range of relevant quantities.
Technical improvements include the full momentum-dependent phononic response, rather than resorting to the long-wavelength limit of Born effective charge models~\cite{Gjerding2020}, which notably misses the correct relative weight of electronic and phononic polarizabilities. This is key to match our results to density functional perturbation theory (DFPT) in the appropriate limits, in terms of both dispersion and intensity of the collective modes.
Conceptual progress stems from a simple yet powerful setup and exploitation of the formalism. 
We first build a matrix describing the response of given layer when a potential perturbation is applied to another layer. The imaginary part of this matrix yields the elementary excitations of the vdWH. From those, the response to an arbitrary external potential or charge density perturbation can be reconstructed and then projected on a given probe. 
When perturbation and probe are uniform over the layers, the model yields the basic spectroscopic properties of the slab, such as the Electronic Energy Loss Spectroscopy (EELS) cross Section in transmission. More complex quantities describing surface-probing experiments can also be extracted.
Importantly, by perturbing and probing with a charge density localized on a given layer, we extract scattering rates for electrons coupling with plasmons, phonons, and hybrid excitations \cite{Kim1978,Ong2012,Hauber2017a,Hai1997}. Such ab-initio calculation of interlayer electron scattering mechanism, including dynamical screening from electrons and phonons, opens exciting perspectives for the study of vdWH. Indeed, they can be used for transport calculations \cite{Sanborn1995,Ponce2021,Fischetti1991,Jhalani2020,Macheda2018,Hauber2017a,Brunin2020a,Macheda2020,Gopalan2022},  Raman scattering \cite{Piscanec2004,Venanzi2023,Gruneis2009}, excited carriers relaxation \cite{Kim2011,Marini2021,Harb2006,Bernardi2014,Betz2013,Tielrooij2018}, and superconductivity \cite{Bardeen1957,Schrieffer1999,Marsiglio2020}. 
\\
We focus on graphene encapsulated with multilayer hexagonal boron nitride, a prototypical system both for polaritonics and electronic transport.
It hosts both charge plasmons from graphene and polar phonons from BN. Their coexistence in the same energy range leads to their mixing and anti-crossing \cite{Brar2014,Dai2015,Ong2012}.
We present an extensive exploitation of the VED framework, discussing spectroscopic spectral functions and electron scattering including contributions from plasmons, phonons, and hybrid excitations.
We analyse the contributions of different layers to gain insight on the nature of the excitations (plasmon, phonon or hybrid), and propose a method to separate the corresponding scattering strengths. We study the impact of two main parameters for BN-capped graphene, i.e. the number of BN layers and the Fermi level (electron doping) of graphene, that can be tuned experimentally \cite{Sonntag2023,Yankowitz2019} to change the properties of the plasmons, phonons, and their scattering of electrons. We focus notably on the nature of the polaritonic states, and the dynamical screening of the remote coupling between graphene's electrons and BN's LO phonons. Gaining access to the microscopic origin of the interactions, we progress over state-of-the-art modelling and show that in realistic systems the remote coupling is mediated by several different LO phonon branches, and assess its impact on the graphene's resistivity value at room temperature.
\\
The paper is structured as follows. Section \ref{sec:VED} describes the theoretical and computational frameworks. Section \ref{Sec:CollModes} describes the collective modes of the BN/graphene systems. Section \ref{sec:BeforeCoupling} considers graphene's plasmons and BN's phonons before coupling, validating the method against known results and DFPT calculations.
Section \ref{sec:interplay} explores the nature of the excitations in the coupled system, as well as their dependence on the number of BN layers and graphene's Fermi level. Finally, Section \ref{sec:elecscatt} studies the coupling of graphene's electrons with those collective modes, and compares the associated scattering rates to those associated with graphene's intrinsic optical modes.

\section{Electrodynamic response of Van der Waals heterostructures}
\label{sec:VED}

The density-density response of a system at momentum $q$ and frequency $\omega$, $\chi(q, \omega)$, enters the determination of many physical quantities, from spectroscopic responses to intrinsic scattering mechanisms. This Section first shows how the Van der Waals ElectroDynamics (VED) model builds the response of a layered heterostructure from single layer responses.
Then, it shows how it can be used to obtain the collective modes dispersions, their EELS and surface-probes responses, and their coupling to electrons to deduce scattering rates.

\subsection{Van der Waals ElectroDynamics (VED) Model} 

\begin{figure}[h]
\includegraphics[width=0.98\columnwidth]{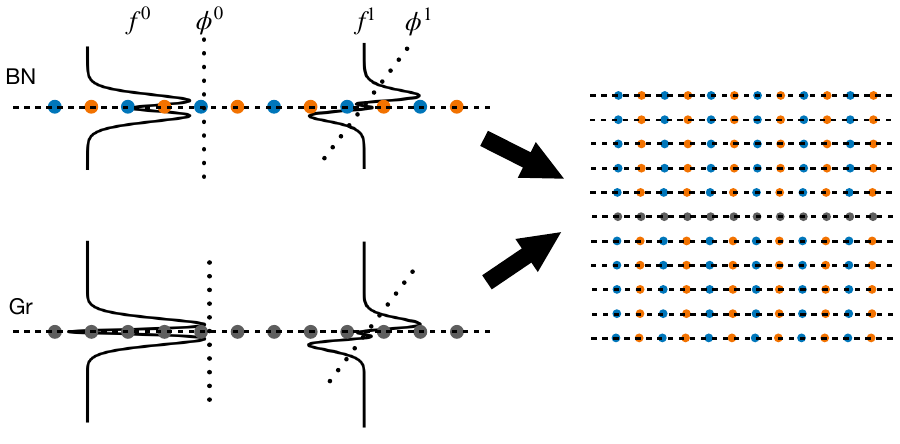}
\caption{Sketch of the Van der Waals electrodynamics (VED) model. Single layer monopole ($f^0$) and dipole ($f^1$) density responses to even ($\phi^0$) and odd ($\phi^1$) potentials are extracted from DFPT. This is then used to build the multilayer response to arbitrary potential or charge perturbation.}
\label{fig:vdWED}
\end{figure}

The VED is the dynamical generalisation of the model developed in Ref. \cite{Sohier2021a}. The general idea, sketched in Fig. \ref{fig:vdWED}, is to build the electrodynamic response of a layered Van der Waals heterostructure (vdWH) from the response of single layers. The following methodology is general, but we will specialize to BN-capped graphene as a prototypical system. The response of each different layer is simulated and parametrized from first principles. The static, clamped-ions electronic response is directly extracted from density functional perturbation theory (DFPT) calculations, therefore including local-fields and exchange-correlation effects. The dynamical contribution from free carriers is modelled within the Random Phase Approximation (RPA), equivalent to a time-dependent Hartree theory without local-fields. The dynamical atomic response is evaluated from DFPT ingredients. 
The ab-initio simulation of the whole vdWH, very costly even in the static Born-Oppenheimer approximation of DFPT, is avoided. Direct ab initio simulations of the interplay of dynamical contributions from plasmons and phonons would further require time-dependent density-functional theory, extremely costly in large supercells. \\
Here, the dynamical response of the full vdWH is obtained by solving a self-consistent set of equations that determine the potentials and charges on each layer. The computational burden of our method is limited to the single layer static DFPT response calculations needed for each different kind of layer, plus a small overhead to solve the coupled equations of the VED model \footnote{The code will be made public on gitlab after acceptance}. The single layer responses are easily parametrized and stored, such that for all vdWH combining layers in the database, only the VED coupled equations are left to solve.
The main approximation of this methodology is to neglect the wavefunction overlap between layers. This essentially comes down to an interlayer coarse-graining approximation where different layers see each other only through average macroscopic potentials. This approximation holds remarkably well in vdWH, where the interlayer bonds are weak by definition.

\subsubsection{General definitions}

The general aim is to simulate the macroscopic charge density response of the vdWH to a macroscopic external potential.
When working with screening/dielectric properties in 2d , it is preferable to Fourier transform only the in-plane space variables ($x,y \rightarrow \mathbf{q}$), while keeping the full out-of-plane $z$ dependence of the response. 
The external potential perturbation $V_{\textrm{ext}}(\mathbf{q}, \omega, z)$ is thus periodic in-plane and monochromatic at momentum $\mathbf{q}$ and frequency $\omega$. The total charge density response $\rho(\mathbf{q}, \omega, z)$ is determined by both the rearrangement of the electronic cloud and the displacement of the point-like ions. For both electrons and atoms, the responses are assumed to be isotropic in the plane, depending only on $q = |\mathbf{q}|$. This is usually a good approximation for many 2d material (e.g. hexagonal materials like BN and graphene). Within linear response theory, the charge density is obtained via the density-density response function $\chi$ as
\begin{align}
\rho(q, \omega, z) = \int dz' \chi(q, \omega,z, z') V_{\textrm{ext}}(q, \omega, z').
\label{eq:chiz}
\end{align}
The integration in the out-of-plane variable is meant to cover all the space, since $z$ is free of periodic boundary conditions. The connection between the total and the external potential is then given by the inverse dielectric matrix
\begin{align}
\epsilon^{-1}(q,\omega,z,z')&=\delta(z-z')+ \nonumber \\
&\frac{2\pi e^2}{q}\int dz'' e^{-q|z-z''|}\chi(q,\omega,z'',z').
\label{eq:epsm1z}
\end{align}
The possibility of having any kind of external potential has to be reduced in order to have a tractable problem, while at the same time maintaining the main physical features of the perturbations. Therefore, we find it useful to restrict the $z$ dependence of the external potential to a simplified functional form.  
As formalized in the following Section, we follow the approach of Ref. \cite{Sohier2021a}, which is an approximation of the exact method of Ref. \cite{Royo2021}.

\subsubsection{Dual basis set}
\label{sec:dualbasis}

To write the response problem in an easily solvable matrix form, the first step is to approximate the continuous out-of-plane variables of the response function of Eq. \ref{eq:chiz} with discrete indices over the layers. 
For each layer, $V_{\textrm{ext}}(z)$ is expanded over $N_b$ elements composing a subset of a complete basis set. The induced density can itself be expressed as a linear combination of (different) $N_b$ basis vectors. Eq. \ref{eq:chiz} then transforms into a matrix equation with  densities and potentials expanded over a dual basis set (in the same spirit as Ref. \cite{Andersen2015}). 
\\
We start by defining the mapping between the out-of-plane coordinate $z$ and the layer index. For $N_l$ layers,  $k\in \{1,..,N_l\}$ and 
\begin{align}
z \in [z_k-d/2,z_k+d/2]\rightarrow k.
\label{eq:boxes}
\end{align}
We assume that each layer has a finite thickness $d$  in the out-of-plane direction around its central coordinate $z_k$, within which the density response is fully contained. $d$ depends on the kind of layer but for simplicity, here, we take it to be layer independent. Within the `boxes' defined in Eq. \ref{eq:boxes}, we define the restricted ($N_b=2$) basis set for the external potential in the form
\begin{align}
\phi^i_k(z)=
\begin{cases}
(z-z_k)^i \quad z \in k \\
0 \quad z \notin k
\end{cases},
\label{eq:basisz}
\end{align}
with $i=0,1$. Analogously, we define the restricted basis set for the densities as 
\begin{align}
f^i_{k}(z)=
\begin{cases}
\mathcal{F}^i_{k}(z-z_k) \quad z \in k \\
0 \quad z \notin k \\
\end{cases},
\label{eq:basisprof}
\end{align}
which we refer to as profile functions. $\mathcal{F}^i_{k}(z)$ are the normalized out-of-plane profiles of the density response to the basis potentials of Eq. \ref{eq:basisz} when the $k$ layer is placed at $z_k=0$, as depicted in Fig. \ref{fig:vdWED}. They separate the density response into monopolar (z-symmetric, $i=0$) and dipolar (z-antisymmetric, $i=1$) components.
In principle, they might depend on $q$ and $\omega$. However, we have observed negligible dependence on $q$ and assume the same for $\omega$, as discussed in Section \ref{sec:Qandf}.
The profile functions (Eq. (A2) of Ref. \cite{Sohier2021a}) are normalized as
\begin{align}
\int dz f^i_{k}(z) \phi^{j}_{l}(z)= \delta_{ij}\delta_{kl}.
\label{eq:orthog}
\end{align}
The admitted potentials in our problem can now be written as
\begin{align}
V(q,\omega,z) = \sum_{ik}V^i_{k}(q,\omega)  \phi^{i}_{k}(z) , \label{eq:ketv}\\
V^i_{k}(q,\omega)=\int dz f^i_k(z)  V(q,\omega,z) . 
\end{align}
Similarly, for the density
\begin{align}
\rho_k(q,\omega,z)& = \sum_{i} \rho^i_k(q,\omega)  f^{i}_{k}(z), \\
\rho^i_k(q,\omega) &= \int dz \phi^i_k(z) \rho(q,\omega,z). 
\end{align}
Notice that in Eq. \ref{eq:chiz} and in the Random Phase Approximation (RPA), the induced density is the sum of two contributions due to the purely electronic polarization (`el') and the atomic (phonon) mediated one (`ph'). Since the observables that we study in this work are integrated quantities along $z$, it is convenient to approximate atomic polarization to have the same $z$-profile of the electronic one, with a small error as discussed later. In formulae
\begin{align}
\rho^i_k(q,\omega) & = \rho^i_{k,\textrm{el}}(q,\omega) + \rho^i_{k,\textrm{ph}}(q,\omega).
\end{align}

\subsubsection{Single layer response function}
\label{app:singlelayerresp}
We here consider the basic building block of our methodology, i.e. the response of each single layer to potential perturbations. As shown in Ref. \cite{Royo2021}, the out-of-plane dependence of the single layer density-density response function can be written in a separable form, i.e. with functions of only $z$ or $z'$. Accordingly, in this work we express the density-density response function for a single layer in the dual basis set as
\begin{align}
\left[\chi_{\textrm{1L}}\right]_k(q,\omega,z, z')=\sum_i Q^i_k(q,\omega)f^i_k(z) f^i_k(z'), \label{eq:chilayer} \\
Q^{i}_k(q,\omega)=\int dz dz' \phi^i_k(z) \left[\chi_{\textrm{1L}}\right]_k(q,\omega,z,z') \phi^i_k(z'),
\end{align}
%\begin{align}
%\chi_l(q,\omega,z, z') &= \chi^0_l(q,\omega,z, z') + \chi^1_l(q,\omega,z, z'), \label{eq:chilayer} \\ 
%\chi^0_l(q,\omega,z, z')&=Q^0_l(q, \omega) f^{0}_l(q,z)f^{0}_l(q,z') , \\
%\chi^1_l(q,\omega,z, z')&=Q^1_l(q, \omega) f^{1}_l(q,z) f^{1}_l(q,z'),
%\end{align}
where $Q^0_k$ and $Q^1_k$ are the amplitudes of the `monopole' and `dipole' response functions \footnote{We can now clarify the meaning of the dual basis set in mathematical terms: at each fixed $z$, $\chi_{\textrm{1L}}$ is an element of the dual space of the potentials. The separability of $\chi_{\textrm{1L}}$ and the property $\chi_{\textrm{1L}}(q,\omega,z,z')=\left(\chi_{\textrm{1L}}(q,\omega,z',z)\right)^*$ imply that the basis set of the densities is the same as of the dual space, justifying our nomenclature choice.}. Note that the off-diagonal matrix elements of $\chi_{\textrm{1L}}$  (i.e. $f^i_k(z) f^j_k(z')$ with $i \ne j$ terms) are null since for the single layers considered in this work we have in-plane mirror symmetry. To simplify notations, we will now drop the $(q,\omega)$ dependence when not essential. The density response for a single layer embedded in an heterostructure reads
\begin{align}
\rho^i_{k}= Q^i_k V^{i}_{k,\textrm{eff}},
\label{eq:rhoVeff}
\end{align}
where $V^{i}_{k,\textrm{eff}}$ is the effective potential felt by the embedded single layer. It is the sum of the actual external potential and of the tails of the macroscopic Hartree potential induced by all the other layers. $V^{i}_{k,\textrm{eff}}$ reduces to just the external potential if the single layer is suspended in vacuum.

\subsubsection{Interaction between layers} 

The process of obtaining the multilayer response from the single layer response functions above is described in Ref. \cite{Sohier2021a}. Only the main steps are summarized here.
The objective is to reduce the problem to a finite and limited number of elements, i.e. the layers, interacting with each other via simple interlayer couplings. Those interlayer couplings contain all the information about the out-of-plane structure of the system. The starting point is the Poisson equation for the potential induced by the charge variations 
\begin{align}
V_{\textrm{ind}}(q, \omega, z) = \frac{2\pi e^2}{q} \int dz'  e^{-q|z-z'|} \rho(q, \omega, z').
\label{eq:Poisson}
\end{align}
In the dual basis set introduced in this work, the matrix elements of the Coulomb interaction read
\begin{align}
\label{eq:CoulKernel}
    v^{ij}_{kl}(q) &= \frac{2\pi e^2}{q} F^{ij}_{kl}(q),  \\
    F^{ij}_{kl}(q) &= \int f^i_k(z) \int e^{-q|z-z'|}  f^j_l(z') dz dz',\nonumber 
\end{align}
which essentially describe the potential generated by a unit monopole ($j=0$) or dipole ($j=1$) charge-density on layer $l$ and projected on layer $k$. Eq. \ref{eq:Poisson} then becomes
\begin{align}
V^i_{k,\textrm{ind}}=\sum_{jl} v^{ij}_{kl}\rho^j_l.
\label{eq:Possionindex}
\end{align}
Notice that in the above equations the Coulomb potential is assumed to be frequency-independent. This assumption is valid if we restrict to wavevectors much larger than the light cone $q \gg \frac{ \omega}{c}$, i.e. if we can disregard relativistic retardation effects. The collective modes discussed in this work, including highly-confined plasmon-phonon-polaritons, are far from the light cone. In general though, relativistic retardation effects are important when discussing phonon-polaritons in 3D materials \cite{Grosso2000}, meaning that there will be a critical threshold for the thickness of any heterostructure after which the approximation of Eq. \ref{eq:CoulKernel} fails to describe the physics of phonon-polaritons, i.e. their wavevectors get near to the light cone.  

\subsubsection{Multilayer response function} 
\label{sec:multiresp}
Given the coupling of Eq. \ref{eq:CoulKernel}, one can build a set of linear equations describing the collective response of the layers \cite{Sohier2021a}.
As anticipated, each individual layer $k$ is assumed to respond to an effective external potential $V_{\textrm{eff}}$ which is the sum of the applied external potential plus the sum of the potentials induced by all other layers through Eq. \ref{eq:Possionindex}, i.e. to 
\begin{align}
V^i_{k,\textrm{eff}} &= V^i_{k,\textrm{ext}}+\sum_{l\neq k, j=0,1} v^{ij}_{kl} \rho^j_l.
\label{eq:Veff} 
\end{align}
Using Eq. \ref{eq:rhoVeff}, we can write a self-consistent system of equations for $\rho^i_k$:
\begin{align}
\rho^i_{k}=Q^i_k \left[ V^i_{k,\textrm{ext}}+\sum_{l\neq k, j=0,1} v^{ij}_{kl} \rho^j_l \right].
\label{eq:VofVmain}
\end{align}
Eq. \ref{eq:VofVmain} is composed of $2\times N_l$ equations to solve for $\rho^i_{k}$ at each $q$ and $\omega$, where $i= 0,1$  and $k = 0,\ldots,N_l$.
\\
Once the system is solved, one can determine the density-density response function of the full heterostructure. In fact, the matrix-form of the multi-layer density-density response function of Eq. \ref{eq:chiz} is
\begin{align}
\chi^{ij}_{kl}(q,\omega)=\int dzdz'\phi^i_k(z) \chi(q,\omega,z,z')  \phi^j_l(z'),
\label{eq:phichipchi}
\end{align}
and its value can be deduced from
\begin{align}
\rho^{i}_{k}=\sum_{jl}\chi^{ij}_{kl}V^j_{l,\textrm{ext}}.
\end{align}
Mind that the above equation is the matrix form of Eq. \ref{eq:chiz} written in the dual basis set introduced in Sec. \ref{sec:dualbasis}, and it is different from the single layer density-density response function of Eq. \ref{eq:chilayer}, because it relates the induced density on each layer to the external potential applied to the full heterostructure, rather than to the effective potential felt by each layer. 

% Finally, the `probe' and `perturbation' labels in the upper panel of Fig. \ref{fig:pertprob} refer to left and right $\phi$ of Eq. \ref{eq:phichipchi} respectively. The same terminology will be used for the couplings, where `probe' and `perturbation' refer to profile functions, as shown in the lower panel of Fig. \ref{fig:pertprob}.
\subsubsection{Computation of single layer response}
\label{sec:Qandf}
While the basis set in which potentials are decomposed are simple analytical functions, the profile functions used as a basis set for densities are material dependent and computed in DFPT, as explained in App. \ref{app:compdets}. As anticipated, their dependence in $q$ is very mild. Including it for graphene and BN brings less than 1\% impact on the observables treated in this work, with respect to taking the $q=0$ value. We therefore neglect it.

The fundamental building block of the VED method is the single layer density response function to a potential perturbation. As anticipated, in RPA it is split in purely electronic (clamped-ion) and atomic-mediated contributions (total minus clamped-ion) as
\begin{align}
    Q^i(q,\omega) &= Q^i_{\textrm{el}}(q,\omega)+Q^i_{\textrm{ph}}(q,\omega),
    \label{eq:Q0split}
\end{align}
where we have dropped the layer index for simplicity, as we will do for the rest of this Section. In this work, we will only consider dynamical contributions from plasmons in metals (doped graphene) and longitudinal polar-optical LO phonons in semiconductors or insulators (BN). They contribute to the monopolar component of the electronic and atomic contributions to the response, respectively. The dipolar atomic contribution is neglected, while the electronic dipolar contribution is evaluated in its static limit. In other words, we assume:
\begin{align}
Q^0(q,\omega) &\simeq 
    \begin{cases}
    Q^0_{\textrm{el}}(q,\omega = 0)+Q^0_{\textrm{ph}}(q,\omega) \text{ (semiconductors)} \\
    Q^0_{\textrm{el}}(q,\omega) \text{ (metals)}
    \end{cases},
    \label{eq:Q0} \\
Q^1(q,\omega) &\simeq Q^1_{\textrm{el}}(q, \omega=0).
\label{eq:Q1}
\end{align}
In the general case, the dipolar atomic contribution stemming from the out-of-plane response of ZO phonons would appear in Eq. \ref{eq:Q1}. Neglecting it is a safe approximation in the main energy range of interest for this work, i.e. around the BN's LO phonons energy. Indeed, $Q^1(q,\omega_{\textrm{LO}})=Q^1_{\textrm{el}}(q,0)\gg Q^1_{\textrm{ph}}(q, \omega_{\textrm{LO}})$, as numerically shown in App. \ref{app:Q1appr}. For energies close to the BN's ZO phonon ($\omega_{\textrm{ZO}} \sim \omega_{\textrm{LO}}/2$), the approximation does not hold, since the response is dominated by ZO phonons. We choose to neglect ZO phonons in this work. However, one might wonder if the ZO phonon's response could lead to significant alterations of the plasmon response. This is not the case for the doping levels studied in this work, since the ZO coupling to graphene's plasmons is much weaker than for LO, as also noted in Ref. \cite{Gjerding2020}. This weaker coupling can be understood as follows. The electric field generated by graphene's plasmon extend to other layers as $e^{-q|z - z_{\textrm{Gr}}|}$. Since the plasmon dispersion is at relatively small momenta, this generated potential is fairly flat and its projections on the dipole profile functions of BN layers are small (compared to the monopole ones). The ZO mode being associated with the dipole part of the atomic response $Q^1_{\textrm{ph}}$, it will only be weakly coupled to the plasmon. Neglecting the atomic dipolar term in Eq. \ref{eq:Q1} is therefore justified also for the determination of the plasmon dispersion.

The static ($\omega = 0$) electronic contributions of Eqs. \ref{eq:Q0} and \ref{eq:Q1} are parametrized from DFPT calculations, as detailed in App. \ref{app:compdets} and Ref. \cite{Sohier2021a}, while the dynamical ones are modelled as follows. For graphene, we need $Q^0_{\textrm{el}}(q, \omega)$. One can include the dynamical contribution from free carriers by using the RPA through the non-interacting (irreducible) polarizability, as
\begin{align}
Q^0_{\textrm{el}}(q,\omega) &= \frac{\chi_{\textrm{irr}}(q,\omega)}{1- v^{00}_{\textrm{Gr,Gr}} \chi_{\textrm{irr}}(q,\omega)}, \label{eq:Q0elrpa} \\
\label{eq:Qelfree}
\chi_{\textrm{irr}}(q,\omega) &= \frac{2}{(2\pi)^2} \sum_{nm}  \int d^2\mathbf{k}  \frac{n^{\textrm{FD}}_{\varepsilon_{n\mathbf{k}}} - n^{\textrm{FD}}_{\varepsilon_{m\mathbf{k}+\mathbf{q}}}}{\varepsilon_{n\mathbf{k}} - \varepsilon_{m\mathbf{k}+\mathbf{q}} + \hbar \omega + i \hbar\eta_{\textrm{pl}}} \\
& \times |\braket{u_{n\mathbf{k}}|{u_{m\mathbf{k+q}}}}|^2, \nonumber
\end{align}
where $n^{\textrm{FD}}_{\varepsilon_{n\mathbf{k}}}=\frac{1}{e^{(\varepsilon_{n\mathbf{k}}-\varepsilon_{\textrm{F}})/k_{\textrm{B}}T}+1}$ is the Fermi-Dirac occupation function for state of momentum $\mathbf{k}$ in band $n$ with energy $\varepsilon_{n\mathbf{k}}$ and Bloch periodic function $u_{n\mathbf{k}}$, $\varepsilon_{\textrm{F}})$ being the Fermi energy, $k_{\textrm{B}}$ the Boltzmann constant and $T$ the temperature. There is a factor 2 for spin degeneracy. The plasmon linewidth $\eta_{\textrm{pl}}$ is an external parameter in the model.
In the case of graphene considered in this work, we simply compute the above $\chi_{\textrm{irr}}$ semi-numerically in the Dirac cone model, with a Fermi velocity that includes GW corrections \cite{Sohier2014}. The wavefunction overlap is $|\braket{u_{n\mathbf{k}}|{u_{m\mathbf{k+q}}}}|^2 = \frac{1+ nm \cos\theta}{2}$ with $n,m=\pm 1$ for $\pi, \pi^*$ bands and $\theta$ the angle between $\mathbf{k}$ and $\mathbf{k}+\mathbf{q}$. Deviations from the Dirac cone model are assumed negligible for the frequencies and doping levels studied in the following. The effect of other band levels beside $\pi$ and $\pi^*$ is also neglected because their contribution to metallic screening is very weak \cite{PhysRevB.91.165428}.
\\
For semiconducting materials with polar phonons, long-range Coulomb interactions induce macroscopic electric fields in the crystal. The microscopic responses to such fields add up in a contribution $Q^0_{\textrm{ph}}(q, \omega)$ that can be expressed in terms of macroscopic quantities only\cite{Royo2021}. We deduce $Q^0_{\textrm{ph}}(q,\omega)$ from the atomic contribution to the inverse dielectric function
\begin{align}
\left[\epsilon^{-1}_{\textrm{1L}}\right]^i(q,\omega)=\int dzdz' \phi^i(z) \epsilon^{-1}_{\textrm{1L}}(q,\omega,z,z') \phi^i(z').
\end{align}
For BN's single layer, the monopolar contribution may be written as (see App. \ref{app:invdieltens2d})
\begin{align}
\label{eq:epsatom} 
\left[\epsilon^{-1}_{\textrm{1L}}\right]^0(q,\omega) = \left[\epsilon^{-1}_{\textrm{1L,el}}\right]^0(q) \left( 1+\frac{\mathbf{e}_{\textrm{LO} q} \mathcal{D}^{\textrm{L}}(q) \mathbf{e}^*_{\textrm{LO} q}}{(\hbar\omega+i\hbar\eta_{\textrm{LO}})^2-\hbar^2\omega^2_{\textrm{LO} q}} \right), 
\end{align}
where $\mathbf{e}_{\textrm{LO}} \mathcal{D}^{\textrm{L}} \mathbf{e}^*_{\textrm{LO}}$ is the long-range, in-plane polar contribution to the dynamical matrix projected on the eigenvector of mode $\textrm{LO}$ \cite{Macheda2023}, and 
\begin{align}
\left[\epsilon^{-1}_{\textrm{1L,el}}\right]^0(q) = 1+ v^{00}_{\textrm{BN,BN}}(q) Q^0_{\textrm{el}}(q)
\end{align} 
is the static electronic dielectric function. The expression for $\mathbf{e}_{\textrm{LO} q} \mathcal{D}^{\textrm{L}} (q) \mathbf{e}^*_{\textrm{LO} q}$ where the profile of the atomic polarizations are approximated with the electronic ones is given in Eq. \ref{eq:exacteDe2}. We found that a simpler yet more accurate approximation (discussed in App. \ref{app:invdieltens2d}) is to deduce $\mathbf{e}_{\textrm{LO} q} \mathcal{D}^{\textrm{L}} (q) \mathbf{e}^*_{\textrm{LO} q}$ from the difference of the squared frequencies of the LO phonon with respect to the TO phonon 
\begin{align}
 \mathbf{e}_{\textrm{LO} q} \mathcal{D}^{\textrm{L}}(q) \mathbf{e}^*_{\textrm{LO} q} = \hbar^2\omega^2_{\textrm{LO}q} - \hbar^2\omega^2_{\textrm{TO}q}. % \times F^{00}_{\textrm{BN}}(q).
\label{eq:omLO}
\end{align}
where $\omega_{\textrm{LO}q}$ and $\omega_{\textrm{TO}q}$ are computed via direct DFPT calculations. In particular, for BN the single layer LO mode has a DFPT frequency of $\omega_{\textrm{LO}q} \sim 1500$ cm$^{-1}$. The phonon linewidth $\eta_{\textrm{LO}}$ is an external parameter for our model, that we take as a constant ($\sim 1$ meV). We finally get
\begin{align}
\left[\epsilon^{-1}_{\textrm{1L}}\right]^0(q,\omega) &= 1+ v^{00}_{\textrm{BN,BN}}(q) \left[Q^0_{\textrm{el}}(q)+ Q^0_{\textrm{ph}}(q,\omega)\right],
\end{align}
with
\begin{align}
Q^0_{\textrm{ph}}(q, \omega) = \frac{\hbar^2\omega^2_{\textrm{LO}q} - \hbar^2\omega^2_{\textrm{TO}q}}{(\hbar\omega+i\hbar\eta_{\textrm{LO}})^2-\hbar^2\omega^2_{\textrm{LO}q}} \nonumber \times \\
\left[\epsilon^{-1}_{\textrm{1L,el}}\right]^0(q) \frac{1}{v^{00}_{\textrm{BN,BN}}(q)}.
\label{eq:Q0phlr}
\end{align}
The above expression contains all in-plane multipolar orders of the charge density expansions, going beyond the leading order from Born effective charges at $q \to 0$, an improvement with respect to Ref. \cite{Gjerding2020}. Notice that Eq. \ref{eq:Q0phlr}, as well as Eq. \ref{eq:Q0elrpa}, contain the Coulomb kernel in the profile basis function, instead of its $q \rightarrow 0$ form $2 \pi e^2/ q$ used in Ref. \cite{Gjerding2020}. While asymptotically equivalent at $q \to 0$, this has significant quantitative impacts on the atomic response, phonon dispersions and electron-phonon couplings as $q$ increases, as shown in Fig. \ref{fig:Qph_compa}.
\\
As a final comment, we remind that the disregard of dynamical effects in Eq. \ref{eq:Q1} means that we cannot study the out-of-plane plasmons of the heterostructures, which we leave for future studies. 

\subsection{Measurable quantities}

\subsubsection{Density-density response}
\label{sec:densdensresp}
\begin{figure}[h]
\includegraphics[width=0.8\columnwidth]{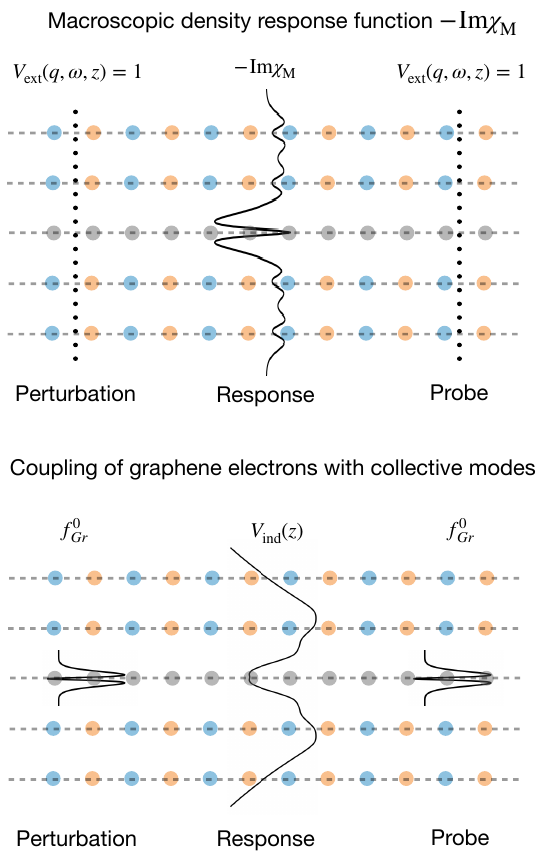}
\caption{Sketch of the computation of the macroscopic $\chi_{\textrm{M}}$ and of the coupling between electrons and collective modes in the VED framework.
On the top, only the monopole part of the density-density response $\chi_{\textrm{M}}$ is used, since the projection of the dipole part on the probe vanishes. On the bottom, the induced potential response is plotted, which drives the coupling of electrons. The plotted curves are computed with the VED method in 2BN/Gr/2BN for relatively large momentum at $\omega\sim\omega_{LO}$.}
\label{fig:pertprob}
\end{figure}

The density-density response function of the heterostructure $\chi$ defined in the previous Section can be used to evaluate several fundamental response functions. For example, in the transmission setup of Electronic Energy Loss Spectroscopy (EELS), one is interested in the scattering of an electron-beam with all the possible excitations of the system. An electron in the beam of energy $E_i=\hbar^2q_i^2/2m_e$ (with a typical order of magnitude of $\sim$ 10 keV) is scattered by the material, that absorbs a quantity of energy $\omega$ through the excitation of internal degrees of freedom. The Stokes EELS cross-section is then proportional to \cite{Nazarov2015,Sturm1993}
\begin{align}
\frac{d^2\sigma}{d\Omega d\omega}(q,\omega)\propto\frac{-1}{\big(q^2+q_z^2\big)^2}\times \nonumber \\
\left[1+n^{\textrm{BE}}_{\omega}\right]\textrm{Im}\Big[\int dzdz' e^{iq_z(z-z')}\chi(q,z,z',\omega)\Big],
\label{eq:scattcross}
\end{align}
where $ q_z=q_z(q,\omega) = q_i -\sqrt{q_i^2-q^2-2m_e\omega/\hbar}$, $m_e$ being the electron mass, and $n^{\textrm{BE}}_{\omega}=\frac{1}{e^{\omega/k_{\textrm{B}T}}-1}$ is the Bose-Einstein statistical distribution. The $q$-dependent prefactors of Eq. \ref{eq:scattcross} are not important to the aims of this work, and we will drop them in the following. We will also drop $n^{\textrm{BE}}_{\omega}$, since it can be easily re-inserted a posteriori in the evaluation of the cross-section. Then, in the approximation where $q_z$ is negligible---as in typical EELS experiments on heterostructures that are not too thick---the cross section is proportional to
\begin{align}
 \frac{d^2\sigma}{d\Omega d\omega}(q,\omega)\propto -\textrm{Im}\left[ \chi_{\textrm{M}}(q,\omega) \right],\\
 \label{eq:chitot}
\chi_{\textrm{M}}(q, \omega) =  \int dz dz'\chi(q, \omega, z, z'). 
\end{align}
`M' stands for `macroscopic', since $\chi_{\textrm{M}}$ is the in-plane average of the full heterostructure's density-density response integrated along the out-of-plane coordinates.
Eq. \ref{eq:chitot} is rewritten, in our formalism, as
\begin{align}
\chi_{\textrm{M}}(q, \omega)=\sum_{kl}  {\chi}^{00}_{kl}(q,\omega),
\end{align}
i.e. as the response to an external potential perturbation of the form $V_{\textrm{ext}}(q,\omega,z) = \sum_k \phi^0_k(z)$. Consistently with the neglect of $q$-dependent prefactors, in this work we will always normalize $\chi$-related quantities at each $q$ to their maximum, unless otherwise stated. In doing so, the plots of  $-\textrm{Im}\chi_{\textrm{M}}$ or EELS scattering cross Sections become equivalent \cite{Senga2019}.
\\
The peaks of $-\textrm{Im}\chi_{\textrm{M}}$ in the $(q,\omega)$ plane determine the collective modes of the system that are symmetrical with respect to the out-of-plane centre of the heterostructure. The antisymmetric modes instead average to zero when integrated in the out-of-plane direction, and are therefore dark in an EELS experiment. The antisymmetric modes are though visible in the total spectral function of the density-density response, defined as
\begin{align}
\chi_{\textrm{Tr}}(q,\omega)=\sum_{ik} \chi^{ii}_{kk}(q,\omega),
\label{eq:spectfunc}
\end{align}
%where we have dropped the specification $i=0,1$, and will do the same for the rest of this work unless otherwise stated. 
Minus the imaginary part of Eq. \ref{eq:spectfunc} in fact contains all the collective excitations of the system, and it is well defined even where modes cross.\\
More complex response functions can be defined from the $\chi$ matrix, in particular for perturbations that are non uniform over the layers or containing finite projections on the dipole response. As an example, one could consider evanescent waves as a way to reveal plasmons of graphene on a substrate \cite{Fei2012,Chen2012}. We here propose a simple modellization for such kind of measurement. We assume that we have a tip that can both induce exponentially suppressed potentials in the heterostructure and measure averages of the induced potential on a given surface layer S. In our formalism, the form of the evanescent (`ev') external potential has components
\begin{align}
V^i_{k,\textrm{ev}}(q,\omega)=\int dz f^i_k(z)  e^{-q|z-z_\textrm{S}|},
\end{align}
where $z_\textrm{S}$ is the central coordinate of the surface layer. Using Eq. \ref{eq:Possionindex}, the average of the potential on the surface layer, i.e. its monopolar component, is proportional to
\begin{align}
V^0_{\textrm{S},\textrm{ind}}=\sum_{ik} v^{0i}_{\textrm{S}k}(q)\rho^i_k(q,\omega)=\sum_{ijkl}v^{0i}_{\textrm{S}k}\chi^{ij}_{kl}(q,\omega) V^j_{l,\textrm{ev}}(q,\omega).
\end{align}
From this induced potential we can define a `local' density-density response via the Coulomb kernel
%, that is a $q$-dependent prefactor that will not show up in our normalized plot, via
\begin{align}
\chi_{\textrm{loc}}(q,\omega)=V^0_{\textrm{S},\textrm{ind}}(q,\omega)/v^{00}_{\textrm{SS}}(q).
\end{align}
More realistic and complicate descriptions are possible, but they go beyond the scope of this work. We end this Section reminding that from $\chi$ one can also derive the sheet optical conductivity, to define figures of merit such as the propagation quality factor, i.e. the distance (in number of wavelength) traveled before decay, and reflectivity \cite{Dai2014,Chen2023}.

\subsubsection{Scattering rates}

The electrons of the heterostructure experience scattering from collective modes. Understanding and quantifying those processes is essential to characterise the relaxation of the electron energy and momentum.  
In the single layer case, the squared modulus of the electron-phonon and/or electron-plasmon couplings describing this scattering are obtained \cite{Mahan1990,Kim1978} via the product of the imaginary part of the inverse total dielectric function, containing electronic and atomic contributions, and the Coulomb kernel, as shown in App. \ref{app:scatt}. For the multilayer, if we only assume that the matrix elements of $(z-z_k)$ between Bloch states is small for every $k$, then the scattering rate for an electron in the Bloch state $n\mathbf{k}$ reads (App. \ref{app:scatt} )
\begin{align}
\tau^{-1}_{n\mathbf{k}}=\frac{2\pi}{\hbar} \frac{1}{N_{\mathbf{q}}}\sum_{\mathbf{q}}\int^{\infty}_{-\infty} d\omega \mathcal{A}(\omega) \delta(\hbar\omega+\varepsilon_{m\mathbf{k+q}}-\varepsilon_{n\mathbf{k}}) \times \nonumber \\
\sum_{kl}\langle u_{n\mathbf{k}}|\phi^0_k|u_{m\mathbf{k+q}}\rangle \langle u_{m\mathbf{k+q}}|\phi^0_l|u_{n\mathbf{k}}\rangle
g^2_{kl}(q,\omega), \nonumber \\
\mathcal{A}(\omega)= \left[n^{\textrm{BE}}_{\omega}+\frac{1}{2}+\textrm{sign}(\omega)\frac{1}{2}-\textrm{sign}(\omega)n^{\textrm{FD}}_{\varepsilon_{m\mathbf{k+q}}}\right],
\label{eq:lifetime}
\end{align}
where $N_{\mathbf{q}}$ is the number of points in the $\mathbf{q}$-grid,  while $g^2_{kl}$ is
\begin{align}
\label{eq:g2main}
g^2_{kl}(q, \omega) = \textrm{sign}(\omega)\sum_{ijk'l'} \frac{-1}{A \pi} \textrm{Im}\left[   v^{0i}_{kk'}(q) \chi^{ij}_{k'l'}(q, \omega)  v^{j0}_{l'l}(q) \right].
\end{align}
$A$ is the area of the unit cell of the single layer, taken equal for all the layers in this work for simplicity. Notice that the $\textrm{sign}(\omega)$ term is included in the definition of $g^2$ so that it doesn't become negative with $\omega\rightarrow -\omega$, due to the property Eq. \ref{eq:retcc}. Scattering times are therefore always positive definite. Classically, the coupling $g^2_{kl}$ is the imaginary part of the potential projected on layer $k$ when layer $l$ is perturbed with a normalized charge density perturbation, as sketched in the bottom panel of Fig. \ref{fig:pertprob}.

The interaction of electrons of the heterostructure with phonons, plasmons, and any hybrid collective modes is entirely contained in the response of the system. In practice, it is mediated via every kind of collective modes contained inside $\chi$.
Only in certain limits, i.e. away from plasmon-phonon hybridization, can the above coupling be formally identified with electron-phonon or electron-plasmon couplings, see discussion in App. \ref{app:scatt}. 
\\
Note also that $g^2_{kl}$ is the interaction stripped of the Bloch functions overlaps, while the scattering rate expression includes the overlaps. For BN, Bloch overlaps can be approximated to unity or zero at small $q$. For graphene, this would be notably wrong. Indeed, the wavefunctions overlap strongly depends on the angle between the momenta of the two electrons. We will nonetheless still study the coupling $g^2_{kl}$ in the following, and use Eq. \ref{eq:lifetime} with the overlap given under Eq. \ref{eq:Qelfree} while studying the impact of remote coupling on graphene's electron scattering rates.

\section{Collective modes}
\label{Sec:CollModes}

In this section, the VED formalism is used to obtain the electrodynamic excitations of graphene and BN, and their interplay within in the heterostructure. The more general term "collective mode" is also used, with the understanding that in the current context, they are the one driven by long-range electromagnetic interactions.
The treatment of the coupling between electrons and collective modes is treated in Sec. \ref{sec:elecscatt}.

\subsection{Graphene's plasmons and BN's phonons before coupling}
\label{sec:BeforeCoupling}
\begin{figure*}[t!]
\includegraphics[width=0.36\textwidth]{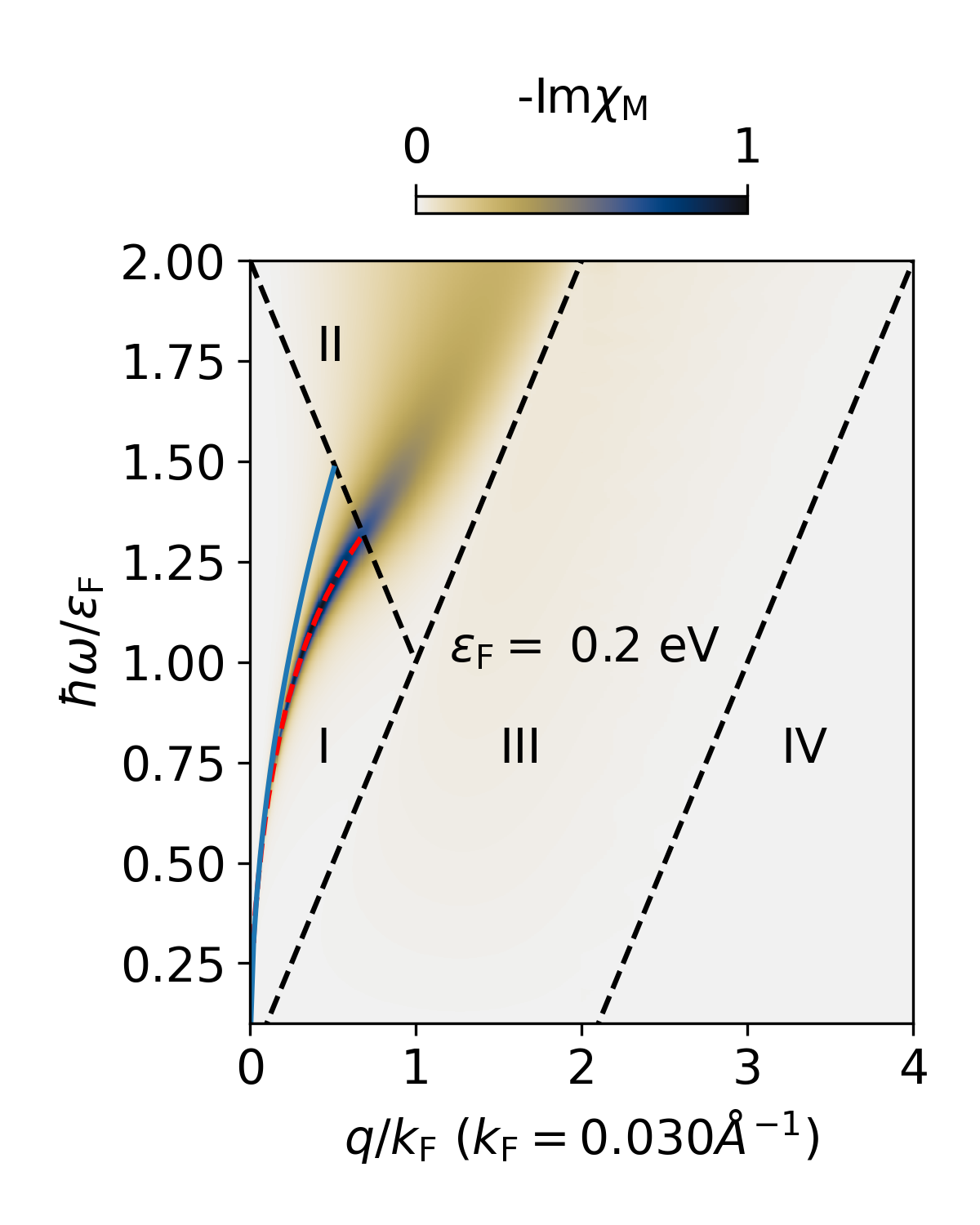}
\includegraphics[width=0.63\textwidth]{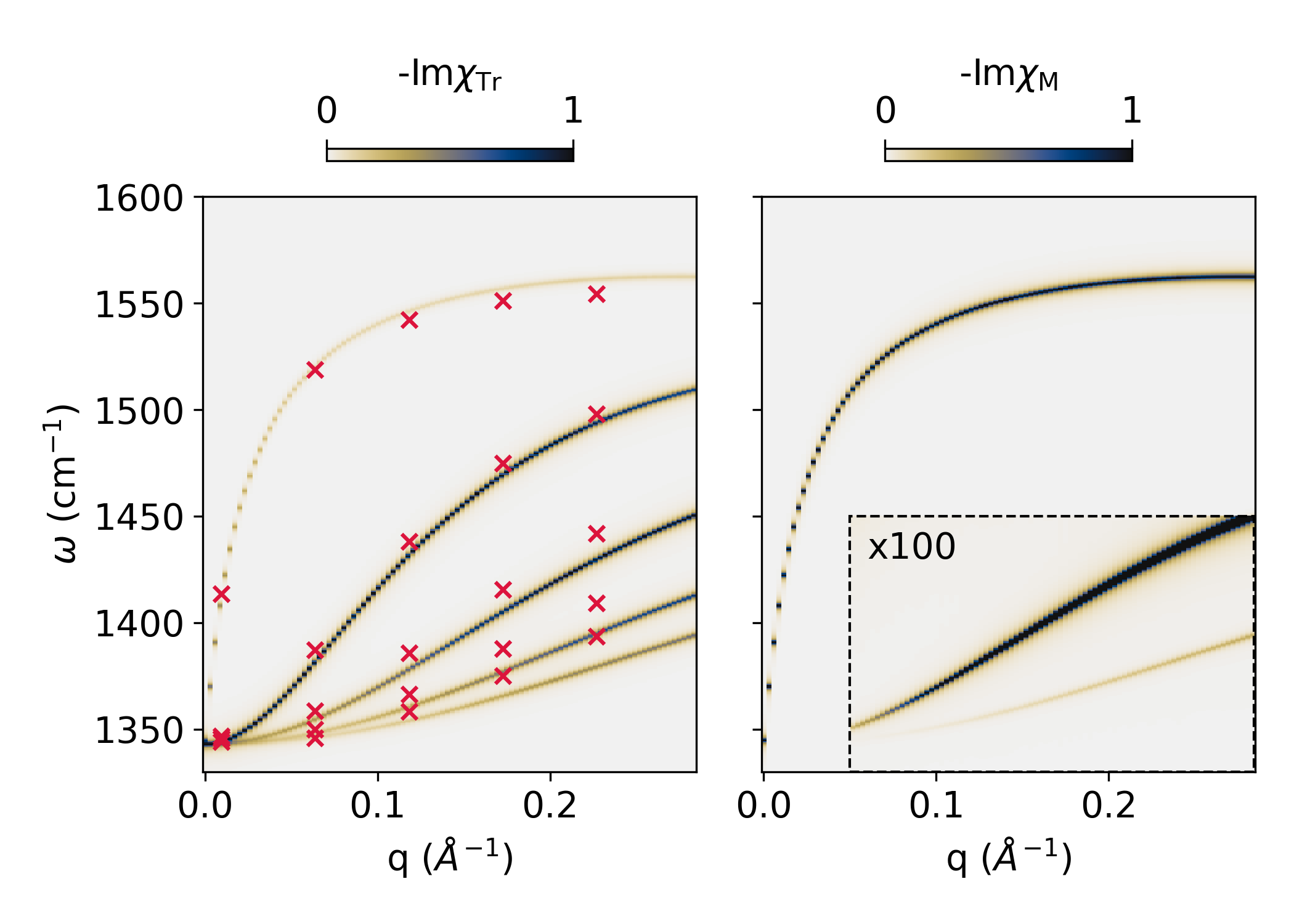}
\caption{(Left) graphene's plasmonic excitation compared to the analytical model of Eq. \ref{eq:AnPlDi} with $\kappa=1$ (blue curve). The colourmap represents $-\textrm{Im}\chi_{\textrm{M}}$, normalized by the global maximum of the intensity. The dashed red curve is the extrapolated maximum of $-\textrm{Im}\chi_{\textrm{M}}$. The black dashed lines separate the following zones. I: no electron-hole transition and undamped/dispersed plasmon, II: interband electron-hole continuum, III: intraband electron-hole continuum, IV: no electron-hole transition.
(Middle) Polar LO phonon modes of 5-BN from the resonances of the full spectral function $-\textrm{Im}\chi_{\textrm{Tr}}$, normalized at each momentum. The relative signs of the layer components of the phonon polarizations (corresponding to the sign of $\textrm{Im}\rho^\textrm{M}_{k}$, defined in the text) are as follows, from highest to lowest energy mode: $(+++++)$, $(--0++)$, $(-+++-)$, $(+-0+-)$, $(+-+-+)$. Red crosses are phonon frequencies computed in DFPT.
(Right) $\textrm{Im}\chi_{\textrm{M}}$ of 5-BN, normalized at each momentum. The highest LO mode is by far the most intense. Increasing the signal by a factor 100 in the inset  reveals that two other modes are active, although much weaker.}
\label{fig:PlasmonandPhonons}
\end{figure*}

In this Section, the collective modes of multilayer BN and graphene are considered separately, before any coupling between the two, as shown in Fig. \ref{fig:PlasmonandPhonons}. We show that the present model reproduces the features of multilayer BN and doped graphene, at the cost of a simple DFPT calculation for a single layer of each material.
Before coupling, those systems are fairly well-known, such that the VED method can be compared to literature and direct DFPT calculations. At the same time, it brings new insight on certain aspects like the relative intensity of the different modes in the EELS intensity.

\subsubsection{graphene's plasmons}

Graphene is a very promising medium for plasmonics \cite{Grigorenko2012,Politano2014}, due to the strength and tunability of the plasmons.
Analytical models \cite{Hwang2007a,Polini2008,Hill2009} of their dispersion are obtained by considering the polarizability of doped graphene---Eq. (17) of Ref. \cite{Hwang2007a}--- and the RPA relation to obtain the plasmon dispersion. The dielectric environment may be taken into account by a phenomenological background dielectric constant $\kappa$. In the small momentum limit, the plasmon dispersion is found to be
\begin{equation}
\label{eq:AnPlDi}
\hbar\omega_{\text{pl}}(q) \underset{q\rightarrow 0}{\sim} \sqrt{\frac{2e^2 \varepsilon_{\textrm{F}}}{\kappa}} \sqrt{q}.
\end{equation}
Two important features of the plasmon dispersion are shown here:
the $\sqrt{q}$ asymptotic behavior, and a dispersion that scales with the Fermi surface, since $\hbar\omega_{\text{pl}}/\varepsilon_{\textrm{F}} \propto \sqrt{q/k_{\textrm{F}}}$, with a proportionality constant that depends only on the dielectric environment.

The left panel of Fig. \ref{fig:PlasmonandPhonons} shows $-\textrm{Im}\left[\chi_{\textrm{M}}(q, \omega)\right]$ for isolated graphene with a Fermi level of $\varepsilon_{\textrm{F}} = 0.2$ eV  and at room temperature. 
%This is modelled with Eq. \ref{eq:Qelfree}, computed in the Dirac cone model, although it can in principle be done with the DFT band structure computed on a very fine grid.
The only chosen parameter is the plasmon linewidth, fixed at $\eta_{\textrm{pl}} = 5$ meV, consistent in order of magnitude with theoretical \cite{Principi2013,Principi2014} and experimental \cite{Yan2013,Woessner2015} investigations. The maxima of $-\textrm{Im}\left[\chi_{\textrm{M}}(q, \omega)\right]$ agree well with the asymptotic expression Eq. \ref{eq:AnPlDi}, as represented by the blue continuous line for $\kappa=1$, at least up to $q/k_{\textrm{F}} \sim 0.2$. The same results (in units of $\hbar \omega/\varepsilon_{\textrm{F}}$ and $q/k_{\textrm{F}}$) are obtained for different Fermi levels. 
The main features of the dispersion are thus recovered. Other spectral properties are also consistent with the literature. Notably, the spectral weight scales with $\varepsilon_{\textrm{F}}$ \cite{Hwang2007a}. 

In the left panel of Fig. \ref{fig:PlasmonandPhonons}, different zones are delineated with dashed lines, corresponding to the boundaries of the intra- and interband particle-hole continua. In practice, energy and momentum conservation allow intra- and interband electronic transitions only in zones III and II, respectively. Note that those boundaries are only strict at zero temperature. Otherwise they are smeared by the Fermi-Dirac occupation function.

\subsubsection{BN's polar-optical phonons}

Multilayer BN is ubiquitous in 2d devices \cite{Zhang2017c}. It is often used as an encapsulator, to protect the active layer from the environment and thus get closer to ideal intrinsic properties. By itself, it is a very promising platform for various photonics applications \cite{Caldwell2019}. In particular, BN's phonon-polaritons \cite{Dai2014,Shi2015,Dai2019,Li2021b} have attracted a lot of interest based on their ability to shape and control light in matter.
In 2d , phonon-polaritons are simply the polar-optical phonons of the system \cite{Rivera2019}, and near-field microscopy techniques developed to probe polaritons in 2d materials are one of the best ways to probe BN's phonon dispersions, e.g. versus number of layers\cite{Dai2014,Shi2015}. In addition to a detailed microscopic understanding, the VED model provides valuable insights to interpret and predict the results of such experiments. In particular, it clarifies which modes are active and their relative intensities.

The middle panel of Fig. \ref{fig:PlasmonandPhonons} shows the total spectral function
$-\textrm{ImTr}\chi$ for 5-layers BN (5-BN). This is computed in the VED model, using inputs from single-layer BN DFPT calculations. The only input parameter is the BN's phonon linewidth $\eta_{\textrm{LO}} = 0.6$ meV, chosen fairly small here in order to obtain separate peaks.
Only phonon excitations are present, and the red crosses indicate phonon frequencies computed directly in DFPT. We observe a fairly good agreement. The error likely comes from modelling approximations, as discussed in App. \ref{app:compdets}; also, in DFPT calculation each BN single layer does not possess anymore mirror symmetry as instead it is enforced in our vdWH. 

Many interesting characteristics of polar phonons in 2d materials are recovered.
The highest branch displays the dimensionality signature of the LO-TO splitting \cite{Sanchez-Portal2002,Michel2009,Sohier2017a}, i.e. the non-analytic and vanishing splitting at $q \rightarrow 0$, marked by the linear increase of the dispersion at small but finite momentum. 
The total number of LO modes is equal to the number of layers. The highest mode corresponds to in-phase contributions from all the layers, while the lower energy modes corresponds to various out-of-phase combinations. The slopes of the latter vanish at $q \rightarrow 0$. 

The right panel of Fig. \ref{fig:PlasmonandPhonons} shows $-\textrm{Im} \chi_{\textrm{M}}$ for 5-layers BN (5-BN). 
As already mentioned, among all the LO modes, only the symmetric ones appear in $-\textrm{Im}\chi_{\textrm{M}}$. Indeed,  the perturbation and probe are symmetric in $z$, implying that the anti-symmetric modes are inactive. The symmetry of the modes is obtained via the relative sign of the imaginary part of the density response $\rho^\textrm{M}_k = \sum_l \chi^{00}_{kl}$ to a uniform external perturbation at the different resonances, and it is indicated in the caption of Fig. \ref{fig:PlasmonandPhonons}. The relative amplitude of the layers' density response decreases from the center to the outside of the stack, consistent with the relative amplitude of phonon displacement obtained in DFPT.
The VED framework provides solid grounds to explain or predict the relative intensity of the modes in, e.g., EELS experiments. In the case of uniform perturbation and probe, note that the intensity of the peaks decreases with energy, and the highest one dominates largely. 

\subsection{Interplay of plasmons and phonons}
\label{sec:interplay}

Beyond the interest of graphene and BN taken individually, it is essential to investigate their interactions. Since BN is often used as an encapsulator to protect graphene and other 2d materials, it is important to understand exactly how it can interact and modify the intrinsic properties of the encapsulated layer. 
The coupling between the two can also be exploited to tailor the polaritons\cite{Brar2014,Dai2015}, thus opening new pathways for electromagnetic waves manipulation; or to better control the energy and momentum relaxation of graphene's electrons \cite{Tielrooij2018}.
For BN-capped graphene, the main tuning parameters are the Fermi level of graphene and the number of BN layers. 
We study the interplay of graphene's plasmons and BN's LO phonons by focusing on two aspects: their dispersion, discussed here, and their scattering of graphene's electrons, discussed in Sec. \ref{sec:elecscatt}. On top of previous analytical works on plasmon-phonon interactions\cite{Kim1978,Ong2012,Principi2014, Hauber2017a}, we bring the quantitative and predictive insight of microscopic ab initio simulations.

\begin{figure}[h!]
\includegraphics[width=0.98\columnwidth]{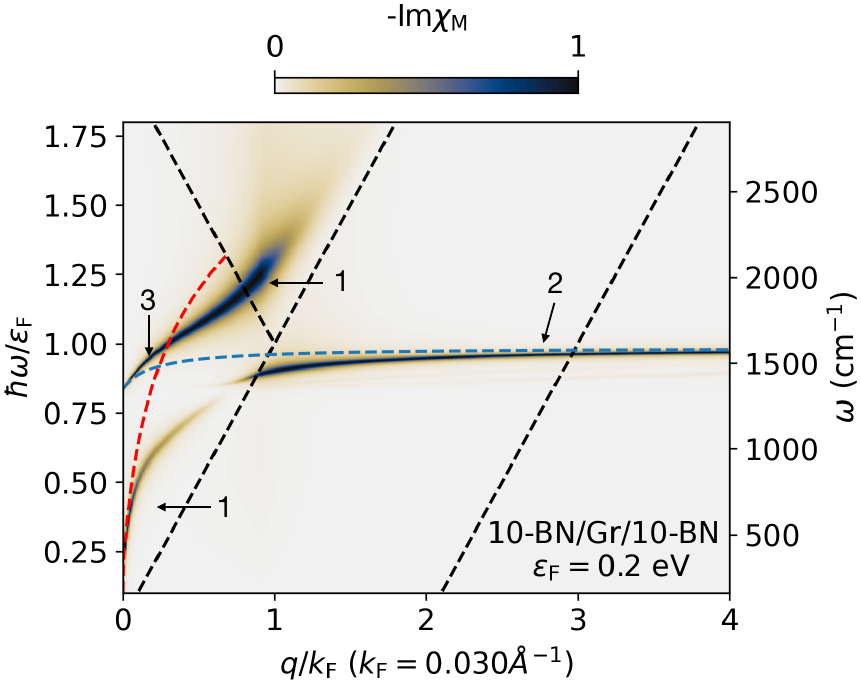}
\caption{ (Top) Macroscopic $-\textrm{Im}\chi_{\textrm{M}}$ for 10-BN/Gr@0.2eV/10-BN, normalized at each momentum. Blue and red dashed lines are the uncoupled excitations: LO dispersion of 20 BN layers and plasmon of free-standing graphene, respectively.
}
\label{fig:chi10BNGr10BN}
\end{figure}

Fig. \ref{fig:chi10BNGr10BN} shows $-\textrm{Im}\chi_{\textrm{M}}$ for the prototypical system made of graphene with a Fermi level of $\varepsilon_{\textrm{F}} = 0.2$ eV encapsulated by 10-BN on each side. The plasmon and phonon linewidths are set to $\eta_{\textrm{pl}} = 5$ meV, $\eta_{\textrm{LO}} = 1$ meV (for numerical efficiency and plotting purposes, within experimental order of magnitude). We clearly observe an anti-crossing that indicates the plasmon-LO phonon coupling. 
Let us already name $(q, \omega)$ regions of interest outside of the anti-crossing, as annotated in Fig. \ref{fig:chi10BNGr10BN}: (1) the plasmon region covers the plasmon dispersion away from the typical LO phonon frequency, $\omega \ne \omega_{\textrm{LO}}$; (2) the phonon region follows the phonon dispersion at large momenta, to the right of the plasmon dispersion $q/k_{\textrm{F}} > \hbar\omega_{\textrm{LO}}/\varepsilon_{\textrm{F}}$ ; (3) the plasmon-phonon hybrid region covers the small momenta $q \rightarrow 0$ around the LO phonon frequency $\omega \sim \omega_{\textrm{LO}}$.

In the plasmon region (1), the plasmon is screened by the presence of BN, pushing its dispersion towards the $\hbar \omega/\varepsilon_{\textrm{F}} = q/k_{\textrm{F}}$ dashed line, as expected from external screening in Eq. \ref{eq:AnPlDi}. 
Along their dispersion, the LO phonons experience different screening regimes from graphene. In the phonon region (2), they experience metallic screening, suppressing the LO-TO splitting and pushing the dispersion downward. Indeed, in the presence of static metallic screening, the slope of the highest mode dispersion at small momenta would vanish. 
Here, however, in the polariton region (3) the free-carrier screening from graphene is inefficient and we recover the finite slope one would obtain in BN alone for $q \rightarrow 0$ and $\omega \sim \omega_{\textrm{LO}}$, and a stiffening of the slope at small but finite momenta. 

Those results are developed in the following. %further analysed and enriched 
First, the nature of the excitations is carefully studied, in order to disentangle, when possible, phonons and plasmons.
Second, we consider the case of a surface probing setup to access other collective modes.
Third, the effect of the number of BN layers is analysed looking at 1, 10, and 30 BN layers on each side.
Finally, the effect of graphene's Fermi level is considered, with two other representative doping: $\varepsilon_{\textrm{F}}=0.1$ and $0.3$ eV.

\subsubsection{Nature of the excitations and layer contributions}
\label{sec:nature}

\begin{figure*}[t!]
\includegraphics[width=0.9\textwidth]{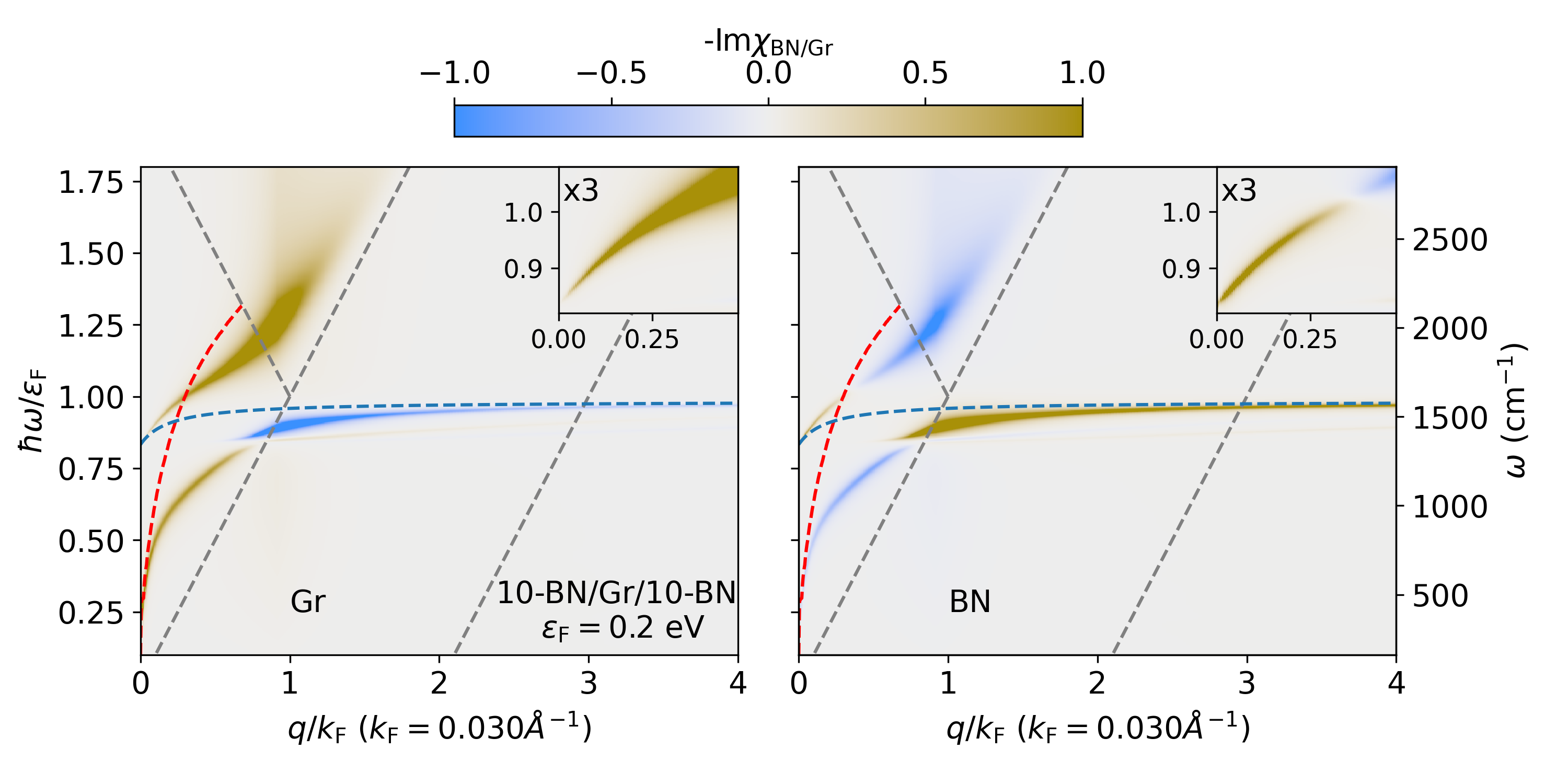}
\caption{10BN/Gr@0.2eV/10BN. Contributions to $-\textrm{Im}\chi_{\textrm{M}}$ from graphene and BN, $-\textrm{Im}\chi_{\textrm{Gr}}$ and $-\textrm{Im}\chi_{\textrm{BN}}$ as described by Eq. \ref{eq:chisplit}, normalized at each q point. The inset focuses on the hybrid plasmon-phonon excitation at small $q$ and around $\hbar \omega_{\textrm{LO}}$, with the intensity magnified by a factor of 3 for better visualization. Blue and red dashed lines are the uncoupled excitations: LO dispersion of 20 BN layers and isolated graphene's plasmon, respectively. 
}
\label{fig:layercontrib}
\end{figure*}

Here, we look at the layers from which the response originates to unravel the nature of the excitations. This procedure provides valuable insights into the interplay of phonons and plasmons, showing which mode drives the excitation, and how electrons screen it. 

In Fig. \ref{fig:layercontrib}, a unit potential perturbation is applied to all layers and the response coming from either graphene or BN are separated as follows
\begin{align}
    \chi_{\textrm{M}} =\chi_{\textrm{BN}} + \chi_{\textrm{Gr}},\\
    \chi_{\textrm{BN}} = \sum_{k = \textrm{BN},l}  \chi^{00}_{kl}, \quad
    \chi_{\textrm{Gr}} = \sum_{k = \textrm{Gr},l}  \chi^{00}_{kl}.
    \label{eq:chisplit}
\end{align}
Notice that the sum over BN runs over all the BN layers. Fig. \ref{fig:layercontrib} shows $-\textrm{Im}\chi_{\textrm{BN}}$ and $-\textrm{Im}\chi_{\textrm{Gr}}$. It clearly reveals 3 types of excitations, clarifying the nomenclature of the 3 regions of $(q, \omega)$ space defined in the previous section. First, in the phonon region (2), we observe a positive contribution from BN layers and a negative one from graphene. Note that the sum of the two is always positive, so that here  
$ |-\textrm{Im}\chi_{\textrm{BN}}| > |-\textrm{Im}\chi_{\textrm{Gr}}|$. Therefore, the excitation is driven by BN's polar phonons, and graphene is simply responding with (static) free-carrier screening. 
Second, in the plasmon region (1), we observe instead a positive contribution from graphene, and a negative one from BN layers. Thus, the plasmon is driving the excitation, and BN's electrons respond by screening it.
Third, in the plasmon-phonon hybrid region (3) we observe positive contributions from both graphene and BN. This is a true hybrid polariton, driven by both plasmonic and phononic responses. 
The insets zoom in on this plasmon-phonon hybrid. The dispersions of the modes before the coupling are also plotted, i.e. graphene's plasmon without BN and the highest LO phonon mode corresponding to the total number of BN layers in absence of graphene.
The slope of the hybrid excitation dispersion (i.e. the group velocity  of the polariton) approaches that of the isolated 20-BN system when $q \rightarrow 0$. In that limit, the polariton gradually becomes pure LO phonon, and contextually $-\textrm{Im}\chi_{\textrm{Gr}}$ vanishes. 
At small but finite $q$, however, the slope remains fairly high, rather than flattening like that of the intrinsic BN's phonons. This is thanks to a gradual shift of the plasmon-phonon hybrid from phonon to plasmon, as $-\textrm{Im}\chi_{\textrm{BN}}$ decreases with increasing $q$ (eventually changing sign) while $-\textrm{Im}\chi_{\textrm{Gr}}$ increases. Eventually, following this branch, we fall back on the plasmon-driven excitation.
We thus have a clear interpretation of all the excitations and the role played by the electrons, phonons, and plasmons.

\subsubsection{Probing other modes}
\label{sec:otherProbes}

\begin{figure}[h!]
\includegraphics[width=0.98\columnwidth]{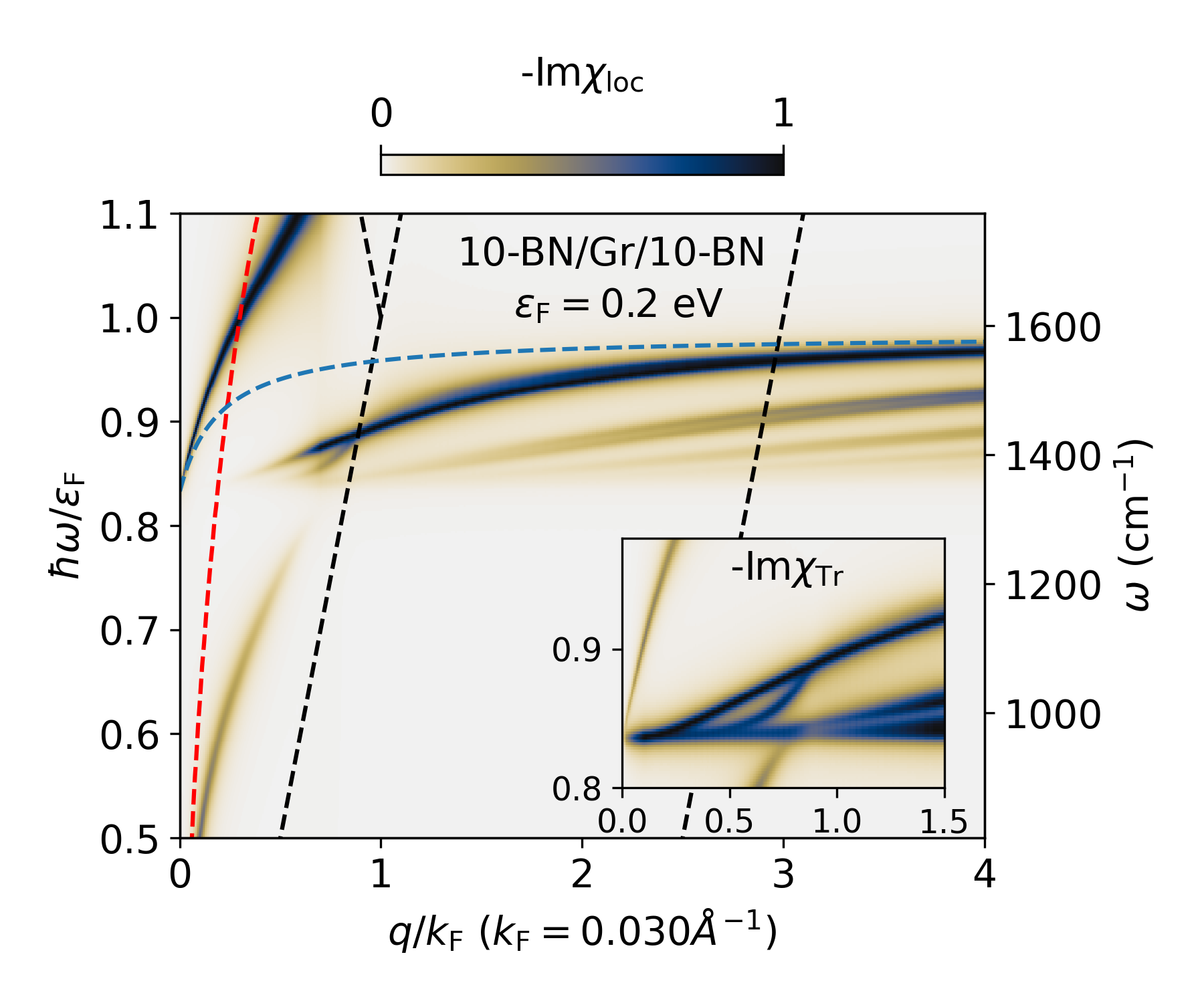}
\caption{ $-\textrm{Im}\chi_{\textrm{loc}}$ normalized at each $q$ in 10-BN/Gr@0.2eV/10-BN, representing the results of a probe measurement at the surface, with the model explained in Sec. \ref{sec:densdensresp}. The inset zooms in on the anti-crossing region, and shows the spectral function $-\textrm{Im}\chi_{\textrm{Tr}}$. 
}
\label{fig:g20010BNGr10BN}
\end{figure}

The macroscopic response function $\chi_{\textrm{M}}$ only probes symmetric modes because perturbation and probe are uniform over the layers. This is not the case of surface spectroscopic techniques, like scattering-type Scanning Near-field Optical Microscopy\cite{Dai2014} (sSNOM) or 2d High-resolution electron energy loss spectroscopy \cite{Li2023} (2d -HREELS). 
A quantitatively accurate simulation of those experimental setups is out of scope here. However, it is informative to consider a setup where perturbation and probe are localised on the surface, and qualitatively assess the consequences on the activity and relative intensity of the collective modes. Using the model explained in Sec. \ref{sec:densdensresp}, we expect the physical observables to be proportional to $\chi_{\textrm{loc}}(q, \omega)$, plotted in Fig. \ref{fig:g20010BNGr10BN}.

The relative intensities of the modes is very different with respect to the plots of $-\textrm{Im}\chi_{\textrm{M}}$, with the out-of-phase LO modes more visible. Interestingly, the most intense phonon-like mode is not the highest LO mode, but the mode corresponding to the BN layers moving in-phase with respect to layers on the same side of the graphene layer, but out-of-phase with respect to the BN layers on the other side of graphene. 
This mode is unaffected by screening from graphene, since any electric field from phonons on one side of graphene would be equal and opposite to the field from the other side. Thus, the net potential from phonons felt by graphene vanishes, and it does not respond. 
This mode actually crosses the dispersion of the in-phase LO mode, becoming higher in energy when the latter is strongly screened by graphene. This raises the question of which mode is actually most seen in surface-probing experiments. Near field spectroscopy would probe the vicinity of the light cone, dominated by the plasmon-phonon hybrid at vanishing momentum. Techniques probing larger momenta like 2d -HREELS \cite{Li2023}, however, might very well be dominated by this anti-symmetric mode.

\subsubsection{Effects of number of BN layers}

The number of BN layers may vary significantly in different realisation of similar devices. Whether it is a parameter one can control precisely to tailor the properties of the system, or a fluctuating number depending on the experimental realization of the sample, it is important to understand its impact on the collective modes of BN-capped systems. Beyond the fairly obvious increase of the plasmon-phonon coupling, the VED framework enables the quantification and a detailed interpretation of the changes in the collective modes dispersions.

\begin{figure}[h!]
\includegraphics[width=0.98\columnwidth]{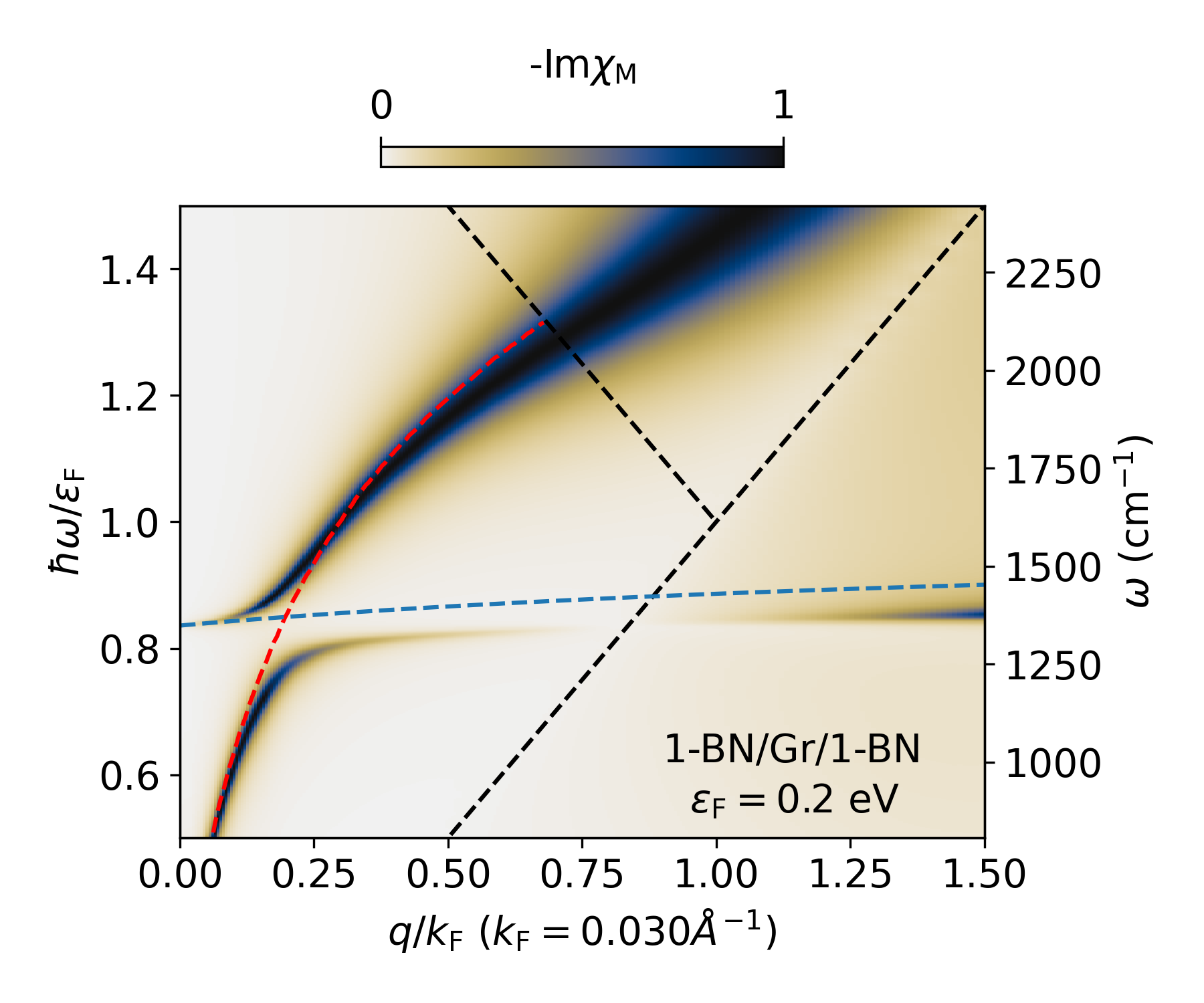}
\includegraphics[width=0.98\columnwidth]{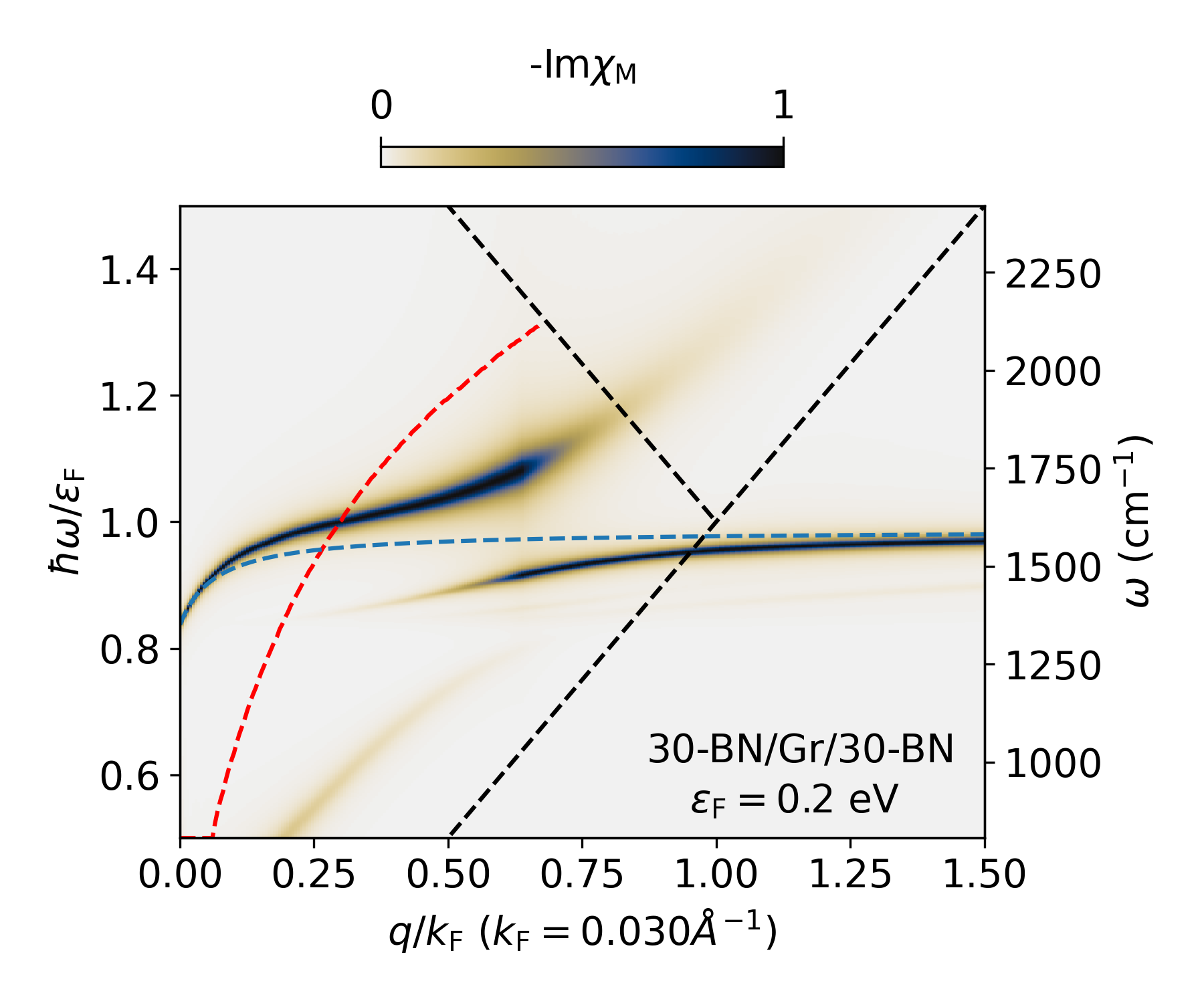}
\caption{$-\textrm{Im}\chi_{\textrm{M}}$ for 1-BN/Gr@0.2eV/1-BN (top) and  30BN/Gr@0.2/30BN (bottom) normalized at each q. The blue curve and dashed red line are the uncoupled modes (LO dispersion of total number of BN layers in blue, isolated graphene's plasmon in red)}
\label{fig:chiNBNGrNBN}
\end{figure}

Fig. \ref{fig:chiNBNGrNBN} shows $-\textrm{Im}\chi_{\textrm{M}}$ for graphene at $\varepsilon_{\textrm{F}} = 0.2$ eV encapsulated by 1 or 30 layers of BN on each side, on the top and bottom panels respectively. This complements the dispersion of Fig. \ref{fig:chi10BNGr10BN}, with 10-BN on each side. As the number of BN layers increases, the features already noted in the previous Sections are enhanced. 
Due to overall stronger phonon responses, the plasmon-phonon coupling and the size of the anti-crossing increase.
Also, the stronger electronic screening from BN pushes the plasmonic dispersion closer to the $\hbar \omega/\varepsilon_{\textrm{F}} = q/k_{\textrm{F}}$ line. 
The slope of the polariton at $q \rightarrow 0 $ and $\omega \sim \omega_{\textrm{LO}}$ increases, following that of the intrinsic LO phonon, as the polar contributions from all the layers add up \cite{Sohier2017a}.

We remind that, as discussed after Eq. \ref{eq:CoulKernel}, our framework can be applied until the slope of the longitudinal phonon is not so steep to enter in the light-cone region, an event that happens for a certain limiting width of the layer (way beyond what is considered here). Increasing the width further causes relativistic retardation effects to become important and we therefore enter the realm of three-dimensional phonon-polaritonics. Notice that this gives a rigorous criteria to distinguish between a slab and a proper three-dimensional materials from the polaritonics point of view.

\subsubsection{Effects of graphene's Fermi level}

Graphene's Fermi level is often tuned electrostatically \cite{Novoselov2004,Schwierz2010,Ferrari2015}. Within a field-effect transistor geometry, free carriers are accumulated in graphene, typically up to the order of $10^{13}$ cm$^{-2}$. This corresponds to Fermi levels of a few hundreds of meV. Thus, in the following, three Fermi levels $\varepsilon_{\textrm{F}} = 0.1, 0.2, 0.3$ eV are considered and the consequences on the collective modes of the BN-encapsulated graphene are studied. 
Since the effect of doping on the plasmon are relatively well-known, we focus on the rich and yet unresolved discussion around the dynamical screening of phonons and their coupling to electrons. 

\begin{figure}[h]
\includegraphics[width=0.48\textwidth]{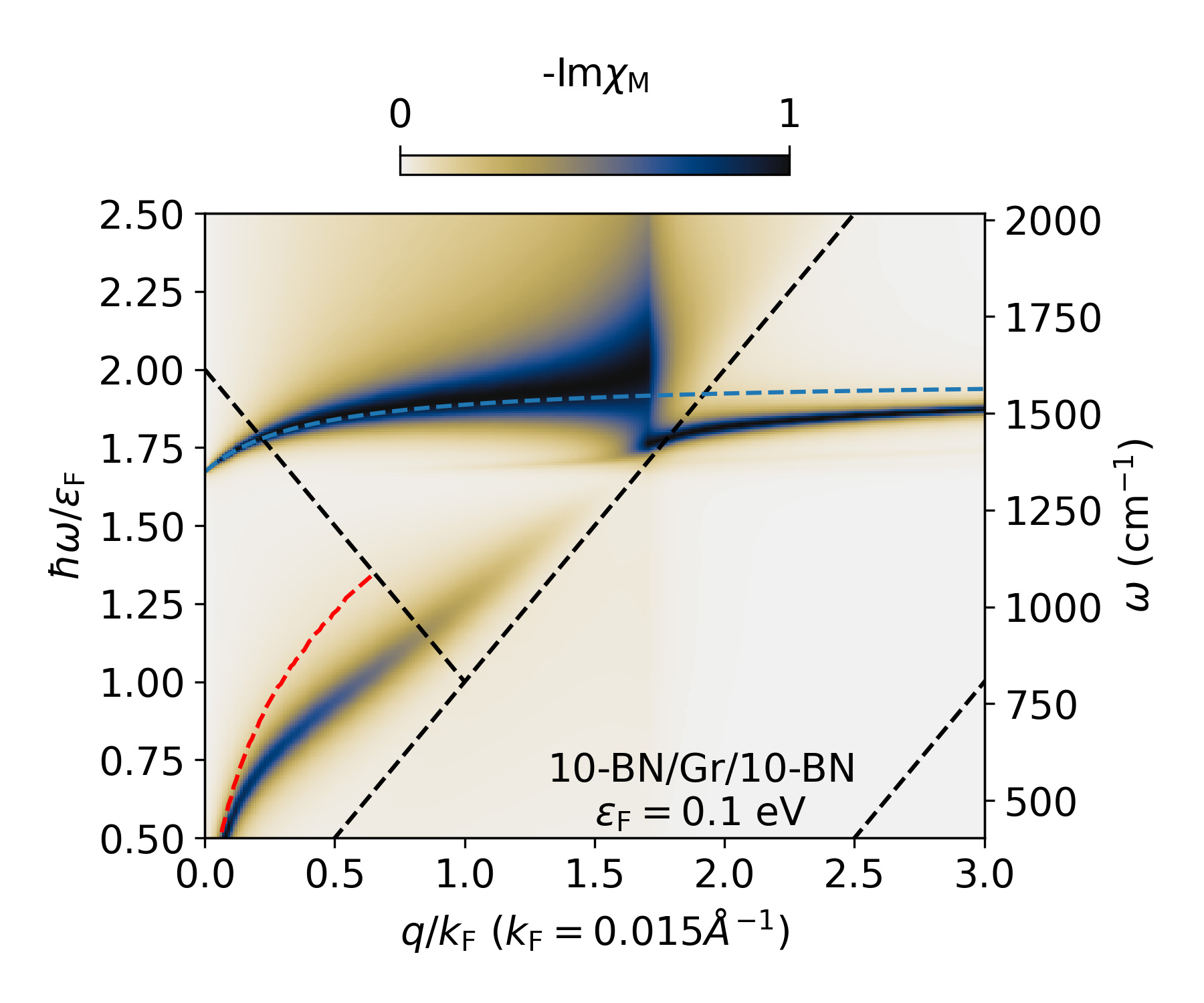}
\includegraphics[width=0.48\textwidth]{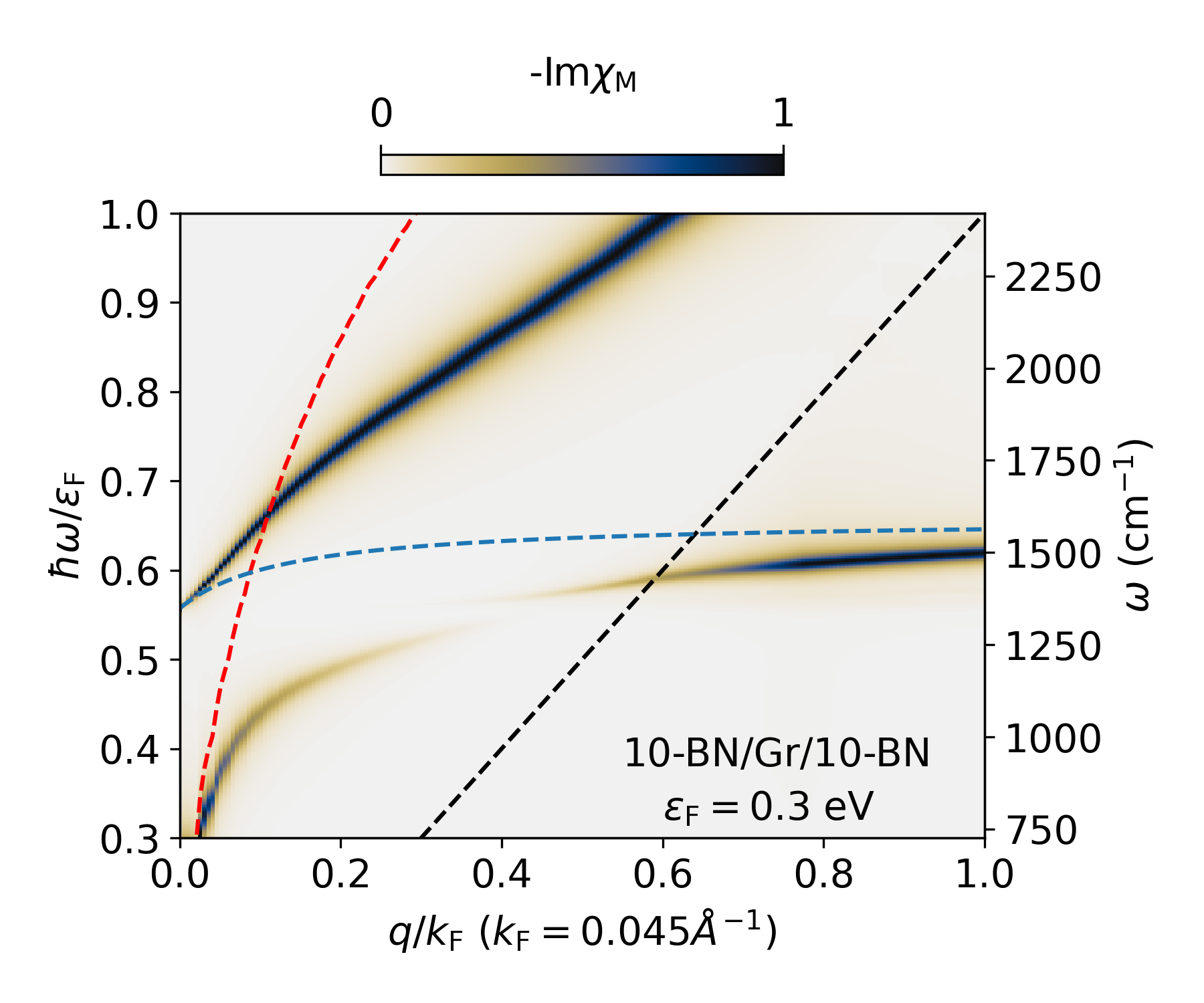}
\caption{ $-\textrm{Im}\chi_{\textrm{M}}$ for 10-BN/Gr@0.1eV/10-BN (top) and 10-BN/Gr@0.3eV/10-BN (bottom), normalized at each q, in the plotted $\omega$ window. See also Fig. \ref{fig:chi10BNGr10BN} for the case of $\varepsilon_{\textrm{F}}=0.2$ eV.
Dashed black lines indicate intra- and inter-band electron-holes continuum. The dashed red line corresponds to the plasmon dispersion in isolated graphene. The blue line indicates the phonon dispersion of the highest LO mode in 20-BN without graphene.}
\label{fig:chi10BNGr10BNdoping}
\end{figure}

Fig. \ref{fig:chi10BNGr10BNdoping} shows $-\textrm{Im}\chi_{\textrm{M}}$ for graphene encapsulated by 10 BN, at Fermi levels $\varepsilon_{\textrm{F}}=0.1$ eV and $\varepsilon_{\textrm{F}}=0.3$ eV, in addition to the $\varepsilon_{\textrm{F}}=0.2$ eV case from Fig. \ref{fig:chi10BNGr10BN}. In all cases, some trends persist. The plasmon-phonon hybrid follows the intrinsic LO phonon dispersion in the $q\rightarrow 0$ limit, while the LO-TO splitting is reduced due to free-carrier screening from graphene in the intraband electron-hole continuum, that is for $q > \frac{\hbar\omega_{\textrm{LO}}}{\varepsilon_{\textrm{F}}} k_{\textrm{F}} = \frac{\omega_{\textrm{LO}}}{v_{\textrm{F}}}$. The behavior between those two regimes ($q\rightarrow 0$ and $q > \frac{\omega}{v_{\textrm{F}}}$), however, is quite sensitive to the Fermi level.
In the case of $\varepsilon_{\textrm{F}}=0.2$ eV, where phonon energy and Fermi level are similar, the plasmon-phonon anti-crossing happens around the middle of the triangular region I. For $\varepsilon_{\textrm{F}}=0.3$ eV, this is shifted towards smaller momenta, and so is the region where we can observe the intrinsic LO phonon dispersion. 
Indeed, assuming the plasmon dispersion to be of the form Eq. \ref{eq:AnPlDi}, the momentum $q$ where the plasmon and phonon cross is proportional to $k_{\textrm{F}}$.
For both $\varepsilon_{\textrm{F}}=0.2$ eV and $\varepsilon_{\textrm{F}}=0.3$ eV, we observe the intrinsic and screened LO dispersion on the left and right of the anti-crossing, respectively. The qualitative behavior starts to change significantly when $\hbar \omega_{\textrm{LO}}>\varepsilon_{\textrm{F}}$, and the crossing of plasmon and phonon dispersions fall into the interband electron-hole continuum. There is no clear anti-crossing then. Instead, a collective mode emerges that roughly follows the intrinsic LO phonon dispersion. A clear free carrier screening only starts in the intraband continuum. Qualitatively, the difference between interband and intraband comes from the possibility in the latter case to have Lindhard/Thomas-Fermi type of screening \cite{Grosso2000,Giuliani2005}, ultimately coming down to the non vanishing Bloch overlaps in the $q \rightarrow 0$ limit \cite{Macheda2023}.

\section{Electron scattering}
\label{sec:elecscatt}
The VED model provides the coupling between electrons and the various collective modes, to be included in the calculations of scattering times. Above, the study of quantities like $\chi_{\textrm{M}}$ or $\chi_{\textrm{Tr}}$ informed us about the dispersion and nature of the modes. In this Section we investigate how those modes may scatter the electrons of the heterostructure.

\subsection{Validation in multilayer BN}

\begin{figure}[h]
\includegraphics[width=0.49\textwidth]{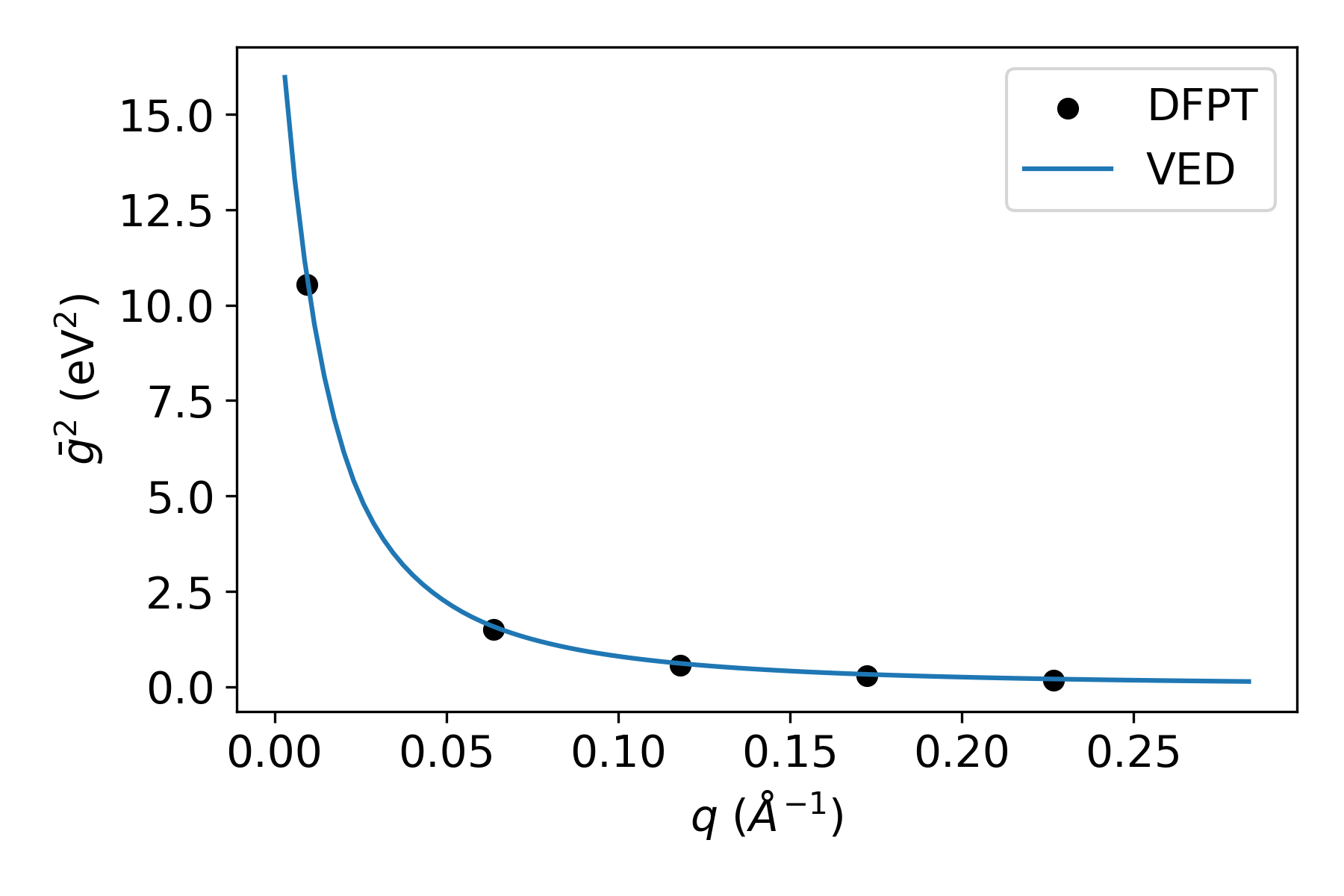}
\caption{Total coupling, for a 5-BN heterostructure, compared to DFPT calculations performed as described in App. \ref{app:GrBNcheck}. The coupling is entirely due to the interaction between electrons and phonons, statically screened by the BN's dielectric function.}
\label{fig:g5BN}
\end{figure}
We first validate the method by comparing the VED results on multilayer BN to the DFPT ones, obtained as described in App. \ref{app:GrBNcheck}. Indeed, since multilayer BN is a large gap semiconductor, all electronic contributions to the response are treated in the static limit within the VED model in that case. This corresponds to the adiabatic Born-Oppenheimer approximation of DFPT, and the results should thus differ only due to the approximations of the VED model (see App. \ref{app:compdets}).

Fig. \ref{fig:g5BN} shows the electron-phonon coupling with the LO modes in 5-BN. Practically, using Eq. \ref{eq:g2main} we compute
\begin{align}
    g^2(q, \omega) &= \frac{1}{N^2_l} \sum_{kl} g^2_{kl}(q, \omega),
    \label{eq:g2total}
\end{align}
which physically is the square of the electron-phonon coupling since other collective modes are absent. We also define
\begin{align}
    \bar{g}^2(q) &= \int_{0}^{\infty} g^2(q, \omega) d\omega.
\end{align}
In the following, the bar sign will be used consistently on all other $g^2$ quantity to signify integration over $\omega$ from zero to infinity. Notice that the sums run on all the BN layers. In fact, in the vdW approximation the Bloch electronic state of the full heterostructure layers are degenerate so that electrons are delocalized over the full structure. The agreement with DFPT is excellent, with the remaining error likely due to aforementioned modelling approximations.

% We can interpret the above equation as perturbing and probing the system with a normalized monopole charge density (as represented in Fig. \ref{fig:pertprob}) delocalized on all layers, because multilayer electrons BN are delocalized. In fact, in the vdW approximation the Bloch electronic state of the full heterostructure layers are degenerate and are a combination of orbital localized on each layer. The agreement with DFPT is excellent, with the remaining error likely due to aforementioned modelling approximations.\\

Note that we do not force any phonon perturbation singularly in the VED method. The external perturbation itself excites all the possible phonons modes. Therefore, we get the sum of the couplings with all phonon modes, so we compare with the sum of the (squared) couplings over DFPT modes.  Nevertheless, as seen in DFPT and expected in VED, the highest mode largely dominates the coupling. This result also validates the  spectral weights of the peaks in the VED method as a function of momentum, which is a significant progress with respect to other ab-initio works \cite{Gjerding2020}. 

\subsection{Scattering from coupled plasmon and phonons}

\begin{figure*}[t!]
\includegraphics[width=0.9\textwidth]{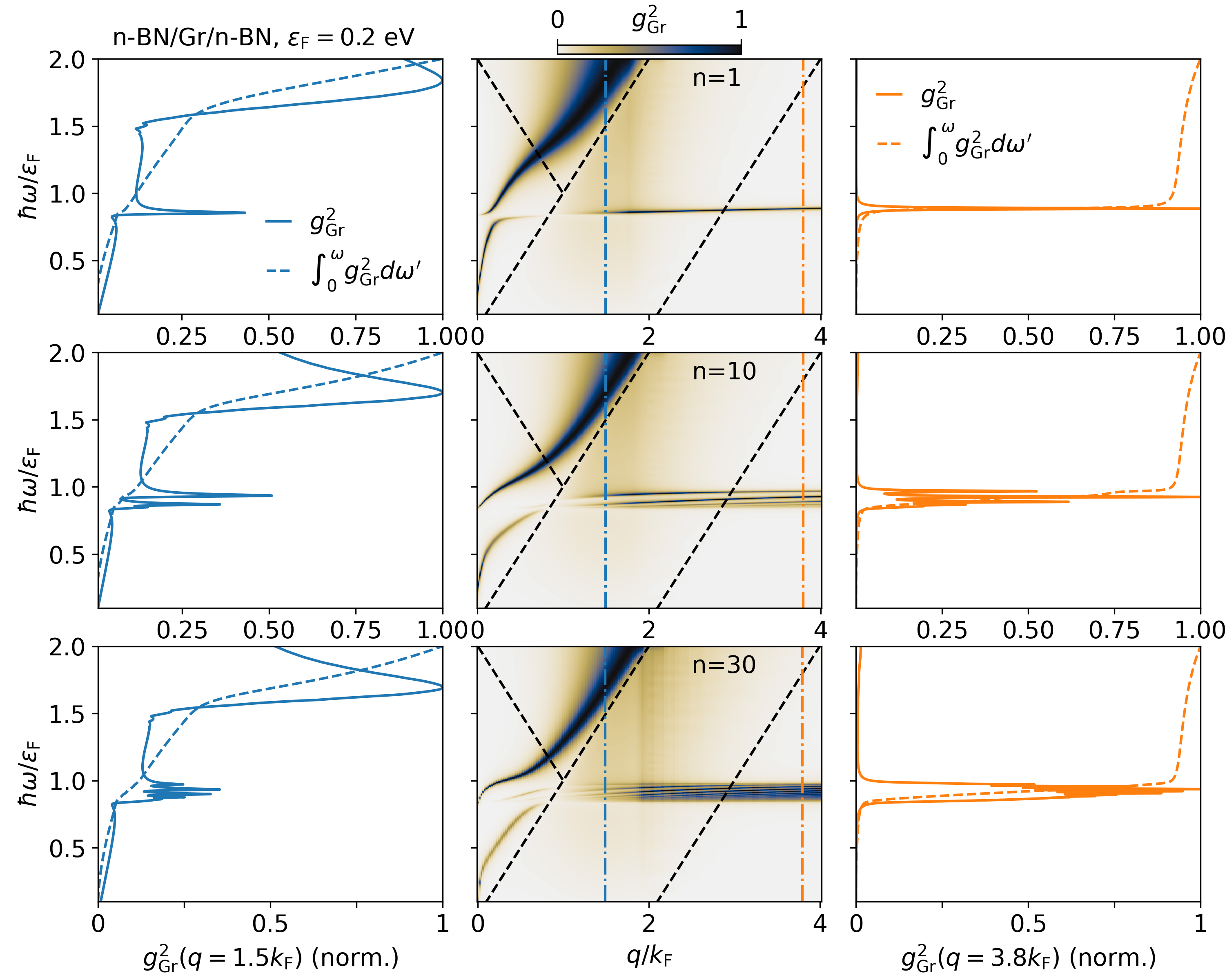}
\caption{(Central) $g^2_{\textrm{Gr}}(q,\omega)$, normalized at each q, for an heterostructure of n-BN/Gr@0.2eV/n-BN, with n=1,10,30. The blue and orange vertical lines represent the $q$ values for which $g^2_{\textrm{Gr}}(\omega)$, and its running integral, are represented in the (left) and (right) panels. In the left panel, the contributions to $g^2_{\textrm{Gr}}(\omega)$ come from the intraband electron-hole continuum ($\hbar \omega/\varepsilon_{\textrm{F}}<1$), from the electron-phonon coupling $\hbar \omega/\varepsilon_{\textrm{F}}\sim 1$, and from the coupling with the dispersed plasmon in the interband region $\hbar \omega/\varepsilon_{\textrm{F}}>1.5$. In the right panel, up to $\hbar \omega/\varepsilon_{\textrm{F}}>1.75$ only electron-phonon is present. Notice that the electron-phonon coupling gets contribution from many different modes, contrary to the picture of one single surface phonon responsible for all the interaction.}
\label{fig:gcuts}
\end{figure*}

We now continue with the prototypical system consisting of graphene sandwiched by $n$ layers of BN on each side. Here we consider the coupling with graphene's electrons, i.e. we compute
\begin{align}
\label{eq:el-EDMcoupling}
    g^2_{\textrm{Gr}}(q, \omega) = \sum_{ij k' l'} \frac{-1}{A \pi} \textrm{Im}{  \left[ v^{0i}_{\textrm{Gr}k'}(q) \chi^{ij}_{k'l'}(q, \omega)  v^{j0}_{l'\textrm{Gr}}(q) \right]}.
\end{align}
The above equation can be interpreted as using normalized charge densities as perturbation and probes of the system, as sketched in Fig. \ref{fig:pertprob}. The coupling of Eq. \ref{eq:el-EDMcoupling} contains  contributions from phonons and plasmons. We plot it in the central column of Fig. \ref{fig:gcuts} in the full $(q,\omega)$ plane, normalized at each $q$, for different $n$ (1,10 and 30 BN per side) at a doping level of $\varepsilon_{\textrm{F}}=0.2$eV.  The left and right panels of Fig. \ref{fig:gcuts} show $g^2_{\textrm{Gr}}(\omega)$ and its running integral $\int_0^{\omega} g^2_{\textrm{Gr}}(\omega')d\omega'$ $q=1.5 k_{\textrm{F}}$ and $q=3.8 k_{\textrm{F}}$. 
In both cases, the number peaks around the phonon energy increases with the number of BN layers. 
Clearly, graphene's electrons couple to several LO modes with different polarizations. This implies that considering a single surface-optical phonon for the remote coupling between electrons and phonons, as routinely done for bulk substrates like SiO$_2$\cite{PhysRevB.6.4517},  is not valid for BN encapsulation and vdWH in general. 
%Also, our treatment is microscopic and does not rely on transfer matrix methods used in the context of continuum electrodynamics \cite{Caldwell2014,Woessner2015}.
At $q=3.8 k_{\textrm{F}}$, up to $\hbar \omega/\varepsilon_{\textrm{F}} \sim 1.75$, no electron-hole excitation is possible, and the coupling is totally determined by the electron-phonon interaction. For $q=1.5 k_{\textrm{F}}$, however, $g^2_{\textrm{Gr}}(\omega)$ contains significant contributions from the low-energy spectrum of the intraband electron-hole continuum for $\hbar \omega/\varepsilon_{\textrm{F}}< 1$, from the different phononic branches of BN around $\hbar \omega/\varepsilon_{\textrm{F}}\sim 1$, and finally from the dispersed plasmon dispersion in the interband region for $\hbar \omega/\varepsilon_{\textrm{F}}> 1.5$.  
\\
While  it contributes to electronic lifetimes detected e.g. via ARPES, the scattering of graphene's electrons coming from graphene's  own electronic excitations (plasmon and electron-hole excitations) does not necessarily contribute to the momentum relaxation responsible for resistivity (depending on the magnitude and origin of plasmon dampening). Indeed, momentum is exchanged with essentially the same entity, i.e. graphene's electrons \cite{Ong2012}. Electron-phonon scattering, on the other hand, does. 
%It is clear that the coupling coming from the electron-hole continuum, and the dispersed plasmon, is momentum conserving. As such,, but it does not directly degrade the electronic current. The electron-phonon coupling instead is directly responsible for the existence of electrical resistivity. 
%Indeed, the 
It is then desirable to separate the different couplings arising from the interaction with different collective modes.

\subsubsection{Separation of the coupling with different modes}

\begin{figure*}[t!]
\includegraphics[width=0.99\textwidth]{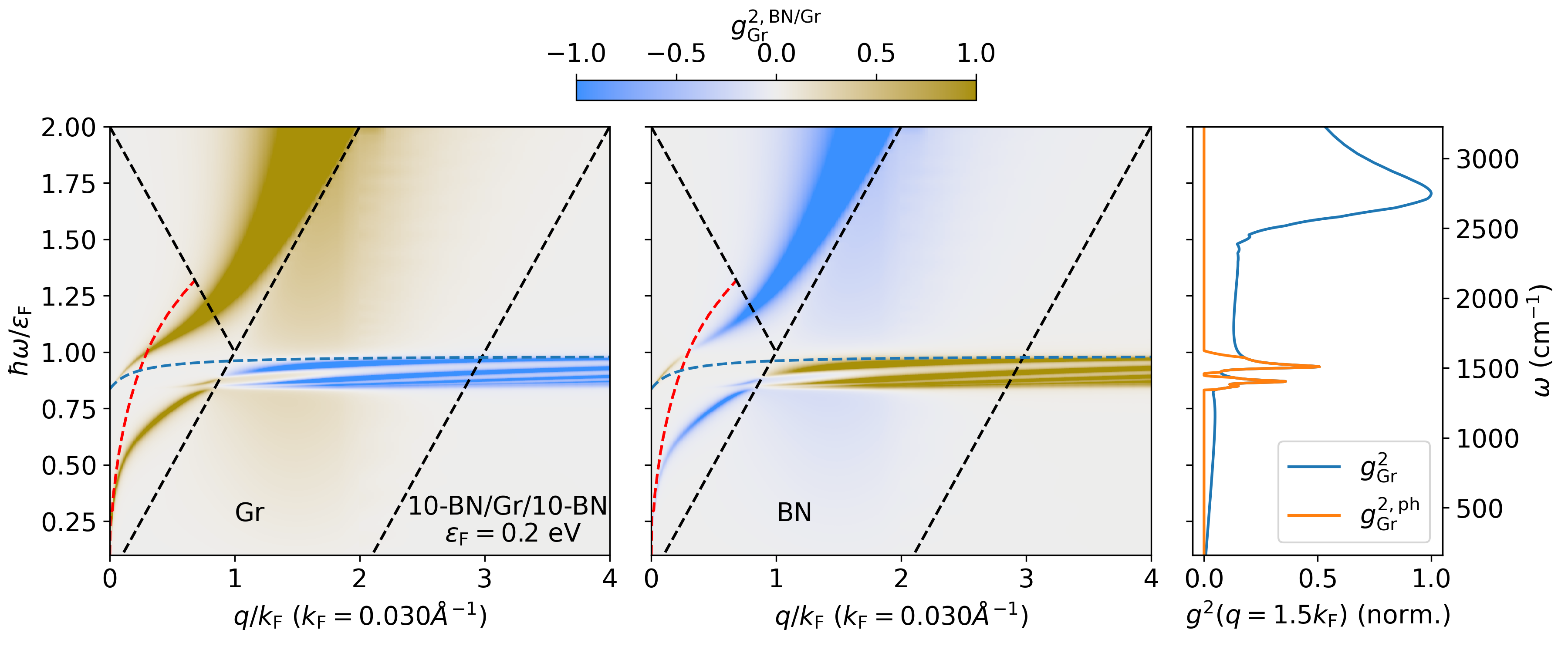}
\caption{Contributions to the coupling of graphene's electrons in 10BN/Gr@0.2eV/10BN, i.e. $g_{\textrm{Gr}}^{2,\textrm{BN}}(q, \omega)$ and $g_{\textrm{BN}}^{2,\textrm{Gr}}(q, \omega)$ as described by Eq. \ref{eq:gsepar} , normalized at each $q$ point. We also represent the phonon contribution to the coupling, i.e. $g^{2,\textrm{BN}}_{\textrm{Gr}}$ of Eq. \ref{eq:gphcases}.}
\label{fig:gcontrib}
\end{figure*}

In Section \ref{sec:nature} we separated the contributions from graphene and BN to gain insight on the nature of the excitations. Fig. \ref{fig:gcontrib} shows the results of a similar procedure applied to interaction between electrons and collective mode interactions. We separate the contributions as follows

\begin{align}
g^2_{\textrm{Gr}}&=g^{2,\textrm{BN}}_{\textrm{Gr}}+g^{2,\textrm{Gr}}_{\textrm{Gr}}, \\
    g^{2,\textrm{BN}}_{\textrm{Gr}}(q, \omega) &= \frac{-1}{A\pi} \sum_{k = \textrm{BN}} \sum_{ijl} v^{0,i}_{\textrm{Gr}, k} \textrm{Im}\left[\chi^{ij}_{kl}\right] v^{j0}_{l, \textrm{Gr}},\nonumber\\
    g^{2,\textrm{Gr}}_{\textrm{Gr}}(q, \omega) &= \frac{-1}{A\pi} \sum_{k = \textrm{Gr}} \sum_{ijl} v^{0i}_{\textrm{Gr}, k} \textrm{Im}\left[\chi^{ij}_{kl}\right] v^{j0}_{l, \textrm{Gr}}.
    \label{eq:gsepar}
\end{align}
Note that even though each of the above terms may be negative, $g^2_{\textrm{Gr}}$ is always positive.
\\
We observe similar trends as for the macroscopic $- \textrm{Im}\chi_M$ of Fig. \ref{fig:layercontrib}. Three types of coupling emerge: one driven by BN's phonons and screened by graphene when $g^{2,\textrm{BN}}_{\textrm{Gr}}>0$ and $g^{2,\textrm{Gr}}_{\textrm{Gr}}<0$ in region (2);
one driven by graphene's plasmon and screened by BN when $g^{2,\textrm{Gr}}_{\textrm{Gr}}>0$ and $g^{2,\textrm{BN}}_{\textrm{Gr}}<0$ in region (1); and one corresponding to a true plasmon-phonon hybrid, when both $g^{2,\textrm{BN}}_{\textrm{Gr}}>0$ and $g^{2,\textrm{Gr}}_{\textrm{Gr}}>0$ in region (3). 
%More visible here, but also present in the case of Fig. \ref{fig:layercontrib}, the scattered halo of graphene intraband electronic excitations in the particle-hole continuum is also screened by BN.

Using the information contained within this separation of the contributions, one may define a dynamically screened remote electron-phonon interaction corresponding to the scattering of graphene's electrons by BN's phonons.  The phonon-driven contribution to scattering can be systematically extracted, with exact asymptotic limits, as 
\begin{align}
g^{2,\textrm{ph}}_{\textrm{Gr}}=
\begin{cases}
0 & \text{  if } g^{2,\textrm{BN}}_{\textrm{Gr}}<0,  \\
g^{2,\textrm{BN}}_{\textrm{Gr}} & \text{ if } g^{2,\textrm{BN}}_{\textrm{Gr}}>0 \text{ and} \ g^{2,\textrm{Gr}}_{\textrm{Gr}}>0  \\
g^2_{\textrm{Gr}} & \text{ if } g^{2,\textrm{BN}}_{\textrm{Gr}}>0 \text{ and} \ g^{2,\textrm{Gr}}_{\textrm{Gr}}<0.
\end{cases},
\label{eq:gphcases}
\end{align}
% The graphene contribution is excluded in the plasmon-phonon hybrid case to extract only the phonon contribution. However, the graphene contribution is included in the phonon-driven case to recover the screening from graphene electrons.

The above procedure of extracting an electron-phonon coupling is arbitrary, but it assures the coupling to be a positive definite quantity. Disentangling plasmon and phonon scattering as such is useful when considering momentum scattering for electronic transport. Despite providing the main contribution, the current procedure does not necessarily provide a full description of the momentum relaxation process. To do so, one could for example consider a system of coupled Boltzmann equations for electrons and collective modes \cite{Hauber2017a} for dynamical couplings, and carefully extract a momentum relaxation time. We leave this for future studies.

\subsubsection{Plasmon and phonon contributions}

\begin{figure}[h!]
\includegraphics[width=0.98\columnwidth]{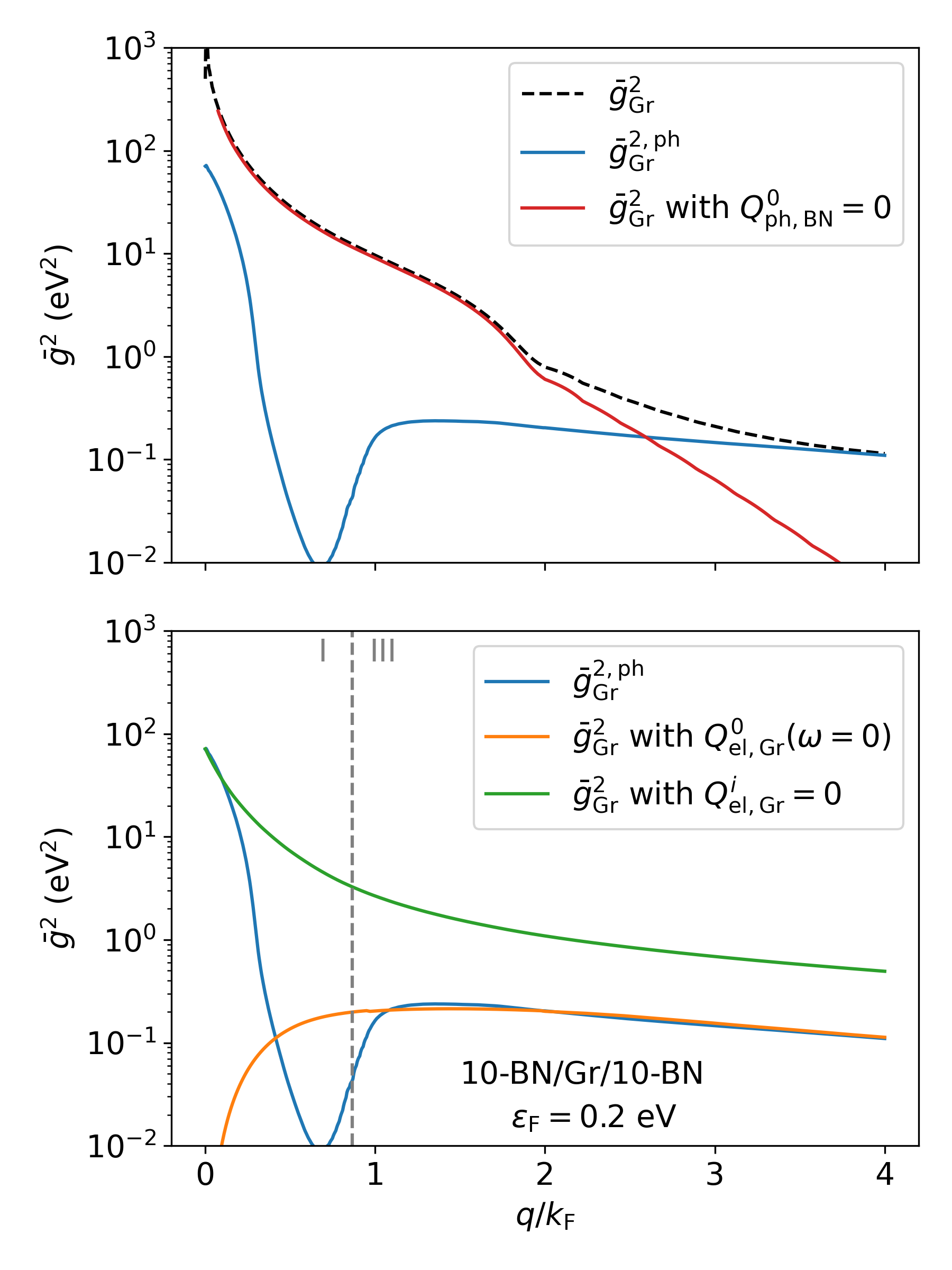}
\caption{Coupling between graphene's electrons and collective modes, integrated over positive frequencies, for different setups. (Upper) $\bar g^2_{\textrm{Gr}}$ : in presence of both dynamical phononic and dynamical plasmonic contributions. $\bar g^2_{\textrm{Gr}}$ with $Q^0_{\textrm{ph,BN}}=0$: in the absence of the dynamical phonon contribution from BN, but including the BN's static electronic response. This corresponds to electron-plasmon interaction screened by BN environment.  $\bar g^{2,\textrm{ph}}_{\textrm{Gr}}$: the contribution to the total coupling due to electron-phonon when the responses are all dynamical. 
(Lower) $\bar g^2_{\textrm{Gr}}$ with $Q^0_{\textrm{el,Gr}}(\omega=0)$ : in the absence of the dynamical contribution from Gr, i.e. by using the static screening from graphene's electrons. This corresponds to remote electron-phonon coupling screened from graphene in the Born-Oppenheimer approximation, as in DFPT calculations.
$\bar g^2_{\textrm{Gr}}$ with $Q^i_{\textrm{el,Gr}}=0$ : the response of graphene layer is set to zero, i.e. it does not respond to any perturbation. This corresponds to remote electron-phonon coupling screened only by BN's dielectric function. The vertical dashed line indicates the passage from the region where no electron-hole excitation is possible, to the intraband region.}
\label{fig:g210BNGr10BN}
\end{figure}

Fig. \ref{fig:g210BNGr10BN} shows the coupling of graphene's electrons with the collective modes for different setups. In the upper panel we plot the total dynamical coupling $\bar g^2_{\textrm{Gr}}$ (black dashed line), compared to electron-phonon part of the coupling $\bar g^{2,\textrm{ph}}_{\textrm{Gr}}$ (blue line), and to $\bar g^2_{\textrm{Gr}}$ with $Q^0_{\textrm{ph,BN}}=0$ (red line), which represents the plasmon coupling screened by BN's electrons.

The lower panel of Fig. \ref{fig:g210BNGr10BN} shows the expected asymptotic behaviors for the electron-phonon contribution. 
At large momenta, $\bar g^{2,\textrm{ph}}_{\textrm{Gr}}$ coincides with $\bar g^2_{\textrm{Gr}}$ computed with graphene's static electronic response $Q^0_{\textrm{el,Gr}}(\omega=0)$. Here, the dynamical nature of the plasmon is not felt, and graphene's free carriers screen the LO phonon statically. We thus recover the static, Born-Oppenheimer approximation of DFPT (see App. \ref{app:GrBNcheck} for a direct comparison of VED and DFPT electron-phonon couplings in a BN/Graphene system in this regime).
In the small momentum limit, the total coupling between electrons and collective modes can be qualitatively divided into two contributions. The dominant part to the total coupling is due to the plasmon ($\bar g^2_{\textrm{Gr}}$ with $Q^0_{\textrm{ph,BN}}=0$). The phonon contribution, $\bar g^{2,\textrm{ph}}_{\textrm{Gr}}$, is seen to approach $\bar g^2_{\textrm{Gr}}$ with $Q^i_{\textrm{el,Gr}}=0$, which represents the electron-phonon coupling obtained by completely suppressing the response from graphene but keeping the static electronic screening from BN. In other words, the phonon is too fast for graphene's electrons to respond. We will refer to the coupling in this region as `intrinsic BN'.

We remind that, following the discussion of App. \ref{app:scatt}, only in these asymptotic regimes the electron-phonon interaction deduced in this work may be rigorously separated from the electron-plasmon scattering and used inside a Fermi golden rule approach to compute electronic scattering rates due to phonons.
In between the asymptotic regimes, following our description there is a region where there is no clear phonon-driven coupling ($g^{2,\textrm{BN}}_{\textrm{Gr}}<0$), which explains the dip in the the electron-phonon coupling curve. Nevertheless, the full coupling with the hybrid plasmon-phonon mode could induce some momentum relaxation, which we leave for future studies. Now that we discussed the division of the couplings, we can study the effect of the number of layers and of the Fermi level on the electron-phonon coupling.

\subsubsection{Effects of number of BN layers}

\begin{figure}[h!]
\includegraphics[width=0.48\textwidth]{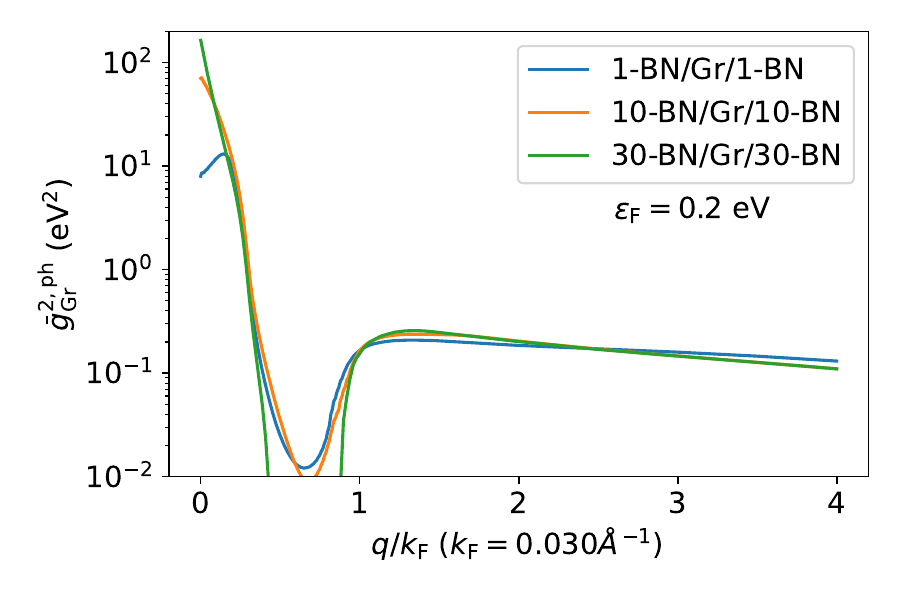}
\caption{Evolution of the electron-phonon coupling as a function of the number of BN layers.}
\label{fig:gNBNGrNBN}
\end{figure}

Fig. \ref{fig:gNBNGrNBN} shows the evolution of the electron-phonon coupling as a function of the number of layers. The `intrinsic BN' $q\rightarrow 0$ limit of the electron-phonon coupling increases as $\frac{n}{\sqrt{n}} = \sqrt{n}$, where $n$ is the number of BN layers. This numerator comes from the sum of the dipole fields over the layers, while the denominator accounts for the the highest LO phonon eigenvector normalization\cite{Sohier2017a}.
The coupling decreases with the number of layers at large $q$. This can be traced back to the intrinsic screening from BN, of the form $\epsilon \sim 1+N\alpha q+o(q^2)$.
There is a saturation in this decrease of the coupling at large $q$. Indeed, the reach of the interlayer Coulomb interactions goes as $e^{-qd}$ where $d$ is the distance between two given layers. At a given $q$, graphene only feels BN layers at distance such that $qd \sim 1$. At large $q$, the threshold $d$ decreases and BN layers added beyond that won't interact with graphene's electrons.
%for wavevector $q \ll \frac{1}{t}$, where $t$ is the full width of the heterostructure, graphene interact with all the BN layers. If $t$ is increased, at fixed wavevector the above condition is weakened and the far away and the effective number of BN layers acting on graphene saturates. 

For the system with only 1 BN on each side and at small q, the coupling can actually be stronger than the `intrinsic BN' coupling, just before reaching it in the $q \rightarrow 0$ limit. The phonon part of the plasmon-phonon hybrid is thus stronger than the intrinsic phonon itself. One may consider that the plasmon is driving the phonon part beyond the normal self-sustained phonon excitation. This is likely due to the relative strength of the plasmon with respect to the phonons.

\subsubsection{Effects of graphene's Fermi level}

\begin{figure}[h]
\includegraphics[width=0.48\textwidth]{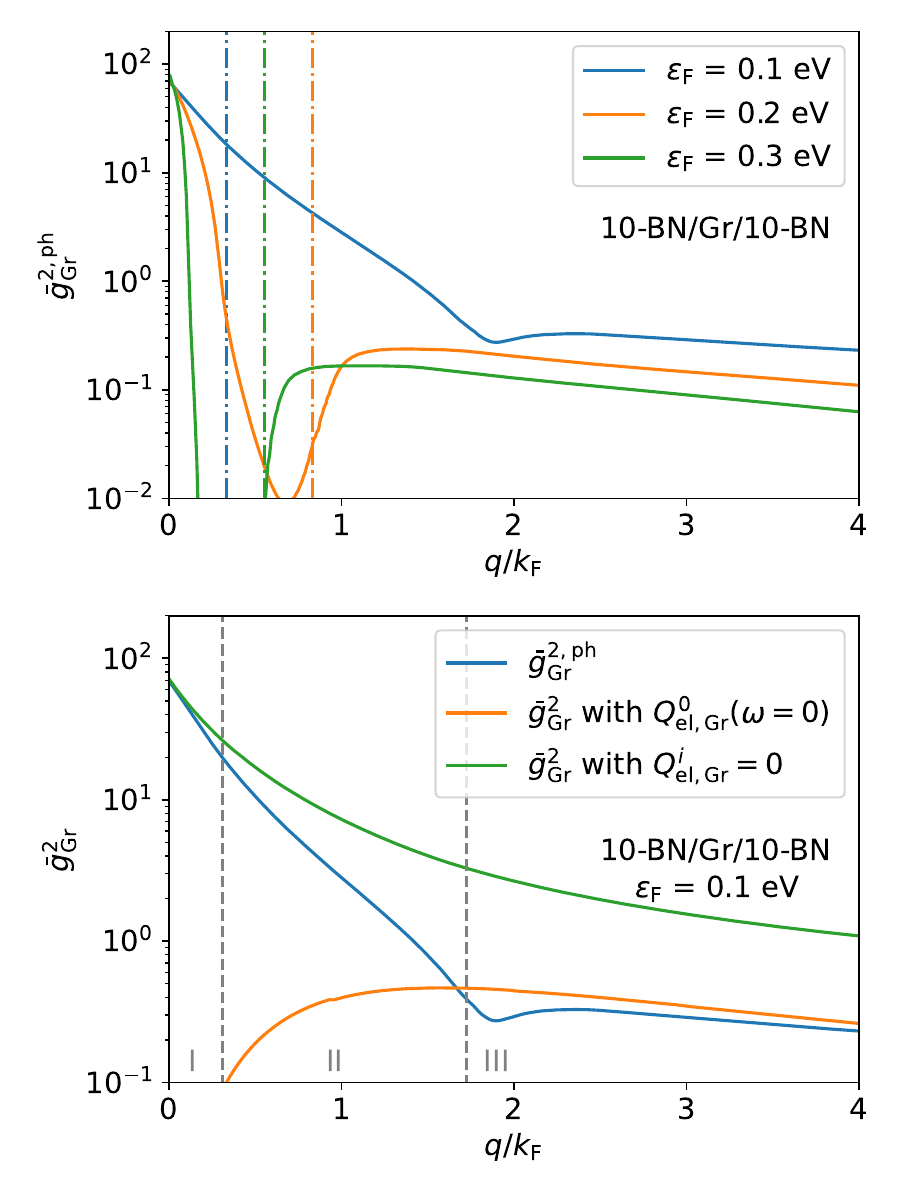}
\caption{(Upper) Evolution of the electron-phonon coupling as a function of doping. The vertical dash-dot lines represents the limit of the electron-hole continuum. For transport, only the momenta larger than this limit mater (not a hard limit, modulated by temperature). (Lower) Asymptotic limits of the electron-phonon coupling, as in Fig. \ref{fig:g210BNGr10BN}, but for $\varepsilon_{\textrm{F}}=0.1$eV. The vertical dashed lines indicate the passage from the region where no electron-hole excitation is possible, to the interband and then intraband regions.}
\label{fig:gvsdoping}
\end{figure}

We verified that the plasmon part of the coupling, which is dominant in the $q\rightarrow 0$ limit, reaches a maximum value independent of $\varepsilon_{\textrm{F}}$. Also, removing the atomic contribution to the screening, the electron-plasmon coupling was checked to be mostly independent of the Fermi level in units of $q/k_{\textrm{F}}$. Thus, differences in the total coupling at different doping levels are mostly due to the phonon contribution, which we study here.
\\
The upper panel Fig. \ref{fig:gvsdoping} shows the electron-phonon coupling as a function of the Fermi level in graphene. The VED framework offers an interesting opportunity to elucidate how the electron-phonon interaction is screened by free-carriers as a function of doping. As discussed further in Section \ref{sec:scattdisc}, this is a non-trivial question in the case of doped 2d semiconductors. 
\\
Neglecting screening from graphene is only valid in the $q\rightarrow 0$ limit, where the electron-phonon coupling tends to the intrinsic LO phonon coupling (see lower panels of Figs.  \ref{fig:g210BNGr10BN} and \ref{fig:gvsdoping}). The vertical dashed lines in Fig. \ref{fig:gvsdoping} show the momenta at which the electron-hole continuum is entered. For $\varepsilon_{\textrm{F}}=0.2$ and $0.3$ eV, this coincides with the onset of strong metallic screening as obtained in the static limit of graphene's response. Only in the case of $\varepsilon_{\textrm{F}}=0.1$ eV does the coupling maintain a large value for a considerable interval of wavevectors when entering the electron-hole continuum.
\\
The reason can be traced back to the crossing the interband region II of the electron-hole continuum.  In fact, for $\varepsilon_{\textrm{F}}= 0.1$ eV (lower panel of Fig. \ref{fig:gvsdoping}) region I is encountered when $0<q/k_{\textrm{F}}<0.4$, region II when $0.4<q/k_{\textrm{F}}<2$ and region III when $q/k_{\textrm{F}}>2$. Instead, for $\varepsilon_{\textrm{F}}= 0.2$ (lower panel of Fig. \ref{fig:g210BNGr10BN}) we have that region II is not encountered, and the phonons directly cross from region I to III at $q/k_{\textrm{F}}\sim 0.9$. At large momenta, for all the Fermi energies the coupling is statically screened by graphene's free carriers. The difference between the curves of Fig. \ref{fig:gvsdoping} at $q>2 k_{\textrm{F}}$  is due to $q/k_{\textrm{F}}$ scale, since graphene's screening is constant and the coupling is independent of $k_{\textrm{F}}$.
%Note that this corresponds to any intraband scattering process where energy and momentum selection rules apply to the initial and final electronic states involved in the coupling. 

We conclude that within the electron-hole continuum, the screening of the electron-phonon interaction by graphene's free carriers is always well approximated by its static limit in the intraband continuum.
%neglecting screening from graphene altogether mostly applies 
Only when the phonon dispersion crosses the interband  continuum does it significantly weakens. In that situation, as shown in Fig. \ref{fig:chi10BNGr10BNdoping}, the anti-crossing disappears to be replaced by a collective mode which mostly follows the intrinsic LO phonon dispersion. 
Note that the case of semi-metallic graphene is peculiar, since interband transitions are easily accessible, as discussed in Sec. \ref{sec:scattdisc} in details.

\subsection{Remote versus intrinsic scattering of graphene's electrons}
\label{app:compaLOTOKBN}

As a final demonstration of the insight brought by the VED framework, we compare the remote scattering from BN's phonons to the intrinsic scattering from graphene's optical phonons. We focus here on momentum scattering, which contributes to the resistivity of graphene. Following Ref. \cite{Sohier2014}, in graphene, one may define an isotropic and energy-dependent transport scattering time $\tau^{\textrm{Tr}}$ from which the resistivity may be expressed as: 
\begin{align}
\frac{1}{\rho}=\frac{e^2}{\pi\hbar^2}\int_{-\infty}^{\infty} d\varepsilon |\varepsilon| \tau^{\textrm{Tr}}(\varepsilon)\left( -\frac{\partial n^{\textrm{FD}}_{\varepsilon}}{\partial \varepsilon}\right).
\end{align}
\\
In general, $\tau^{\textrm{Tr}}(\varepsilon)$ is obtained by solving the full Boltzmann transport equations, i.e. a set of coupled equations relating $\tau^{\textrm{Tr}}$ at different energies\cite{Sohier2014}. All phonons contribute to each of the coupled equations, and it is not straightforward to separate their respective contributions to $\tau^{\textrm{Tr}}(\varepsilon)$ . However, within the relaxation time approximation ($\tau^{\textrm{Tr}}(\varepsilon)$ weakly dependent on energy), the equations decouple and the scattering rate $\tau^{\textrm{Tr}, -1}(\varepsilon)$ (i.e. the inverse of the scattering time) is trivially expressed as a sum over phonon modes. The quantitative impact on the intrinsic resistivity is negligible at room temperature and in the range of Fermi levels studied here (see Fig.6 of Ref.\cite{Sohier2014}). Within this approximation, it is thus reasonable to compute and compare scattering rates from each type of phonons.  $\tau^{\textrm{Tr},-1}$ then differs from the self-energy/ARPES scattering rates $\tau^{-1}$ (Eq. \ref{eq:lifetime}) by an additional $\left(1-\cos\theta\right)$ term inside the angular integration, where $\theta$ is the scattering angle between $\mathbf{k}$ and $\mathbf{k+q}$. That term gives more importance to the back-scattering transitions \cite{Sohier2014} ($\theta=\pi$).
Finally, notice that those approximations preserve the electron-hole symmetry of scattering rates.

\textit{For intrinsic scattering mechanisms} in graphene, we use results from App. B of Ref. \cite{Sohier2014}.  %$\tau^{\textrm{Tr},-1}$ represent the exact solution of the Boltzmann equation in the case of quasi-elastic scattering. In our case, we expect them to be a good approximation to the exact solution when the phonon energy is small compared to the Fermi energy. Nonetheless, even when the phonon frequency is comparable to the Fermi energy, the ratio between transport scattering rates due to different optical phonons is expected to be comparable to the ratio of the exact solutions of the Boltzmann equation. 
%In all this section, we will express scattering rates for a fixed initial electron at energy $\varepsilon$ at a quasi-momentum $\mathbf{k}$ (taken with respect to $\mathbf{\textrm{K}}$) oriented as in Fig. \ref{fig:qselect}. Due to isotropy, this choice doesn't impact the scattering rate value. Every quantity is then accordingly written as a function of energy, rather than momentum. The resistivity $\rho$ is written as
%\begin{align}
%\frac{1}{\rho}=\frac{e^2}{\pi\hbar^2}\int_{-\infty}^{\infty} d\varepsilon |\varepsilon| \tau^{\textrm{Tr},-1}(\varepsilon)\left( -\frac{\partial n^{\textrm{FD}}_{\varepsilon}}{\partial \varepsilon}\right).
%\end{align}
We focus on the scattering rates coming from LO and TO phonons at zone center ($q \sim \Gamma$) with frequency $\omega_{\textrm{LO}\Gamma}=\omega_{\textrm{TO}\Gamma}=\omega_{\textrm{O}}\sim 0.2$eV, and A$_1$ phonons at zone border ($q \sim \textrm{K}$) with frequency $\omega_{K}\sim 0.15$eV.
Keeping only phonon-dependent quantities, the transport scattering rate from the sum of LO and TO phonons is proportional to
\begin{align}
\tau^{\textrm{tr},-1}_{\textrm{O}}(\varepsilon) =  \frac{\hbar \beta^2_{\textrm{O}}}{2M\omega_{\textrm{O}}} \mathcal{W}_\pm(\hbar \omega_{\textrm{O}},\varepsilon),\\
\mathcal{W}_\pm( \hbar \omega_{\textrm{O}},\varepsilon) =  \left( n^{\textrm{BE}}_{ \omega_{\textrm{O}}}+\frac{1}{2} \mp \frac{1}{2} \right) \times \nonumber \\ \frac{|\varepsilon \pm \hbar \omega_{\textrm{O}}|}{(\hbar v_{\textrm{F}})^2(1-n^{\textrm{FD}}_{\varepsilon})} (1-n^{\textrm{FD}}_{\varepsilon\pm\hbar \omega_{\textrm{O}}}),
\label{eq:scattBN}
\end{align}
where the summation over both the $\pm$ terms is intended (`+' stands for absorption and `-' for emission). $\beta_{\textrm{O}}$ is the strength of the optical electron-phonon coupling, with $\frac{\hbar \beta^2_{\textrm{O}}}{2M\omega_{\textrm{O}}} = 0.11$ eV$^2$ as obtained from GW calculations. $M$ is the mass of carbon atoms, and $n^{\textrm{BE}}$ and $n^{\textrm{FD}}$ are taken at room temperature here. The scattering rate for the zone-border phonon at K reads
\begin{align}
\tau^{\textrm{tr},-1}_{\textrm{K}}(\varepsilon) =  \frac{\hbar \beta^2_{\textrm{K}}}{2M\omega_{\textrm{K}}} \mathcal{W}_\pm(\hbar \omega_{\textrm{K}},\varepsilon)  \int \frac{d\theta}{2\pi} \Big(1- s_{\textrm{K}}
\cos\theta\Big) \times \nonumber \\
(1 - \cos\theta) = \frac{\hbar  \beta^2_{\textrm{K}}}{2M\omega_{\textrm{K}}} \mathcal{W}_\pm(\hbar \omega_{\textrm{K}},\varepsilon)   (1 + s_{\textrm{K}} \frac{1}{2}),
 \label{eq:Kscatt}
\end{align}
where $s_{\textrm{K}}= \textrm{sign}(\varepsilon) \textrm{sign}(\varepsilon\pm \hbar \omega_{\textrm{K}})$ is positive (negative) for intraband (interband) transitions, and $\frac{\hbar \beta^2_{\textrm{K}}}{2M\omega_{K}} = 0.18$ eV$^2$ as obtained from  GW calculations at large doping \cite{Attaccalite2010}. The calculation of the correct value for $\beta_{K}$, in particular at small doping, is still matter of active theoretical and experimental research \cite{Venanzi2023,PhysRevB.109.075420,Graziotto2024}. The value used here is a lower limit, such that the scattering from those phonons is likely much stronger in practice. 
The first factor in the integral comes from the angular dependence of the wavefunctions bracketing the electron-phonon interaction, with a different sign of the cosine for intraband and interband transitions. The second is the usual angular term in transport giving more importance to the back-scattering transitions.

\textit{For remote scattering from BN's phonons}, we compute the scattering time as explained in App. \ref{app:scattBNremote}. In particular, we start from Eq. \ref{eq:lifetime} and insert the $(1-\cos\theta)$ term. Then, as for the intrinsic scattering, we assume that the relevant BN's phonons are dispersionless phonons of frequency $\sim \omega_{\textrm{BN}}=0.18$eV. We then consider only contributions coming from electron-phonon scattering, i.e. we consider only the coupling $\bar g^{2,\textrm{ph}}_{\textrm{Gr}}$, which is isotropic in momentum space to a very good approximation. Energy conservation is imposed, and the sum over $\mathbf{q}$ points transform in an angular integral. %The radial integral in fact disappears because for each $\theta$ there exist only one $\mathbf{q}$ that is admitted by energy conservation. 
Since energy conservation changes from emission to absorption, we define the admitted scattering wavevectors, at a given $\theta$, as $q^{\pm}_{\theta}$ (see Fig. \ref{fig:qselect}). As detailed in App. \ref{app:scattBNremote} we obtain that the remote phonon scattering rate is proportional to
\begin{align}
\tau^{\textrm{tr},-1}_{\textrm{BN}}(\varepsilon) =  \mathcal{W}_\pm(\hbar \omega_{\textrm{BN}},\varepsilon)   \int \frac{d\theta}{2\pi}  \bar g^{2,\textrm{ph}}_{\textrm{Gr}}(q^{\pm}_{\theta}) \frac{1 +s_{\textrm{BN}} \cos\theta}{2}  \times \nonumber \\
\left(1-\cos\theta\right) , 
\label{eq:BNscatt}
\end{align}
where $s_{\textrm{BN}} = \textrm{sign}(\varepsilon)\textrm{sign}(\varepsilon\pm \hbar \omega_{\textrm{BN}})$. The value of $\bar g^{2,\textrm{ph}}_{\textrm{Gr}}(q^{\pm}_\theta)$ is taken from Fig. \ref{fig:gvsdoping}. Note that the sign of the cosine in the first term of in the angular integral of Eq. \ref{eq:BNscatt}, still depending on whether the transition is intra- or interband, is reversed with respect to Eq. \ref{eq:Kscatt}. This is easily understood noting that the zone border phonon at K leads to an intervalley transition, while BN's phonons scatter within the same K valley.  
\\

% raw values in case anyone complains that some rounded numbers are the same in the table
%0.1 1 2.2411465534375687 0.2008076199654559
%0.1 10 2.7960620081170697 0.25052826476903706
%0.1 30 2.229870114042279 0.1997972465237104
%0.2 1 0.8660936800896221 0.051734891470070384
%0.2 10 0.91805282155897 0.054838597924222764
%0.2 30 0.9209524589231062 0.05501180369605177
%0.3 1 0.6719504675692133 0.040138018914257265
%0.3 10 0.6713084580604082 0.04009966937652903
%0.3 30 0.7793575393191261 0.04655382972992416

\begin{table}
    \centering
    \begin{tabular}{|c|c|c|c|}
    \hline
        $\varepsilon_{\rm F}$ (eV) & n &    $\tau^{\textrm{tr},-1}_{\textrm{BN}}(\varepsilon_{\textrm{F}})/\tau^{\textrm{tr},-1}_{\textrm{O}}(\varepsilon_{\textrm{F}})$ & $\tau^{\textrm{tr},-1}_{\textrm{BN}}(\varepsilon_{\textrm{F}})/\tau^{\textrm{tr},-1}_{\textrm{K}}(\varepsilon_{\textrm{F}})$\\
        \hline
            & 1 &  2.24 & 0.20  \\
        0.1 & 10 &  2.80 & 0.25 \\
            & 30 &  2.23 & 0.20  \\
        \hline
            & 1 & 0.87 & 0.052\\
        0.2 & 10 & 0.92 & 0.055\\
            & 30 & 0.92 & 0.055 \\
        \hline
            & 1 &  0.67 & 0.04\\
        0.3 & 10 & 0.67 & 0.04\\
            & 30 & 0.78 & 0.05 \\
        \hline        
    \end{tabular}
    \caption{Ratio of remote transport scattering rates $\tau^{\textrm{tr},-1}_{\textrm{BN}}(\varepsilon_{\textrm{F}})$ and the intrinsic zone-center $\tau^{\textrm{tr},-1}_{\textrm{O}}(\varepsilon_{\textrm{F}})$ or zone-boundary $\tau^{\textrm{tr},-1}_{\textrm{K}}(\varepsilon_{\textrm{F}})$ ones, for a temperature of $T=300$K.}
    \label{tab:scattrates}
\end{table}

We evaluate the transport scattering rates for $\varepsilon=\varepsilon_{\textrm{F}}=0.1,0.2,0.3$ eV for heterostructures with n-BN layers on each side, n=1,10,30, at room temperature (300K). The results are presented is Tab. \ref{tab:scattrates}.
For $\varepsilon_{\textrm{F}}= 0.2,0.3$ eV, we find that $\tau^{\textrm{tr},-1}_{\textrm{BN}}(\varepsilon_{\textrm{F}})/\tau^{\textrm{tr},-1}_{\textrm{O}}(\varepsilon_{\textrm{F}}) <1$ while  $\tau^{\textrm{tr},-1}_{\textrm{BN}}(\varepsilon_{\textrm{F}})/\tau^{\textrm{tr},-1}_{\textrm{K}}(\varepsilon_{\textrm{F}}) \ll 1$.
For these doping levels $\varepsilon_{\textrm{F}} > \hbar \omega_{\textrm{BN}}$, therefore the admitted scattering is intraband. The remote coupling with BN's LO phonon is always screened by graphene. To understand why in this case remote scattering is smaller than the intrinsic zone center optical scattering, we remark that the electron-phonon coupling for graphene's intrinsic LO and TO phonons is $\frac{\hbar \beta^2_{\textrm{O}}}{2M\omega_{O}} = 0.11$ eV$^2$, independent of the scattering angle\cite{Sohier2014}. Fig. \ref{fig:gvsdoping} shows coupling strengths around the same value for BN's polar phonons. However, the scattering rates also include graphene's electronic wavefunction overlap. Since for intraband transitions this factor vanishes in back-scattering geometry ($s_{\textrm{BN}}=1$), i.e. the most important for transport, one obtains a contribution smaller than graphene's zone center optical phonons in transport, themselves scattering much less than zone border $A_1$ phonons. This holds also for large number of BN layers, since the coupling is essentially independent of the number of layers for large enough momenta, see Fig. \ref{fig:gNBNGrNBN}. Increasing doping further only weakens the relative contribution from BN's phonons.
\\
For $\varepsilon_{\textrm{F}}= 0.1$ eV, the situation is different. We find that $\tau^{\textrm{tr},-1}_{\textrm{BN}}(\varepsilon_{\textrm{F}})/\tau^{\textrm{tr},-1}_{\textrm{O}}(\varepsilon_{\textrm{F}}) > 1$, but still  $\tau^{\textrm{tr},-1}_{\textrm{BN}}(\varepsilon_{\textrm{F}})/\tau^{\textrm{tr},-1}_{\textrm{K}}(\varepsilon_{\textrm{F}}) < 1$. 
To understand why now remote phonons are more important than intrinsic zone-center phonons, we notice that at small doping the possibility of unscreened interband transitions increases the scattering rate due to a strong coupling, as shown in Fig. \ref{fig:gvsdoping}. Further, for interband process the overlap of graphene's wavefunctions now reaches a maximum in the back-scattering geometry ($s_{\textrm{BN}}=-1$). The contribution of BN's LO phonons to graphene's resistivity is thus larger and comparable to the intrinsic contribution from graphene's LO and TO phonons. However, we find that the scattering from BN's phonons remains smaller than the contribution of intrinsic optical $A_1$ phonons at $\mathbf{K}$ at room temperature, despite the overall larger coupling. This is due to the strong reduction of the remote phonon scattering at small $\theta$, and therefore small $q$, due to the $(1-\cos\theta)^2$ angular factor, and to the role played by phonon occupation factors at room temperature. For the intraband contribution (related to phonon absorption), the final electronic states are at $\varepsilon_{\textrm{F}}= 0.3$ eV, and the norm of the corresponding momenta go from $2k_{\textrm{F}}$ to $4k_{\textrm{F}}$, where the remote coupling is screened.
Let us underscore the use of a conservative estimate for coupling to the optical $A_1$ phonons at $\mathbf{K}$, such that they would scatter \textit{at least} 4 times more than remote BN phonons. 

%The dominance of zone border intrinsic optical phonons of graphene mostly comes from the absorption process. To understand that result, first note that the product $(n^{\textrm{BE}}_{\omega}+\frac{1}{2} \mp \frac{1}{2}) (1-n^{\textrm{FD}}_{\varepsilon_{\textrm{F}}\pm\hbar \omega})$ is equal for absorption and emission processes. The most influential factor in $\mathcal{W}_{\pm}$ is thus $|\varepsilon_{\textrm{F}}\pm \hbar \omega|$, which clearly is larger for absorption. Then, the angular integral is larger for absorption of zone border phonons. 
% In practice, the very large remote coupling with BN's phonons at small momenta is never effective. Considering the most critical case of $\varepsilon_{\textrm{F}}=0.1$ eV, 
%Thus, we expect the transport properties of graphene to remain largely dominated by its intrinsic phonons. 

In conclusion, remote electron-phonon coupling with BN's LO phonons does not seem to explain the discrepancies seen between ab initio simulations of intrinsic phonon-limited transport in graphene and experiments in BN-capped graphene at room temperature and higher \cite{Sohier2014,Banszerus2019,Wang2013a}. However, while we don't expect the dominance of the $A_1$ phonons at $\mathbf{K}$ to be challenged, a more accurate assessment of BN's remote contributions to graphene's resistivity is left for future investigations. Indeed, BN's out-of-plane ZO phonons would also participate to electron scattering and, as mentioned before, the definition of electron-phonon coupling proposed here does not describe the full momentum relaxation process from the plasmon-phonon excitation.

\subsection{Discussion: screening of electron-phonon interactions in doped 2d semiconductors}
\label{sec:scattdisc}

In this work we specifically quantify the remote scattering of graphene's electrons from BN's phonons. More generally, qualitative conclusions can be drawn on the scattering from long-range polar phonons in doped 2d semiconductors. In particular, the VED framework sheds some light on the screening of those phonons from the free carriers added by doping. 
\\
Without computing the full dynamical response as done here, there are two easily accessible limits used in the literature: static screening or no screening. It is general knowledge that static screening should be used at high doping when the plasma frequency (which increases with free carrier density) is expected to be well above the phonon frequency, i.e. phonons are much slower than free carriers. In the opposite small doping limit, one may neglect free-carrier screening if the plasma frequency is much smaller than the phonon, i.e. phonons are much faster than free carriers. 
In the case of 2d materials, the universal square root behavior of the plasmon dispersion $\omega_{\textrm{pl}}  \propto \sqrt{q}$  (as opposed to a constant in 3D) somewhat complicates the picture, as there is no single plasma frequency to compare to. The systems studied in this work are then an instructive example for the 2d case.

We only consider scattering rates that involve real states, i.e. within the electron-hole continua. Since the scattering mechanisms sensitive to screening mostly come from polar-optical phonons, we can assume the existence of a constant typical phonon frequency $\omega_{\textrm{ph}}$, despite the presence of a linear dispersion at small momenta. 
If $\omega_{\textrm{ph}}$ is lower than the onset of the interband electron-hole continuum $\omega_{\textrm{inter}}$, i.e. if $\omega_{\textrm{ph}}<\omega_{\textrm{inter}}$, then the screening of scattering from polar phonons by free carriers is always effective, as in the case of the lower panel of Fig. \ref{fig:g210BNGr10BN}. Indeed, in this case the phonon dispersion only crosses the electron-hole continuum in the intraband Landau damping region. When $\omega_{\textrm{ph}}> \omega_{\textrm{inter}}$, the situation is similar to the lower panel of Fig. \ref{fig:gvsdoping}. While screening remains effective in the intraband continuum, the phonon also crosses the interband continuum,  where free-carrier screening vanishes or at least is weaker. 
\\
Graphene being gapless, $\omega_{\textrm{inter}}$ can be made very small with respect to $\omega_{\textrm{ph}}$ by lowering the doping. Typical high-mobility doped semiconductors, however, have a well-isolated transport band \cite{Sohier2020},  and $\omega_{\textrm{inter}}$ remains finite even at vanishing doping. Further, other bands are far enough that $\omega_{\textrm{ph}}< \omega_{\textrm{inter}}$, in which case the strong screening from free carriers is effective. 
Of course, the assessment of a realistic situations is more complex, with multiple phonon frequencies and multiple onsets of interband transitions to be considered.
\\
Nevertheless, given the above arguments, we conclude that simply neglecting free-carrier screening in 2d doped semiconductors is not valid, even in the low doping limit. Yet, this approximation is often done in state-of-the-art first-principles electronic transport calculations\cite{Kaasbjerg2012a,PhysRevLett.130.166301,PhysRevLett.130.087001,CHENG2021110468,D3CP00036B,10.1063/5.0015102, Cheng2018,PhysRevB.107.155101,Li2019a}. Instead, without access to the full dynamical screening, using the static free-carrier screening might be a more reasonable and robust approximation \cite{Ma2014a,Sohier2019,Sohier2020,Sohier2021a,PhysRevMaterials.2.114010, Macheda2023}.

\section{Conclusions}
In summary, we have developed an ab initio computational technique capable of describing plasmons, phonons and their mixing in Van der Waals heterostructures. We avoid the computational effort of simulating the full heterostructure. Instead, single layer response functions are computed and inserted in a system of equations that takes into account the electrodynamic interactions between different layers. 
Despite the reduced computational cost, our method is able to reproduce the results obtained by full ab initio calculations in the static limit of electronic responses, and extends to the dynamical case. Those developments are applied to BN-capped graphene, where we study the features of the collective excitations and their evolution when changing the number of BN layers and the doping level. In particular, we show that the phonon and plasmon mode mixing generates interesting observable results, such as the stiffening of the plasmon-phonon-polariton group velocity due to both vanishing remote screening of BN's phonons from graphene and the interplay with the plasmon. 
We also obtain the coupling of the collective modes with electrons of the heterostructure.
In particular, we investigate the role played by the plasmon in screening the remote electron-phonon coupling between graphene's electrons and BN's LO phonons. We quantitatively describe the crossover between the regions where the coupling is screened by graphene's free carriers (intraband region), to the one where it is not (interband region).
%We quantitatively describe the crossover between the regions where the coupling is screened by graphene free carriers ($\omega_{\textrm{ph}}(\mathbf{q})\ll \omega_{\textrm{pl}}(\mathbf{q})$), to the one where it is not  ($\omega_{\textrm{ph}}(\mathbf{q})\gg \omega_{\textrm{pl}}(\mathbf{q})$). This picture works surprisingly well even when the plasmon enters in the interband continuum and is therefore dispersed into the electron-hole continuum. 
This implies a contribution to graphene's low doping resistivity from remote BN's LO phonons that is similar to graphene's zone-center optical phonons, but it is still negligible with respect to the scattering from intrinsic zone-border optical phonons. Thus, it is not sufficient to explain the large discrepancy between experiments and state-of-the-art ab-initio simulations\cite{Sohier2014,Park2014,Banszerus2019,Wang2013a}. 
For a general heterostructure, one expects the interplay between plasmons and phonons to increase in complexity. This may be due to a complex band structure of the metallic layer presenting e.g. subbands \cite{Lee1999}, intersubband or multi-valley \cite{Sohier2019a} plasmons, or to an increased importance of cross-talks between in-plane and out-of-plane electrostatics \cite{Royo2021}. Both of these cases are exciting perspectives for this work.

\begin{acknowledgements}
We thank A. Guandalini, P. Barone, M. Stengel, M. Gibertini and G. Cassabois for useful discussions. F. Mauri and F. Macheda acknowledge the MORE-TEM ERC-SYN project, grant agreement no. 951215. F. Mauri and F. Macheda acknowledge the EuroHPC Joint Undertaking for awarding this project access to the EuroHPC supercomputer LUMI, hosted by CSC (Finland) and the LUMI consortium through a EuroHPC Regular Access call. First-principles calculations were performed with the support of MESO@LR-Platform at the University of Montpellier.
\end{acknowledgements}
\clearpage

\appendix

\section{Theoretical aspects of dispersions, couplings and scattering rates}
\subsection{Basic definitions}
\label{app:Green}
In these Sections we restore the possible anisotropy of the system using the vectori symbols (as e.g. $\mathbf{q}$ and $\mathbf{r}$) for momentum and space related quantities. We define a generic retarded response as
\begin{align}
f^{\textrm{R}}(\mathbf{r},\mathbf{r'},t)=-\frac{i}{\hbar}\theta(t)\langle [\hat A(\mathbf{r},t),\hat A(\mathbf{r})]\rangle
\end{align}
where $\langle \rangle$ indicates the average over the ground state (or the thermal average), $\theta$ is the Heaviside function and $\hat A$ is a generic operator. Its Fourier transform satisfies the property
\begin{align}
f^{\textrm{R}}(\mathbf{q},\omega)=\left[f^{\textrm{R}}(\mathbf{q},-\omega)\right]^{\textrm{*}}.
\label{eq:retcc}
\end{align}
The retarded density-density response function in the random phase approximation (RPA) is expressed as a function of the irreducible independent-particle density-density response $\chi^{\textrm{R}}_{\textrm{irr}}$ as in Eq. \ref{eq:Qelfree}. The non-interacting retarded Green function for a phonon of branch $\mu$ at zero temperature may be written as a function of the harmonic oscillator ladder operators $a,a^{\dagger}$ in real time as
\begin{align}
D^{\textrm{R}}_{\mu}(\mathbf{q},t)=-i\theta(t)\langle \hat a^{\dagger}_{\mu\mathbf{q}}(t)a^{\dagger}_{\mu\mathbf{q}}+a_{\mu\mathbf{-q}}(t)a^{\dagger}_{\mu\mathbf{-q}}\rangle,
\end{align}
and in the frequency space as 
\begin{align}
D_{\mu}^{\textrm{R}}(\mathbf{q},\omega)=\frac{2\hbar\omega_{\mu\mathbf{q}}}{(\hbar\omega+i\hbar\eta)^2-\hbar^2\omega_{\mu\mathbf{q}}^2} = \\
-\frac{1}{\hbar\omega+\hbar\omega_{\mu\mathbf{q}}+i\hbar\eta}+\frac{1}{\hbar\omega-\hbar\omega_{\mu\mathbf{q}}+i\hbar\eta}.
\label{eq:Dret}
\end{align}

As shown later, the expression of the total inverse dielectric function contains the density-density response and the phonon propagator, both in the retarded flavour. We drop the retarded apex everywhere in the text, so that any Green/response function is to be intended as retarded. We also remind that to deduce the finite temperature expression of the retarded Green/response functions, one computes their time ordered expression in the Matsubara formalism and then performs the substitution $i\hbar\omega_{n}\rightarrow \hbar\omega+i\hbar\eta$ ~\cite{Mahan1990}. As a consequence of using retarded Green functions, all the computed quantities will be analytical and will therefore respect the Kramers-Kronig relations \cite{DeL.Kronig1926}.

\subsection{Long-range atomic-mediated inverse dielectric function/screened Coulomb}
\label{app:invdieltens}

The goal of this Section is to justify the form of Eq. \ref{eq:epsatom} for a wide gap semiconductor such as BN. Therefore,  we here perform the derivation of the long-range atomic-mediated inverse dielectric function and the related screened Coulomb interaction. We start from the derivation for the 3d case. In this case, following Ref. \cite{Macheda2022} we use the RPA approximation for the electronic response and obtain that the atomic-mediated inverse dielectric function can be expressed in terms of effective charges. We then use the conclusions and procedures of Ref. \cite{Macheda2023} to extend the results to the case of a 2d single layer of small width.
\subsubsection{3d expression for the inverse dielectric function}
\label{app:invdieltens3d}

The screened electronic interaction is defined as \cite{Keldysh2014}
\begin{align}
v(\mathbf{r})=\frac{4\pi e^2}{|\mathbf{r}|},\\
w(41)=\int d\mathbf{r}''\epsilon^{-1}(4,\mathbf{r''},t_1)v(\mathbf{r''}-\mathbf{r}_1),
\label{eq:weps}
\end{align}
where the compact notation $1=(\mathbf{r}_1,t_1)$ has been introduced. The total inverse dielectric function---and consequently the Coulomb potential---is usually split into `electronic' (clamped-ion) and `phononic' (total minus clamped-ion) contributions as
\begin{align}
\epsilon^{-1}=\epsilon^{-1}_{\text{el}}+\epsilon^{-1}_{\text{ph}}, \label{eq:epsm1split}\quad w=w_{\textrm{el}}+w_{\textrm{ph}}.
\end{align}
Such separation bring great simplification if vertex corrections that couple the systems are small \cite{Mahan1990}, as assumed here. The phononic part of the screened Coulomb interaction may be written, following Refs. \cite{Hedin1970,Dolgov1981,osti_6556355}, as
\begin{align}
\epsilon^{-1}_{\textrm{ph}}(12)=\sum_{\substack{s\alpha p\\s'\alpha' p'}}\int d(34) \epsilon^{-1}_{\text{el}}(13) \left(\nabla_{\mathbf{r}_3}\right)_\alpha V_s(\mathbf{r}_3-\mathbf{u}^{p}_{s}) \times \nonumber \\
D_{s\alpha p,s'\alpha'p'}(t_3t_4)\frac{\nabla^2_{\mathbf{r}_4}}{4
\pi e^2}\epsilon^{-1}_{\text{el}}(42)\left(\nabla_{\mathbf{r}_4}\right)_{\alpha'}V_{s'}(\mathbf{r}_4-\mathbf{u}^{p'}_{s'}).
\label{eq:wphdirect}
\end{align}
where $D$ is the dressed phonon propagator. $\mathbf{u}^{p}_s=\mathbf{R}_p+\boldsymbol{\tau}_s$ are the atomic equilibrium positions, which in reciprocal space representation read as
\begin{align}
\mathbf{u}^{\mathbf{q}}_{s}=\sum_{p}e^{i\mathbf{q}\cdot(\mathbf{R}+ \boldsymbol{\tau}_s)} \mathbf{u}^{p}_{s}.
\end{align}
$V_s$ indicates, in the all-electron formalism, the ion-electron interaction due to the ion at site $s$, and its Fourier transform is
\begin{align}
V_{s\alpha}(\mathbf{q+G})= \frac{1}{e}v(\mathbf{q+G})\rho^{\textrm{ext}}_{s\alpha}(\mathbf{q+G}), \\
\rho^{\textrm{ext}}_{s\alpha}(\mathbf{ q}+\mathbf{ G})=-i\frac{\mathcal{Z}_s e}{V}\left[( q_{\alpha}+ G_{\alpha})e^{-i\mathbf{ G}\cdot\boldsymbol{ \tau}_{s}}\right],
\end{align}
where $\mathcal{Z}_s$ are the ionic charges. We assume to be looking at times large enough that all the response functions depend only on relative time, the retardation condition restoring the causality at the end. We express the phonon propagator as
\begin{widetext}
\begin{align}
D_{s\alpha p,s'\alpha'p'}(\omega)=\frac{1}{\sqrt{M_{s}M_{s'}}}\sum_{\mu}\int \frac{d\mathbf{q}}{V_{\text{BZ}}}\frac{e_{\mu\mathbf{q}}^{s \alpha}e^{i\mathbf{q}\cdot (\mathbf{R}_p+\boldsymbol{\tau}_s)}e_{\mu\mathbf{q}}^{s' \alpha'* }e^{-i\mathbf{q}\cdot (\mathbf{R}_{p'}+\boldsymbol{\tau}_{s'})}}{(\hbar\omega+i\hbar\eta)^2-\hbar^2 \omega^2_{\mu\mathbf{q}}},
\end{align}
\end{widetext}
where $e_{\mu\mathbf{q}}^{s\alpha}$ are eigenvectors of the dynamical matrix and $\omega_{\mu\mathbf{q}}$ its eigenvalues. Notice that we have written the phonon propagator in the `clean limit', i.e. we assume a delta-like spectral function. In BN (bulk or single layer) the clean limit is respected. Notice also that the dynamical frequencies coincide with the adiabatic ones, since it is a large gap semiconductor. We now introduce the in-plane screened and unscreened charge densities, and related effective charges, in the RPA approximation for $\epsilon^{-1}_{\textrm{el}}$
\cite{Macheda2022,Macheda2023,Binci2021,Marchese2023,PhysRevLett.128.095901,Wang2022a}. These read as
\begin{align}
\rho_{s\alpha}(\mathbf{q},\omega)=\epsilon^{-1}_{\textrm{el}}(\mathbf{q},\omega) \bar \rho_{s\alpha}(\mathbf{q},\omega),  \\
\bar Z_{s\alpha}(\mathbf{q},\omega)=i\frac{V}{eq} \bar \rho_{s\alpha}(\mathbf{q},\omega) .
\end{align}
An explicit expression for the effective charges is found as
\begin{align}
\bar Z_{s\alpha}(\mathbf{q},\omega)=\frac{\mathcal{Z}_{s}}{q v(\mathbf{q})}\nonumber \times \\
\sum_{\mathbf{G}} \frac{\epsilon^{-1}_{\text{el}}(\mathbf{ q},\mathbf{ q}+\mathbf{G},\omega)}{\epsilon^{-1}_{\text{el}}(\mathbf{q},\omega)}
(q_{\alpha}+G_{\alpha}) v(\mathbf{q+G}) e^{-i\mathbf{G}\cdot\boldsymbol{\tau}_s},
\label{eq:effcharge}
\end{align}
where $\mathbf{G}$ are reciprocal lattice vectors. Using all the above expressions, and
Fourier transforming Eq. \ref{eq:wphdirect} we find, for the atomic-mediated macroscopic component of the Coulomb interaction
\begin{widetext}
\begin{align}
w_{\textrm{ph}}(\mathbf{q},\omega)=-\frac{1}{V}q^2 \left[w_{\textrm{el}}(\mathbf{q},\omega)\right]^2\frac{1}{\sqrt{M_{s}M_{s'}}}\sum_{\mu} \frac{\left(\sum_{s\alpha}\bar Z_{s\alpha}(\mathbf{q},\omega) e_{\mu\mathbf{q}}^{s \alpha}\right)\left(\sum_{s'\alpha'}\bar Z_{s'\alpha'}(-\mathbf{q},\omega)e_{\mu\mathbf{-q}}^{s' \alpha' }\right)}{(\hbar\omega+i\hbar\eta)^2-\hbar^2 \omega^2_{\mu\mathbf{q}}}=\nonumber \\
\frac{V}{e^2} \left[w_{\textrm{el}}(\mathbf{q},\omega)\right]^2\frac{1}{\sqrt{M_{s}M_{s'}}}\sum_{\mu} \frac{\left(\sum_{s\alpha}\bar \rho_{s\alpha}(\mathbf{q},\omega) e_{\mu\mathbf{q}}^{s \alpha}\right)\left(\sum_{s'\alpha'}\bar \rho_{s'\alpha'}(-\mathbf{q},\omega)e_{\mu\mathbf{-q}}^{s' \alpha' }\right)}{(\hbar\omega+i\hbar\eta)^2-\hbar^2 \omega^2_{\mu\mathbf{q}}},
\label{eq:wphorders}
\end{align}
\end{widetext}
where
\begin{align}
e_{\mu\mathbf{-q}}^{s' \alpha' }=e_{\mu\mathbf{q}}^{s' \alpha' *}.
\end{align}
For an isolant such as BN, for frequencies much smaller than the band gap, we have 
\begin{align}
\epsilon^{-1}_{\textrm{el}}(\mathbf{q},\omega)\sim \epsilon^{-1}_{\textrm{el}}(\mathbf{q}), \quad 
\bar Z(\mathbf{q},\omega)\sim \bar Z(\mathbf{q}).
\end{align}
The unscreened effective charges have an expansion of the form \cite{Macheda2022,Macheda2023}
\begin{align}
\bar Z_{s\alpha}(\mathbf{q})=\frac{q_{\beta}}{q}\bar Z_{s\alpha\beta}-\frac{i}{2} \frac{q_{\beta}}{q}q_{\gamma}Q_{s\alpha\beta\gamma}+\nonumber -\frac{1}{3!}\frac{q_{\beta}}{q}q_{\gamma} q_{\delta} O_{s\alpha\beta\gamma\delta}\\
+\mathcal{O}(q^3)\nonumber ,
\end{align}
where the various orders of the expansion are the Born effective charges, dynamical quadrupoles and octupoles, as so on.
With the above expansion, at leading order in $q$, one can rewrite Eq. \ref{eq:wphorders} as
\begin{align}
w_{\text{ph}}(\mathbf{q},\omega)=\frac{1}{V}\sum_{s\alpha\beta ,s'\alpha'\beta'}\left[w_{\textrm{el}}(\mathbf{q})\right]^2\frac{1}{\sqrt{M_{s}M_{s'}}}\times \nonumber \\
\sum_{\mu} \frac{\left(\sum_{s\alpha\beta}\bar{Z}_{s\alpha\beta}q_{\beta}e_{\mu\mathbf{q}}^{s \alpha}\right)\left(\sum_{s'\alpha'\beta'}\bar{Z}_{s'\alpha'\beta'}q_{\beta'}e_{\mu\mathbf{q}}^{s' \alpha'* }\right)}{(\hbar\omega+i\hbar\eta)^2-\hbar^2 \omega^2_{\mu\mathbf{q}}},
\label{eq:epsphcompact}
\end{align}
which is consistent with the form obtained in Ref. \cite{Keldysh2014}. Eq. \ref{eq:epsphcompact} includes only the leading order Fr\"olich coupling \cite{doi:10.1080/00018735400101213,PhysRevB.92.054307,PhysRevLett.115.176401,PhysRevB.13.694,Pick1970}, whereas Eq. \ref{eq:wphorders} include quadrupoles and higher multipolar order expansions \cite{PhysRevX.11.041027,PhysRevLett.130.166301,PhysRevB.107.155424}. Using the form of Eq. (28) of Ref. \cite{Macheda2023}, one can rewrite the equivalent of Eq. \ref{eq:wphorders} for the inverse dielectric matrix in the static case as a function of the  long-range component of the dynamical matrix:
\begin{align}
\epsilon^{-1}_{\textrm{ph}}(\mathbf{q},\omega) = \epsilon^{-1}_{\textrm{el}}(\mathbf{q}) \sum_{\mu}\frac{\mathbf{e}_{\mu \mathbf{q}} \mathcal{D}^{\textrm{L}}(\mathbf{q}) \mathbf{e}^*_{\mu \mathbf{q}}}{(\hbar \omega+i\hbar \eta)^2-\hbar^2 \omega^2_{\mu \mathbf{q}}}.
\label{eq:epsphcompact2}
\end{align} 
\subsubsection{2d case for a single BN single layer}
\label{app:invdieltens2d}
As in the rest of this work, for a BN single layer we only consider in-plane displacements of atoms, and suppose that there are no cross-talks between in-plane and out-of-plane components of the response functions. The error of this assumption, since our single layers always present in-plane mirror symmetry, is of order $\mathcal{O}(q^2z^2)$. As explained in Ref. \cite{Macheda2023}, if one also assume that the material thickness is much smaller than the wavelength of the perturbation (thin limit), one can then forget about the out-of-plane direction of the electrostatic problem and, for the formulae of App. \ref{app:invdieltens3d}, replace the three-dimensional Coulomb kernel with its two dimensional counterpart, and volume- intensive quantities with the area-intensive ones. In this work, we extend the validity of 2d formulae beyond the thin limit by using Eq. \ref{eq:CoulKernel} to describe the 2d Coulomb interaction. We finally obtain for the LO phonon, assuming a small finite lifetime
\begin{align}
\left[\epsilon^{-1}_{\textrm{1L,ph}}\right]^0(\mathbf{q},\omega) = \left[\epsilon^{-1}_{\textrm{1L,el}}\right]^0(\mathbf{q}) \frac{\mathbf{e}_{\textrm{LO} \mathbf{q}} \mathcal{D}^{\textrm{L}}(\mathbf{q}) \mathbf{e}^*_{\textrm{LO} \mathbf{q}}}{(\hbar \omega+i\hbar \eta_\textrm{LO})^2-\hbar^2 \omega^2_{\textrm{LO} \mathbf{q}}},
\label{eq:epsphcompact3}\\
\left[\epsilon^{-1}_{\textrm{1L,el}}\right]^0(\mathbf{q}) = 1+ v^{00}_{\textrm{BN, BN}}(\mathbf{q}) Q^0_{\textrm{el}}(\mathbf{q}), \\
\left[\epsilon^{-1}_{\textrm{1L,ph}}\right]^1(\mathbf{q},\omega)=0.
\label{eq:epsphcompact3.2} 
\end{align}
The above equations are the one used, for the isotropic case, in Sec. \ref{sec:Qandf}. The explicit expression of the numerator of Eq. \ref{eq:epsphcompact3} for the thin limit is \cite{Macheda2023}
\begin{align}
\mathbf{e}_{\textrm{LO} \mathbf{q}} \mathcal{D}^{\textrm{L}}(\mathbf{q}) \mathbf{e}^*_{\textrm{LO} \mathbf{q}}=\frac{2\pi A}{q}\times \nonumber \\
\left(\sum_{s\alpha} \rho_{s\alpha}(\mathbf{q}) e_{\textrm{LO}\mathbf{q}}^{s \alpha}\right)\left(\sum_{s'\alpha'}\bar \rho_{s'\alpha'}(\mathbf{q})e_{\textrm{LO}\mathbf{q}}^{s' \alpha' }\right)^*.
\label{eq:exacteDe}
\end{align}
In the approximation of this work that the atomic profile density is the same of the electronic one, we extend the above expression to
\begin{align}
\mathbf{e}_{\textrm{LO} \mathbf{q}} \mathcal{D}^{\textrm{L}}(\mathbf{q}) \mathbf{e}^*_{\textrm{LO} \mathbf{q}}=Av^{00}_{\textrm{BN,BN}}\times \nonumber \\
\left(\sum_{s\alpha} \rho_{s\alpha}(\mathbf{q}) e_{\textrm{LO}\mathbf{q}}^{s \alpha}\right)\left(\sum_{s'\alpha'}\bar \rho_{s'\alpha'}(\mathbf{q})e_{\textrm{LO}\mathbf{q}}^{s' \alpha' }\right)^*.
\label{eq:exacteDe2}
\end{align}
Using the Poisson equation we obtain an equivalent formula
\begin{align}
\label{eq:eDe}
\mathbf{e}_{\textrm{LO} \mathbf{q}} \mathcal{D}^{\textrm{L}} (\mathbf{q})\mathbf{e}^*_{\textrm{LO} \mathbf{q}} = \frac{A}{v^{00}_{\textrm{BN,BN}}(\mathbf{q})} &
\int f^0_{\textrm{BN}}(z) \bar{V}_{\textrm{LO}}(\mathbf{q},z) dz  \\
& \times  \int f^0_{\textrm{BN}}(z) V_{\textrm{LO}}^*(\mathbf{q},z) dz . \nonumber
\end{align}
In the above equation we have used the abbreviation $\bar{V}_{\textrm{LO}} (\mathbf{q},z) =  \sum_{s, \alpha} \frac{e_{\textrm{LOq}}^{s, \alpha}}{\sqrt{M_s}}  \frac{\partial \bar{V}_{\textrm{KS}}(\mathbf{q},z)}{\partial u^{\mathbf{q}}_{s, \alpha}}$ is the projection on the LO mode of $\frac{\partial\bar{V}_{\textrm{KS}}(\mathbf{q},z)}{\partial \mathbf{u}^{\mathbf{q}}_s}$ (and the same for non-barred potential;`KS' indicates the Khon-Sham potential). $\bar V_{\textrm{LO}}$ is computed in this work as explained in Ref. \cite{Macheda2023}. The approximation of \ref{eq:eDe} can be bettered assuming that the difference between longitudinal and tranverse optical phonon frequencies is all attributable to the long-range component of the dynamical matrix---i.e. short range interactions differences between LO and TO are negligible. This amounts to writing
\begin{align}
\mathbf{e}_{\textrm{LO} \mathbf{q}} \mathcal{D}^{\textrm{L}}(\mathbf{q}) \mathbf{e}^*_{\textrm{LO} \mathbf{q}}=\hbar^2 \omega^2_{\textrm{LO}\mathbf{q}}-\hbar^2\omega^2_{\textrm{TO}\mathbf{q}},
\label{eq:omLOapp}
\end{align}
as it is customarily done e.g. for 3d cases. This is the same approximation included in Eq. \ref{eq:omLO} in the isotropic case. The difference between using Eq. \ref{eq:omLOapp} and Eq. \ref{eq:eDe} is commented in App. \ref{app:compdets}. \\
Inserting Eq. \ref{eq:omLOapp} in Eq. \ref{eq:epsphcompact3}, we end up with
\begin{align}
\left[\epsilon^{-1}_{\textrm{1L,ph}}\right]^0(\mathbf{q},\omega)=\left[\epsilon^{-1}_{\text{1L,el}}\right]^0(\mathbf{q})\frac{\hbar^2\omega^2_{\textrm{LO}\mathbf{q}}-\hbar^2\omega^2_{\textrm{TO}\mathbf{q}}}{(\hbar\omega+i\hbar\eta_{\textrm{LO}})^2-\hbar^2\omega^2_{\textrm{LO}\mathbf{q}}},
\label{eq:epsm1tot}
\end{align}
which in the isotropic case directly leads to Eq. \ref{eq:Q0phlr}.
 If both the positive and negative phonon spectral functions are very sharp, for positive frequencies we have
\begin{align}
-\textrm{Im}\left[\epsilon^{-1}_{\textrm{1L,ph}}\right]^0(\mathbf{q},\omega)\underset{\omega\sim \omega_{\textrm{LO}\mathbf{q}}}{\sim}
\pi\left[\epsilon^{-1}_{\textrm{1L, el}}\right]^0(\mathbf{q})\times \nonumber \\
\frac{\hbar^2\omega^2_{\textrm{LO}\mathbf{q}}-\hbar^2\omega^2_{\textrm{TO}\mathbf{q}}}{2\hbar\omega_{\textrm{LO}\mathbf{q}}}\delta(\hbar\omega-\hbar\omega_{\textrm{LO}\mathbf{q}}).
\label{eq:vchi}
\end{align}
In the above derivations we have always treated the phonon system as non-interacting; this hypothesis may be lifted by dressing the phonon propagator with the appropriate self-energies. We simplified the treatment by using for $\eta$ a finite constant value, i.e. $\eta_{\textrm{LO}}$.
\subsubsection{Rewriting atomic-mediated screened Coulomb as a function of electron-phonon in the static case}
In the same approximations of App. \ref{app:invdieltens3d} and \ref{app:invdieltens2d}, Eq. \ref{eq:wphorders} can be rewritten in an interesting fashion for the static case via the Fr\"olich electron-phonon coupling stripped of the Bloch functions overlaps. In fact, using Eq. (36) of Ref. \cite{Macheda2023} we define
\begin{align}
g^{\textrm{Fr}}_{\textrm{LO}\mathbf{q}}= \frac{i}{A} w_{\textrm{el}}(\mathbf{q})\sum_{s\alpha\beta}q_{\beta}\bar{Z}_{s\alpha\beta}e_{\textrm{LO}\mathbf{q}}^{s\alpha}\times \left(\frac{\hbar}{2M_s\omega_{\textrm{LO}\mathbf{q}}}\right)^{1/2}.
\label{eq:gfrnobloch}
\end{align}
Eq. \ref{eq:wphorders} then becomes
\begin{align}
\label{eq:wphgfr}
w_{\textrm{ph}}(\mathbf{q},\omega)=A  |g^{\textrm{Fr}}_{\textrm{LO}\mathbf{q}}|^2\frac{ 2\hbar\omega_{\textrm{LO}\mathbf{q}}}{(\hbar\omega+i\hbar\eta_{\textrm{LO}})^2- \hbar^2\omega^2_{\textrm{LO}\mathbf{q}}},
\end{align}
with
\begin{align}
-\textrm{Im}[w_{\textrm{ph}}(\mathbf{q},\omega)]\underset{\omega\sim \omega_{\textrm{LO}\mathbf{q}}}{\sim} \pi A|g^{\textrm{Fr}}_{\textrm{LO}\mathbf{q}}|^2\delta(\hbar\omega-\hbar\omega_{\textrm{LO}\mathbf{q}}).
\label{eq:wphdelta}
\end{align}
This means that the weight of the poles of $w_{\textrm{ph}}(\mathbf{q},\omega_{\textrm{LO}})$ is proportional to the squared modulus of the Fr\"olich coupling.
\\
In the general case where the static approximation for the electronic screening cannot be performed, the imaginary part of the screened Coulomb interaction will contain both electron-electron and electron-phonon contributes. The interplay of the electronic system and the phononic one determines the dispersion of plasmon and phonon poles, and their coupling.

\subsection{Scattering rates and $g^2$}
\label{app:scatt}

The scattering rate $\tau^{-1}$ of electrons relates to the retarded electronic self-energy as
\begin{align}
-\text{Im} \Sigma_{\textrm{el}} = \frac{\hbar}{2}\tau^{-1}.
\label{eq:gammaselfen}
\end{align}
Following Refs. \cite{Kim1978} and \cite{Hauber2017a}, one can deduce the depopulation rate of electrons, i.e. scattering times, using the structure factor and the fluctuation-dissipation theorem, obtaining an expression proportional to $-\textrm{Im} \left[\epsilon^{-1}(\mathbf{q})\right]v(\mathbf{q})$. We generalize such 3d expressions to the 2d case. We take into acocunt the matricial ordering of the inverse dielectric matrix and Coulomb kernel (not relevant in the purely 3d case) by noting that the electronic self-energy reads \cite{Giustino2017}
\begin{align}
\Sigma_{\textrm{el}}(12)=i\hbar\int d(34) G(13) \Gamma(324) w(41^+).
\label{eq:selfenint}
\end{align}
The scattering rate for a Bloch electron $n\mathbf{k}$ is then obtained as
\begin{align}
k_\textrm{B}T\frac{\partial f_{n\mathbf{k}}}{\partial \epsilon_{n\mathbf{k}}}\tau^{-1}_{n\mathbf{k}}=\frac{2\pi}{\hbar}\frac{1}{AN_{\mathbf{q}}}\sum_{\mathbf{q}}\int_{-\infty}^{\infty} d\omega \int d \mathbf{r}dz d \mathbf{r'}dz' d\bar z  \times \nonumber \\ 
n^{\textrm{FD}}_{\varepsilon_{n\mathbf{k}}}\left(1-n^{\textrm{FD}}_{\varepsilon_{m\mathbf{k+q}}}\right)\left(1+n^{\textrm{BE}}_{\omega}\right) \times \nonumber \\
 u^*_{n\mathbf{k}}(\mathbf{r},z)u_{n\mathbf{k}}(\mathbf{r'},z')u_{m\mathbf{k+q}}(\mathbf{r},z)u^*_{m\mathbf{k+q}}(\mathbf{r'},z')  \times \nonumber \\  
\textrm{Im}\left[-\epsilon^{-1}(\mathbf{q},z,\bar z,\omega)\frac{v(\mathbf{q},\bar z,z')}{\pi} \right]\delta(\hbar\omega+\varepsilon_{m\mathbf{k+q}}-\varepsilon_{n\mathbf{k}}).
\label{eq:Pz0}
\end{align}
where we have assumed that final states are represented by non-interacting electrons. Depending on the derivation, the overlap factor comes from bringing the self-energy in the basis of the Bloch basis functions, and taking its diagonal components (see e.g. Eq. (157) of Ref. \cite{Giustino2017}), or from the connection between the scattering rate and the structure factor (see e.g. Eq. (5) of Ref. \cite{Kim1978}). The integral over positive frequencies will produce the scattering due to phonon/plasmon emission, while the one over negative frequencies is related to absorption because $1+n^{\textrm{BE}}_{-\omega}=-n^{\textrm{BE}}_{\omega}$, where the minus sign is compensated by the property \ref{eq:retcc} applied on the inverse dielectric matrix. To connect with usual formulae, by using the algebraic relations discussed in the Supplementary Material of Ref. \cite{Macheda2020}, one finds
\begin{align}
\tau^{-1}_{n\mathbf{k}}=\frac{2\pi}{\hbar}\frac{1}{AN_{\mathbf{q}}}\sum_{\mathbf{q}}\int_{-\infty}^{\infty} d\omega \int d \mathbf{r}dz d \mathbf{r'}dz' d\bar z  \textrm{sign}(\omega)\mathcal{A}(\omega) \times \nonumber \\ 
 u^*_{n\mathbf{k}}(\mathbf{r},z)u_{n\mathbf{k}}(\mathbf{r'},z')u_{m\mathbf{k+q}}(\mathbf{r},z)u^*_{m\mathbf{k+q}}(\mathbf{r'},z')  \times \nonumber \\  
\textrm{Im}\left[-\epsilon^{-1}(\mathbf{q},z,\bar z,\omega)\frac{v(\mathbf{q},\bar z,z')}{\pi} \right]\delta(\hbar\omega+\varepsilon_{m\mathbf{k+q}}-\varepsilon_{n\mathbf{k}}), \nonumber \\
\mathcal{A}(\omega)= \left[n^{\textrm{BE}}_{\omega}+\frac{1}{2}+\textrm{sign}(\omega)\frac{1}{2}-\textrm{sign}(\omega)n^{\textrm{FD}}_{\varepsilon{m\mathbf{k+q}}}\right].
\label{eq:Pz}
\end{align}
In the dual basis set used in this work we have
\begin{align}
 \int d\bar z \epsilon^{-1}(\mathbf{q},z,\bar z,\omega)v(\mathbf{q},\bar z,z')= \nonumber \\
  \sum_{\substack{ii'j'j \\ kk'l'l}} \phi^i_{k}(z)v^{ii'}_{kk'}(\mathbf{q}) \chi^{i'j'}_{k'l'}(\mathbf{q},\omega)v^{j'j}_{l'l}\phi^{j}_l(z').
 \label{eq:imvchiv}
\end{align}
In the scattering rate expression we have pairs of cell-periodic Bloch functions which are building blocks for the density, which gets coherently bracketed with the basis set of Eq. \ref{eq:basisz}. In fact, Eq. \ref{eq:Pz} becomes
\begin{align}
\tau^{-1}_{n\mathbf{k}}=\sum_{\substack{ii'j'j\\kk'l'l}}\frac{2\pi}{\hbar}\frac{1}{AN_{\mathbf{q}}}\sum_{\mathbf{q}}\int_{-\infty}^{\infty} d\omega \bra{u_{n\mathbf{k}}} \phi^{i}_{k}\ket{u_{m\mathbf{k+q}}} \times \nonumber \\
\bra{u_{m\mathbf{k+q}}} \phi^{j}_{l}\ket{u_{n\mathbf{k}}} \textrm{sign}(\omega)\mathcal{A}(\omega) \times \nonumber \\
 \textrm{Im}\Big[-\frac{1}{\pi}v^{ii'}_{kk'}(\mathbf{q})\chi^{i'j'}_{k'l'}(\mathbf{q},\omega)  v^{j'j}_{l'l}(\mathbf{q})  \Big]\delta(\hbar\omega+\varepsilon_{m\mathbf{k+q}}-\varepsilon_{n\mathbf{k}}).
 \label{eq:Pfinal}
\end{align}
\subsubsection{BN single layer in the static case: electron-phonon}
The scattering rate due to electron-phonon coupling for a BN single layer is obtained by using that the density-density response function is of the form of Eq. \ref{eq:chilayer}, where the atomic-mediated contribution to the response is obtained from Eqs. \ref{eq:epsphcompact3} and \ref{eq:epsphcompact3.2}. Due to in-plane mirror symmetry, the unperturbed Bloch wavefunctions are even with respect to the layer position, so that in Eq. \ref{eq:Pfinal} one has the selection rule $i=j=0$. The scattering rate may then be written as
\begin{align}
\tau^{-1,\textrm{LO}}_{n\mathbf{k}}=\frac{1}{A}\frac{2\pi}{\hbar}\int_{-\infty}^{\infty} \hbar d\omega |\braket{u_{n\mathbf{k}}|{u_{m\mathbf{k+q}}}}|^2 \textrm{sign}(\omega)\mathcal{A}(\omega)\times \nonumber \\
\textrm{Im}\left[-\left[\epsilon^{-1}_{\textrm{1L,ph}}\right]^0(\mathbf{q},\omega)\frac{ v^{00}_{\textrm{BN,BN}}(\mathbf{q})}{\pi} \right] \delta(\hbar\omega+\varepsilon_{m\mathbf{k+q}}-\varepsilon_{n\mathbf{k}}),
\label{eq:transprobmonol}
\end{align}
which using Eqs. \ref{eq:wphgfr} and \ref{eq:Dret} becomes the standard expression
\begin{align}
\tau^{-1,\textrm{LO}}_{n\mathbf{k}}=\frac{2\pi}{\hbar}\frac{1}{N_{\mathbf{q}}}\sum_{\mathbf{q}}  |g^{\textrm{Fr}}_{\textrm{LO}\mathbf{q}}|^2|\braket{u_{n\mathbf{k}}|{u_{m\mathbf{k+q}}}}|^2 \times \nonumber \\ 
\Big[\left(1+n^{\textrm{BE}}_{\omega_{\textrm{LO}\mathbf{q}}}-n^{\textrm{FD}}_{\varepsilon{m\mathbf{k+q}}} \right)\delta(\hbar\omega_{\textrm{LO}\mathbf{q}}+\varepsilon_{m\mathbf{k+q}}-\varepsilon_{n\mathbf{k}}) + \nonumber \\
\left(n^{\textrm{BE}}_{\omega_{\textrm{LO}\mathbf{q}}}+n^{\textrm{FD}}_{\varepsilon{m\mathbf{k+q}}}\right) \delta(-\hbar\omega_{\textrm{LO}\mathbf{q}}+\varepsilon_{m\mathbf{k+q}}-\varepsilon_{n\mathbf{k}})\Big].
\label{eq:Psingle}
\end{align}

\subsubsection{Heterostructures}

In general, one can express the imaginary part of the screened Coulomb as a squared modulus times a delta (with a minus sign), i.e. as in Eq. \ref{eq:wphdelta}, only when one the electronic dielectric function can be treated as static. In the heterostructures studied in this work, this is not always the case, due to the coupling of plasmons and phonons. Nevertheless, there exist limits where one can use $w_{\textrm{el}}(\mathbf{q},\omega)\sim w_{\textrm{el}}(\mathbf{q},0)$. They correspond to the cases where the coupling is screened statically by graphene and BN (large $|\mathbf{q}|$ limit), or just by BN (small $|\mathbf{q}|$ limit). Only in these asymptotic limits, decoupling the electron-phonon and electron-plasmon contribution to the total coupling is not ambiguous. E.g., one could determine determine a screened electron-phonon coupling directly using $\epsilon^{-1}_{\textrm{el}}$ \cite{Verdi2017}. The same conclusions can be obtained looking at the electronic contribution to the screened Coulomb kernel, for a system without atomic polarization. In that case, one finds that the pole residue is related to the electron-plasmon coupling \cite{PhysRevB.94.115208}. 

The general expression Eq. \ref{eq:Pfinal} can nevertheless be simplified in some situations. Bloch overlaps are important to understand selection rules and the localization of the indexes $k$ and $l$.  If e.g. a Bloch electron is delocalized in the out-of-plane direction over the whole structure, such as in the case of a purely BN heterostructure, brackets with $k\neq l$ are all relevant. If instead we are interested in the electron-phonon coupling between graphene's Bloch electrons, which are localized on the graphene's plane, then we must have $k=l=\textrm{Gr}$, since the overlap vanish for the other cases. Moreover, as done for the single layer, we still assume the unperturbed Bloch wavefunctions are even with respect to the layer position, so that in Eq. \ref{eq:Pfinal} one has $i=j=0$. The result of this operation is
\begin{align}
\tau^{-1}_{n\mathbf{k}}=\frac{2\pi}{\hbar}\frac{1}{AN_{\mathbf{q}}}\sum_{\mathbf{q}}\sum_{\substack{i'j'\\kk'l'l}} \langle u_{n\mathbf{k}}|\phi^0_k|u_{m\mathbf{k+q}}\rangle \langle u_{m\mathbf{k+q}}|\phi^0_l|u_{n\mathbf{k}}\rangle \times \nonumber \\
\int_{-\infty}^{\infty} d\omega \textrm{sign}(\omega) \mathcal{A}(\omega) \textrm{Im}\Big[-\frac{1}{\pi}v^{0i'}_{kk'}(\mathbf{q})\chi^{i'j'}_{k'l'}(\mathbf{q},\omega)  v^{j'0}_{l'l}(\mathbf{q})  \Big] \times \nonumber \\ \delta(\hbar\omega+\varepsilon_{m\mathbf{k+q}}-\varepsilon_{n\mathbf{k}}).
 \label{eq:Pgolden}
\end{align}
In the above expression, in the asymptotic limits where we can decouple the electron-phonon from the electron-electron interaction, $-\textrm{Im}v\chi v$ reduces to the squared modulus of the electron-phonon coupling. Motivated by this  observation, we recast the above expression with the following notation
\begin{align}
\tau^{-1}_{n\mathbf{k}}=\frac{2\pi}{\hbar} \frac{1}{N_{\mathbf{q}}}\sum_{\mathbf{q}}\int^{\infty}_{-\infty} d\omega \mathcal{A}(\omega) \delta(\hbar\omega+\varepsilon_{m\mathbf{k+q}}-\varepsilon_{n\mathbf{k}}) \times \nonumber \\
\sum_{kl}\langle u_{n\mathbf{k}}|\phi^0_k|u_{m\mathbf{k+q}}\rangle \langle u_{m\mathbf{k+q}}|\phi^0_l|u_{n\mathbf{k}}\rangle
g^2_{kl}(\mathbf{q},\omega),
 \label{eq:Pgolden2}
\end{align}
where $g^2_{kl}(\mathbf{q},\omega)$ is given by Eq. \ref{eq:g2main}.

\subsection{Practical expression for the remote phonon transport scattering rates}
\label{app:scattBNremote}
Eq. \ref{eq:Pfinal} is the most general expression for the scattering rate of a heterostructure. Neglecting the scalar product of Bloch functions with $\phi^1$, one obtains Eq. \ref{eq:Pgolden2}. \begin{figure}[h]
\includegraphics[width=0.8\columnwidth]{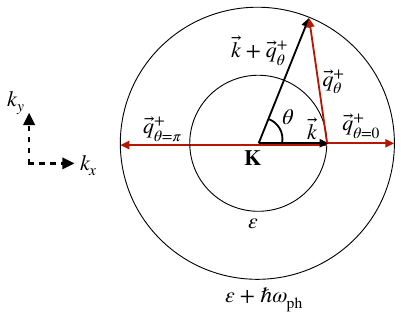}
\caption{Scattering process for an electron with an initial energy $\varepsilon$. Graphical representation  of the isoenergetic surface at $\varepsilon$ and $\varepsilon+\hbar\omega_{\textrm{ph}}$, on which the electron ends up after an absorption of a phonon of frequency $\omega_{\textrm{ph}}$. In this case, we are representing an electron-doped Dirac cone. Several different setups of the Fermi surface and the isoenergetic surfaces are possible in dependence of the relative magnitude of $\varepsilon$ and $\omega_{\textrm{ph}}$, and considering also the emission case. }
\label{fig:qselect}
\end{figure}

Here, we are interested in the scattering rates where the initial and final states are electrons of graphene, and the scattering is mediated by remote phonons. In the Dirac cone approximation, we take $\mathbf{k}$ to be a vector around $\mathbf{\textrm{K}}$. Without loss of generality due to isotropy, we consider a $\mathbf{k}$ on the horizontal axis, as depicted in Fig. \ref{fig:qselect}. In Fig. \ref{fig:qselect} we consider initial electrons belonging to isosurface $\varepsilon$. We then define the $\theta$ angle as the scattering angle between $\mathbf{k}$ and $\mathbf{k+q}$, i.e. $\theta=\theta(\mathbf{q})$. The transport scattering time due to remote phonon scattering is then obtained as
\begin{align}
\tau^{\textrm{tr},-1}_{n\mathbf{k}}=\frac{2\pi}{\hbar} \frac{1}{N_{\mathbf{q}}}\sum_{\mathbf{q}}\int^{\infty}_{-\infty} d\omega \mathcal{A}(\omega) \delta(\hbar\omega+\varepsilon_{m\mathbf{k+q}}-\varepsilon_{n\mathbf{k}}) \times \nonumber \\
\sum_{kl}\langle u_{n\mathbf{k}}|\phi^0_k|u_{m\mathbf{k+q}}\rangle \langle u_{m\mathbf{k+q}}|\phi^0_l|u_{n\mathbf{k}}\rangle g^{2,\textrm{ph}}_{\textrm{Gr}}(\mathbf{q},\omega) \left[1-\cos\theta(\mathbf{q})\right],
 \label{eq:Pgoldentr}
\end{align}
where $g^{2,\textrm{ph}}_{\textrm{Gr}}$ is defined by Eq. \ref{eq:gphcases}. To have more manageable expressions, we now approximate the phonon frequency to be dispersionless, i.e. we assume that $g^{2,\textrm{ph}}_{\textrm{Gr}}$ is peaked at $\omega=\pm\omega_{\textrm{ph}}$ and approximate
\begin{align}
g^{2,\textrm{ph}}_{\textrm{Gr}}(\mathbf{q},\omega)\sim \bar g^{2,\textrm{ph}}_{\textrm{Gr}}(q) \delta(\omega\pm\omega_{\textrm{ph}}).
\end{align}
We remind that the above coupling is of long-range nature, and it is in very good approximation isotropic, as we used in the right hand side of the equation. Given this coupling, if we consider the special case $\varepsilon=\varepsilon_{\textrm{F}}$ the two energy conserving Dirac delta functions of Eq. \ref{eq:Pgoldentr} (for positive and negative frequencies, i.e. emission of absorption) imply that $\mathbf{k+q}$ lies on the isoenergetic lines $|\varepsilon_{\textrm{F}}\pm \hbar \omega_{\textrm{ph}}|$. As presented in Fig. \ref{fig:qselect} for an absorption process, $q$ and $\theta$ are now related geometrically, i.e. we can write $q_{\theta}$. Since for a given $\theta$ we have two different values of $q$ for emission and absorption, we will call them $q^{\pm}_{\theta}$ respectively. They respect the following relations
\begin{align}
q^{\pm}_{\theta=0} &\leq q^{\pm}_{\theta} \leq q^{\pm}_{\theta=\pi}, \\
\frac{q^{\pm}_{\theta=0}}{k_{\rm F}} &= \frac{\big| |\varepsilon_{\rm F} \pm \hbar \omega_{\rm ph}| - |\varepsilon_{\rm F}| \big|}{|\varepsilon_{\rm F}|}, \\
\frac{q^{\pm}_{\theta=\pi}}{k_{\rm F}} &= \frac{ \big||\varepsilon_{\rm F}| + |\varepsilon_{\rm F} \pm \hbar \omega_{\rm ph}| \big|}{|\varepsilon_{\rm F}|}.
\end{align}
Now, the sum over $\mathbf{q}$ can then be transformed in just an angular integration, keeping in mind that the volume element $qdq$ will bring a factor $|\varepsilon_{\textrm{F}}\pm \omega_{\textrm{ph}}|$. The Bloch overlaps of graphene can be written in the form given below Eq. \ref{eq:Qelfree}. With algebraic rewriting of the statistical occupation factors, at last we obtain Eq. \ref{eq:scattBN}, if we choose $\omega_{\textrm{ph}}=\omega_{\textrm{BN}}$.

\section{Computational details}
\label{app:compdets}
We here detail the numerical computation of the ingredients of the VED method, and compare our results to previous approaches in Fig. \ref{fig:Qph_compa}.
\subsubsection{DFT calculations}

DFT and DFPT calculations of ground states, dielectric responses and phonons are carried out with the QUANTUM ESPRESSO package \cite{Giannozzi2009,Giannozzi2017}, with 2d boundary conditions \cite{Sohier2017}, and optimized norm-conserving Vanderbilt pseudopotentials \cite{Hamann2013} from the pseudo-DOJO library \cite{vanSetten2018}.

\subsubsection{Single-layer static electronic response from DFPT}

The parametrization of single layer static electronic responses is  detailed in \cite{Sohier2021a}. In the framework of DFPT, the layer is perturbed with a monochromatic in-plane periodic function of the form $V^i(q, z) = V_0 \phi^i$. The density response $\rho^i$ is extracted from DFPT, renormalized by $V_0$ and integrated to obtain $Q^i(q)$. Examples of plots of $Q^i(q)$ can be found in Fig. 9 of Ref. \cite{Sohier2021a}. The profile function is then defined as $\rho^i(q, z)/Q^i(q)$. A small $q$-dependence of the profile function exists in practice, but has a small impact on the observables. 

\subsubsection{In-plane contribution to atomic response }

We here compare the approximation of Eq. \ref{eq:eDe} to the one of Eq. \ref{eq:omLO}. In Eq. \ref{eq:eDe} the potential is computed using a procedure similar to Ref. \cite{Macheda2023}---an exact procedure may be developed following Ref. \cite{Royo2021}, but we leave it for further developments. 
% Namely, we run a DFPT self-consistent cycle in which we set to zero the macroscopic in-plane  components ($G^2_x+G^2_y = 0$) of: i) the change of the local part of pseudopotential and ii) the change of the Hartree potential. This DFPT cycle provides the density response without macroscopic ($G^2_x+G^2_y = 0$) screening. The associated Hartree potential is then combined with the full change of the potential generated by the ions to build $\frac{\partial \bar{V}_{\textrm{KS}}(q,z)}{\partial \mathbf{u}_{q,s}}$.
As anticipated in the main text, this assumes that the LO potentials are generated by a charge density with the same profile as the electronic response $f^0_{\textrm{BN}}$. In practice that is not exactly true. We find that the difference of implementing Eq. \ref{eq:eDe} instead of Eq. \ref{eq:omLO} (which is expected to be a better approximation) is order of 5\%.

\begin{figure}[h]
\includegraphics[width=0.98\columnwidth]{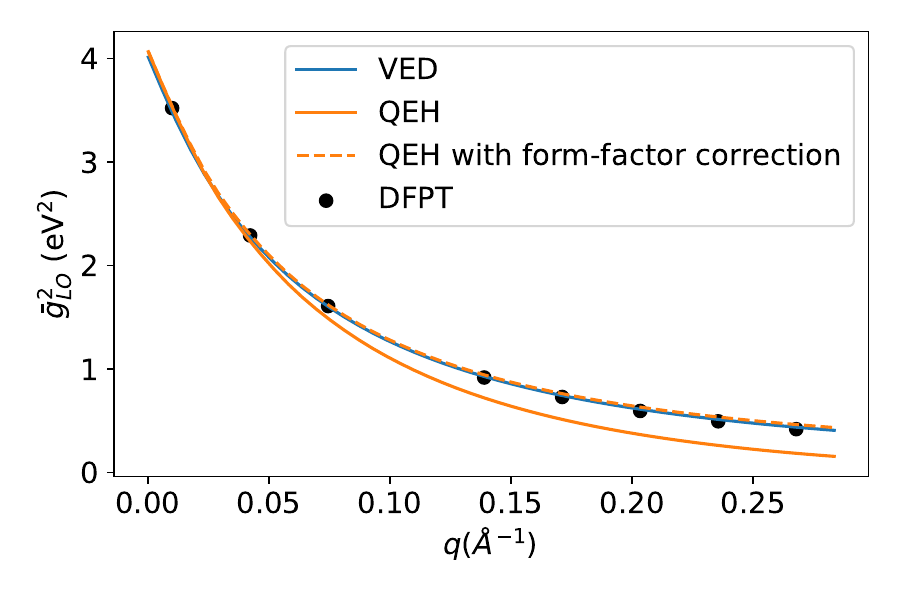}
\includegraphics[width=0.98\columnwidth]{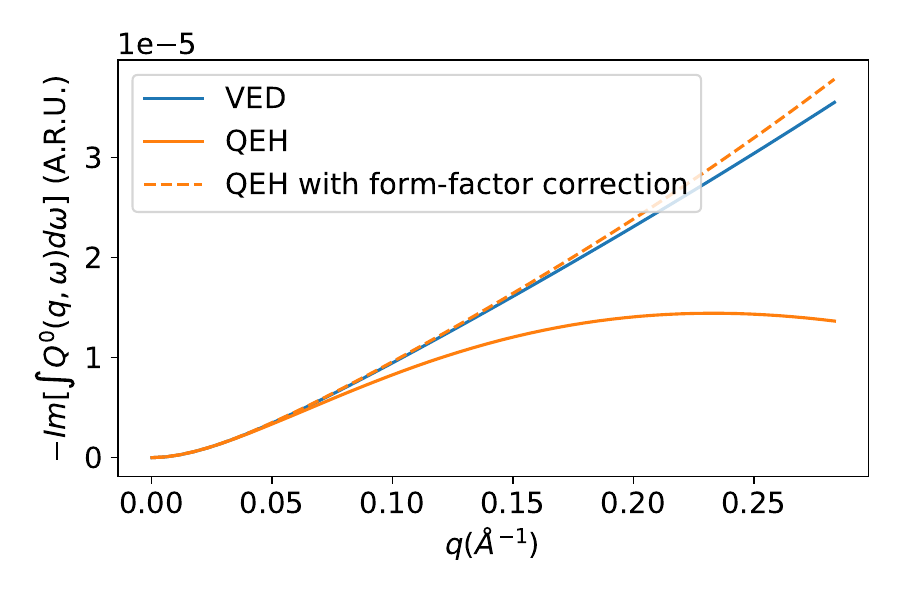}
\caption{(Top) Comparison of $\bar g^2$ in single layer BN, obtained via the VED model and the QEH model of Ref. \cite{Gjerding2020}, reducing to the LO phonon coupling. The discrepancy is attributed to the different evaluation of $-\textrm{Im}\left[\int_{0}^{\infty}Q^0(q, \omega) d\omega \right]$ (Bottom). Notice that the QEH error is  up to $38\%$ at the largest value of q with respect to DFPT values, that are on top of the results of this work. %There are two sources of errors: i) the use of $\Gamma$ Born effective charges rather that the full momentum dependent effective charges as done in the VED; ii) the use, in the QEH method, of the strictly 2d Coulomb kernel $\frac{2\pi}{q}$ rather than the more appropriate $v^{00}_{\textrm{BN,BN}}$ (which includes form factors) to deduce the interacting response function $Q^0(q, \omega)$ from the irreducible response functions of electrons and phonons.
There main source of error is the use, in the QEH method, of the strictly 2d Coulomb kernel $\frac{2\pi}{q}$ rather than the more appropriate $v^{00}_{\textrm{BN,BN}}$ to deduce the interacting response function $Q^0(q, \omega)$ from the irreducible response functions of electrons and phonons.
Correcting this error, the QEH methodology is very near to our results.}
\label{fig:Qph_compa}
\end{figure}

\subsection{Neglect of out-of-plane atomic contribution}
\label{app:Q1appr}

\begin{figure}[h]
\includegraphics[width=0.98\columnwidth]{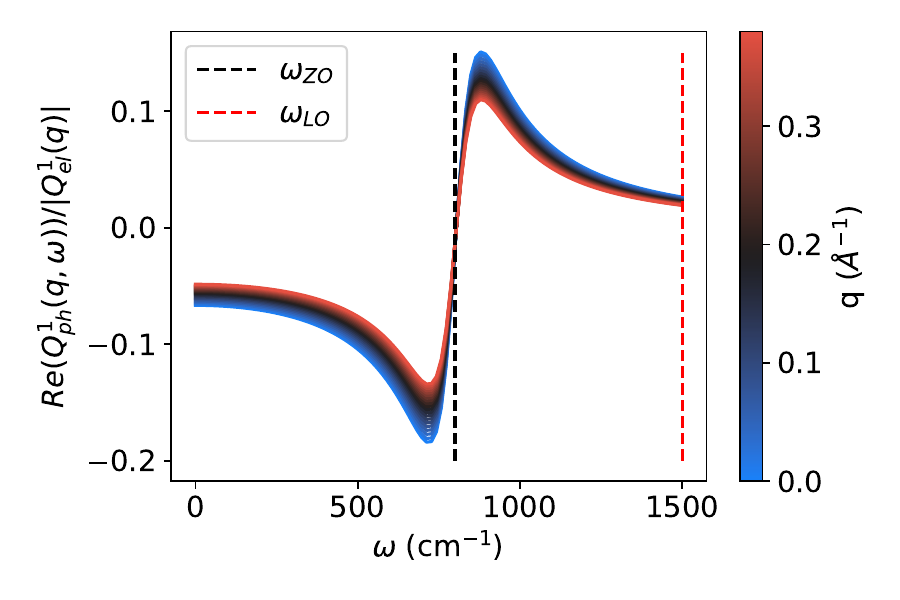}
\caption{ 
$Re(Q^1_{\textrm{ph}}(q,\omega))/|Q^1_{\textrm{el}}(q)|$ 
as a function of $\omega$, for the wavevectors relevant for this work. For the ZO mode at $\sim 800$cm$^{-1}$ (black line), $\eta_{\textrm{ZO}}\sim80$cm$^{-1}$ is employed for representation purposes, the values at $\omega=\omega_{\textrm{LO}}$ (red line) not being influenced by this choice. We notice that the estimated upper limit of the ratio between the atomic and electronic contribution reaches a small value of $\sim 5\%$ at $\omega=\omega_{\textrm{LO}}$.}
\label{fig:Q1ph}
\end{figure}

An upper limit estimation of the atomic contribution to Eq. \ref{eq:Q1} in the long wavelength limit is given by
\begin{align}
Q^1_{\textrm{ph}}(q,\omega)\sim \frac{1}{AM^*}\frac{|Z^{\perp}|^2}{(\omega+i\eta_{\textrm{ZO}})^2-\omega^2_{\textrm{{ZO}}}},
\label{eq:Q1ph}
\end{align}
where $M^*$ is the reduced mass, we have neglected the wavevector dependence of the phonon frequency $\omega_{\textrm{ZO}}\sim 800$cm$^{-1}$, and $|Z^{\perp}|=0.245$ is the absolute value of the Born effective charges of BN as computed from DFPT calculations performed at $q=0$. With respect to the expression for the in-plane atomic response Eq. \ref{eq:Q0phlr}, in Eq. \ref{eq:Q1ph} we have approximated the out-of-plane dielectric function to unity, thus expecting an overestimation of $Q^1_{\textrm{ph}}(q,\omega)$, in line with what discussed in the previous Section in Fig. \ref{fig:Qph_compa}. The smearing is taken as at an unphysical value of $\eta_{\textrm{ZO}}\sim80$cm$^{-1}$, for representation aims. In Fig. \ref{fig:Q1ph} we plot the real part of the ratio $Q^1_{\textrm{ph}}(q,\omega)/|Q^1_{\textrm{el}}(q)|$.

As seen from the plot, Eq. \ref{eq:Q1ph} has a 5\% influence on the total value of $Q^1$ at $\omega=\omega_{\textrm{LO}}$ (vertical red line), meaning Eq. \ref{eq:Q1} is a good approximation for the BN single layer. Nevertheless, contributions may add up when adding several layers in a vdWH, i.e. going towards the hexagonal bulk limit. Even in this case though, the atomic contribution to $Q^1_{\textrm{ph}}$ is negligible. In fact, we can write
\begin{align}
\epsilon^{zz}_{0}=\epsilon^{zz}_{\infty}+\frac{4\pi e^2}{V}\frac{|Z^{zz}|^2}{M^*\omega^2_{\textrm{ZO}}},
\end{align}
where $|Z^{zz}|=0.76$ and $\epsilon^{zz}_{\infty}=2.729$ as from DFPT calculations at zone centers, concluding that $(\epsilon^{zz}_0-\epsilon^{zz}_{\infty})/\epsilon^{zz}_{\infty}\sim 7\%$. Even in this case, the effect of the ZO degrees of freedom is small on the total value of the dielectric function. Since the ZO phonon frequency is roughly equidistant from $\omega=0$ and $\omega_{\textrm{LO}}$, the approximation of Eq. \ref{eq:Q1} remains good for the determination of LO dispersions even when stacking multiple BN layers.

\section{Comparison of screened electron-phonon coupling in VED and DFPT}
\label{app:GrBNcheck}

When considering only the static limit of electronic responses, the VED method reproduces electron-phonon interactions obtained in DFPT calculations in the Born-Oppenheimer approximation. The $g$ coupling is extracted from DFPT for a given mode $\mu$ on layer k as 
\begin{align}
g^{\textrm{DFPT}}_{\mu, k}(\mathbf{q})=\Big{|}\int dz f_{k}(z)V_{\mu}(\mathbf{q},z)\Big{|},
\label{eq:average}
\end{align}
where $\bar{V}_{\mu} (\mathbf{q},z) =  \sum_s \frac{\mathbf{e}_{\mu \mathbf{q}}^{s}}{\sqrt{M_s}} \cdot \frac{\partial \bar{V}_{\textrm{KS}}(\mathbf{q},z)}{\partial \mathbf{u}^{\mathbf{q}}_{s}}$. In the above expression we can take the modulus because the coupling is always positive. From a practical point of view, it is implemented to remove the phase uncertainty of the phonon polarization vectors coming out from first-principles calculations. Notice that in DFPT the perturbation is explicitly inserted in the system via a displacement of the atoms, therefore we cannot choose the layer on which the external perturbation is applied. This is the reason why Eq. \ref{eq:average} presents only one layer index, that is treated as the first index (probe index) of $g^2_{kl}$.
\\
In this Section, we consider a simple system made of BN and graphene. Ground state DFT calculations are performed with a non-uniform grid of momenta to properly sample the small Fermi surface of graphene and be able to use a electronic smearing corresponding to the Fermi-Dirac distribution at room temperature. The non-uniform grid varies from $12 \times 12$ to the equivalent of $196 \times 196$ around the Fermi surface.
Graphene is doped with an extra density of electrons $n = 10^{13}$ cm$^{-2}$, corresponding to a Fermi level $\sim 0.3$ eV. Phonons are simulated at 5 momenta in the $\Gamma - \textrm{M}$ direction. We then select the electron-phonon coupling coming from the (only) LO mode of BN.

To compute the comparable quantity in the VED method, we use the static limit of the electronic response of graphene's electrons, computed directly in the single layer graphene's system with the same parameters as the BN/Graphene system. We then compute $\bar g^2_{\textrm{Gr}}(q)$ and $\bar g^2_{\textrm{BN}}(q)$ from Eq. \ref{eq:el-EDMcoupling}.

\begin{figure}[h]
\includegraphics[width=0.98\columnwidth]{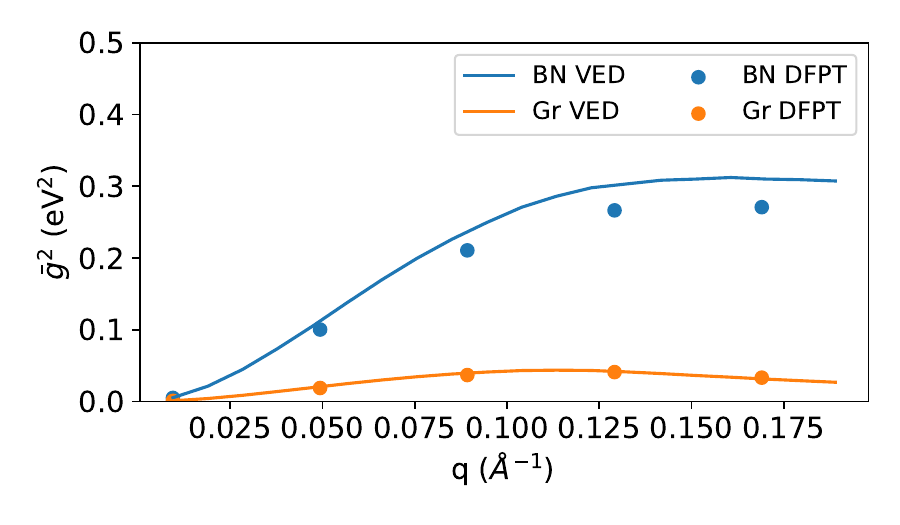}
\caption{Comparison of electron-phonon coupling matrix elements in VED ($\bar  g^2_{\textrm{Gr}}$ and $\bar g^2_{\textrm{BN}}$) and DFPT results in BN/Gr. In the VED method, the static limit of graphene's electronic response is used, so that the coupling is only of electron-phonon type. }
\label{fig:GrBNcheck}
\end{figure}

Fig. \ref{fig:GrBNcheck} shows a good agreement between direct DFPT and the VED method for the main quantity of interest in this work $\bar g^2_{\textrm{Gr}}(q)$.
Along with Fig. \ref{fig:g5BN}, this confirms that the VED method reproduces DFPT in the static limit of electronic responses even when metallic screening is present. 
A larger error is observed for $\bar g^2_{\textrm{BN}}(q)$ in this case, with respect to Fig. \ref{fig:g5BN} . This may be explained by the fact that the total effective potential, i.e. the potential generated by LO phonons and screened by electrons, displays much sharper variations in the out-of-plane direction due to the strong metallic screening from graphene. This would enhance errors related to the approximation of atomic profile functions with the electronic ones in the VED model.

\bibliography{biblio}
\end{document}